\documentclass[preprint]{aastex}
\usepackage{apjfonts}
\usepackage{epsfig}
\usepackage{multirow}
\usepackage{subfigure}
\makeatletter

\renewcommand{\@thesubfigure}{(\alph{subfigure})\hskip\subfiglabelskip}
\renewcommand{\@@thesubfigure}{(\alph{subfigure})}
\makeatother
\usepackage{graphicx}

\slugcomment{Draft version - \today}
\begin{document}
\title{Physical Properties of the Narrow-Line Region of Low-Mass Active Galaxies}
\author{Randi R. Ludwig\altaffilmark{1}}
\author{Jenny E. Greene\altaffilmark{2}}
\author{Aaron J. Barth\altaffilmark{3}}
\author{Luis C. Ho\altaffilmark{4}}
\affil{$^1$University of Texas at Austin, Department of Astronomy, 1 University Station, C1400
Austin, TX   78712, USA}
\affil{$^2$Department of Astrophysical Sciences, Princeton University,
Princeton, NJ 08544, USA}
\affil{$^3$ Department of Physics and Astronomy, 4129 Frederick Reines Hall, University of California, Irvine, CA 92697-4575, USA}
\affil{$^4$The Observatories of the Carnegie Institution for Science, 813 Santa Barbara Street, Pasadena, CA 91101, USA}
\shorttitle{Physical Properties of the NLR of Low-Mass AGN}
\shortauthors{LUDWIG et al.}

\email{randi@astro.as.utexas.edu}
\newcommand{\ledd}{\ensuremath{L_{\mathrm{Edd}}}}
\newcommand{\mbh}{\ensuremath{M_\mathrm{BH}}}
\newcommand{\msun}{\ensuremath{{M}_{\odot}}}
\newcommand{\zsun}{\ensuremath{{Z}_{\odot}}}

\newcommand{\per}{\ensuremath{^{-1}}}
\newcommand{\persq}{\ensuremath{^{-2}}}
\newcommand{\percu}{\ensuremath{^{-3}}}

\begin{abstract}


We present spectroscopic observations of 27 active
galactic nuclei (AGN) with some of the lowest black hole (BH) masses known.  We use 
the high spectral resolution and small aperture of our Keck data, taken with the Echellette Spectrograph and Imager, to isolate the narrow-line regions (NLRs) of these low-mass BHs.  We investigate their emission-line properties and compare them with those of AGN with higher-mass black holes.  While we are unable to determine absolute metallicities, some of our objects
plausibly represent examples of the low-metallicity AGN described by
\cite{2006MNRAS.371.1559G}, based on their [\ion{N}{2}]$/$H$\alpha$
ratios and their consistency with the \cite{2008ApJ...681.1183K}
mass-metallicity relation.  We find tentative evidence for steeper far-UV spectral slopes in lower-mass systems.  Overall, NLR emission lines in 
these low-mass AGN exhibit trends similar to those seen in AGN with higher-mass BHs, such as increasing blueshifts and broadening with increasing ionization potential.  Additionally, we see evidence of an intermediate line region whose intensity correlates with $L/\ledd$, as seen in higher-mass AGN.  We highlight the interesting trend that, at least in these low-mass AGN, the [\ion{O}{3}] equivalent width (EW) is highest in symmetric NLR lines with no blue wing.  This trend of increasing [\ion{O}{3}] EW with line symmetry could be explained by a high covering factor of lower ionization gas in the NLR.  In general, low-mass AGN preserve many well-known trends in the structure of the NLR, while exhibiting steeper ionizing continuum slopes and somewhat lower gas-phase metallicities.
\end{abstract}

\section{Introduction}
\label{sec:intro}
Active galactic nuclei (AGN) with relatively small black hole (BH)
masses ($\mbh \lesssim 10^{6}~M_{\odot}$) comprise a demographic that until
recently went relatively unexplored.  It is difficult to find and observe such low-luminosity objects
\citep{2003ApJ...588L..13F, 2004ApJ...607...90B}.  While luminous
quasars with massive central BHs have been studied intensively for
decades \cite[e.g.][]{1992ApJS...80..109B}, only with the advent of
large surveys like the Sloan Digital Sky Survey
\citep[SDSS;][]{2000AJ....120.1579Y} has it become possible to search
for large numbers of low-mass active galaxies
\citep{2004ApJ...610..722G, 2007ApJ...670...92G}.  The number density
and radiative properties of these accreting, low-mass BHs, which we
refer to as low-mass AGN, are vital to understanding the accretion
history of the Universe.  By understanding their radiative properties,
we can constrain the spectral energy distributions used in models of
primordial seed BHs \citep{2009MNRAS.400.1911V}, which may have prompted early galaxy formation \citep[e.g.][and references
therein]{2011ARA&A..49..373B} and were possibly a significant
participant in the reionization of the Universe
\citep{2009ApJ...696L.146M}.  In addition, investigating these
low-mass AGN is necessary to understand the overall demographics of
BHs at low masses, including the low-mass end of the
$M_{\rm BH}-\sigma^{\ast}$ relation, which can further inform our understanding
of galaxy formation and evolution \citep{2002ApJ...574..740T,
  2009ApJ...698..198G}.

We focus on the sample of 174 low-mass AGN selected by
\citet[][; hereafter GH07]{2007ApJ...670...92G} from the SDSS.  Note that these objects can be
technically classified as narrow-line Seyfert 1 (NLS1) galaxies based
on their broad H$\beta$ widths \citep{1985ApJ...297..166O}.  These low-mass AGN and their host
galaxies are well-studied, with stellar velocity dispersion
measurements presented in \cite{2005ApJ...619L.151B} and
\cite{2011ApJ...739...28X}, and host galaxy analysis with the
\textit{Hubble Space Telescope} presented by
\cite{2008ApJ...688..159G} and \cite{2011ApJ...742...68J}.  In
addition, their spectral slopes from the UV to the X-ray have been 
used to assess whether the
spectral energy distribution gets harder when the BH mass is lower
\citep[][Dong et al. in prep]{2007ApJ...656...84G,
  2009ApJ...698.1515D}.  In this paper we use high-resolution
spectroscopy from the Keck telescope to
specifically study the emission-line properties of 27 of these objects.  
The Keck spectra are observed with a smaller aperture than the SDSS data, using a 0.75" x 1" aperture extraction versus 3" fibers used in SDSS, with $\sigma\sim 22~$km~s$^{-1}$ instrumental resolution. This improves our ability to distinguish emission from the narrow-line region (NLR) of the AGN compared to \ion{H}{2} regions in the host galaxy and makes it possible to decompose the broad and narrow components of the permitted lines more clearly than is possible in the SDSS data.  We make comparisons between AGN with low- and high-mass BHs, and investigate the metallicity of the
NLR, the structure of the NLR, and the radiative
properties of the low-mass AGN themselves.

In Sections \S\ref{sec:sample} and \S\ref{sec:method} we present our
high resolution, high signal-to-noise (S/N) observations of a sample
of 27 low-mass AGN and our fitting methodology and measurements.  In
Section \S\ref{sec:metal}, we investigate the gas-phase metallicities
of these AGN using various emission line diagnostics in pursuit of
rare, low-metallicity AGN.  In Section \S\ref{sec:SED}, we discuss the
inferred far-UV continuum slope and discuss implications for the
spectral energy distributions of accreting low-mass BHs in AGN.  In
Section \S\ref{sec:comps}, we investigate low-contrast, weak emission
lines through use of composite spectra.  In particular, we focus on
high-ionization Fe lines and their behavior relative to more
prominent, lower-ionization NLR emission lines.  In Section
\S\ref{sec:concl}, we present a summary of our conclusions.

\section{The Sample}
\label{sec:sample}
The galaxies studied here are a subset of those presented in
\citet{2004ApJ...610..722G,2007ApJ...670...92G}, with the exception of J2156+1103, which was drawn from the sample in \cite{2005ApJ...619L.151B}.  The low-mass
candidates are AGN with broad H$\alpha$ emission ($z<0.352$)
originally selected from the SDSS data releases one
\citep{2000AJ....120.1579Y} and four \citep{2006ApJS..162...38A}.
So-called virial BH masses, $\mbh=fRv^{2}/G$, are derived indirectly
using the dense gas orbiting the BH (the broad-line region or BLR) as
a dynamical tracer.  The full-width at half-maximum (FWHM) of
H$\alpha$ indicates the velocity of the BLR gas and $L_{\mathrm
  H\alpha}$, in conjunction with the radius-luminosity relation
\citep{2006ApJ...644..133B}, gives the radius of the BLR.
Finally, $f$ is a scaling factor that accounts for the unknown
geometry of the BLR, here assumed to be $f=0.75$
\citep{1990agn..conf...57N}.  Our low-mass AGN have been selected to
have an estimated $\mbh<2\times10^{6} M_{\odot}$.  As naively
expected, these low-mass AGN have been found to inhabit low-luminosity
($\sim$ 1 mag below $L^{*}$), late-type host galaxies
\citep{2008ApJ...688..159G, 2011ApJ...742...68J}. 

The spectra in this paper are drawn from the spectra presented in \cite{2005ApJ...619L.151B} and  \cite{2011ApJ...739...28X}, which are subsets of the galaxies from GH07. In those papers the focus was on the $M_{\rm BH}-\sigma^{\ast}$ relation for these systems, while here we focus on the emission line properties.  Note that our paper represents only a subset of the targets presented in \cite{2011ApJ...739...28X} that were available when this project began.  The general properties of each sample are the same as the parent sample in GH07.  The median BH masses for our sample, GH07, and Xiao et al. are as follows:  $\mbh=3.5\times10^{5} \msun$, $1.3\times10^6 \msun$, $9.5\times10^5 \msun$, respectively.  The median H$\alpha$ luminosities are log $L_{H\alpha}=40.7$, 41.1, and 41.1, while the median Eddington ratios are $L/\ledd=0.3$, 0.4, 0.3, respectively.  We calculate the bolometric luminosity by applying the correction in GH07:  $L_{\mathrm{bol}} = 2.34\times10^{44}(L_{H\alpha}/10^{42})^{0.86}$~ergs~s\per, while $\ledd=1.26\times10^{38}$~($\mbh/\msun$).  Using a Kolmogorov-Smirnov (KS) test, we find that our 27 low-mass AGN have statistically indistinguishable redshifts and luminosities compared to \cite{2011ApJ...739...28X}, ($P=0.05$ and $P=0.03$, respectively).  Compared to GH07, our objects have H$\alpha$ luminosities that are likely drawn from the same parent distribution ($P=0.03$).  Finally, the redshift distribution of our sample is statistically different than that of GH07 ($P<0.01$), such that our objects are at somewhat lower redshift (median $z=0.054$ compared to the median $z=0.086$ for GH07).  Note, however, that the lower redshift objects have more reliable BH mass estimates.  In light of these overall similarities, we conclude that our objects are representative of the parent sample of low-mass AGN from GH07.  

Observations were taken on the Echellette Spectrograph and Imager
\citep[ESI;][]{2002PASP..114..851S} on the Keck telescope during
observing runs in 2003, 2004, 2006, and 2008.  The 0.75" slit width and 1" extraction of these Keck spectra potentially allow us to isolate the NLR
better than in the SDSS data by excluding contributions from \ion{H}{2} regions
at larger radii.  The observed wavelength
range for the spectra is 3900 - 10900 \AA{} over 10 echelle orders.
Exposure times ranged from 900 -- 1800 seconds.  Note that the
absolute flux scale of the spectra is uncertain due to slit losses and
non-photometric conditions on some nights, but the relative flux
calibration across the spectrum should be robust since the AGN and standard stars were observed at the parallactic angle.  The average S/N of
our sample at 5500 \AA{} is S/N$=26$, with an instrumental resolution,
$\sigma\sim 22~$km~s$^{-1}$.  The  full data reduction process is
described in more detail in \cite{2008AJ....136.1179B} and
\cite{2011ApJ...739...28X}, which is identical to that used here.  Table~\ref{observed} lists object
information for our low-mass AGN sample.

\section{Methodology}
\label{sec:method}
\subsection{Fitting Procedure}

\begin{figure}[!tp]
\centering
\includegraphics[trim=0mm 0mm 0mm 0mm, clip, scale=0.65,angle=90]{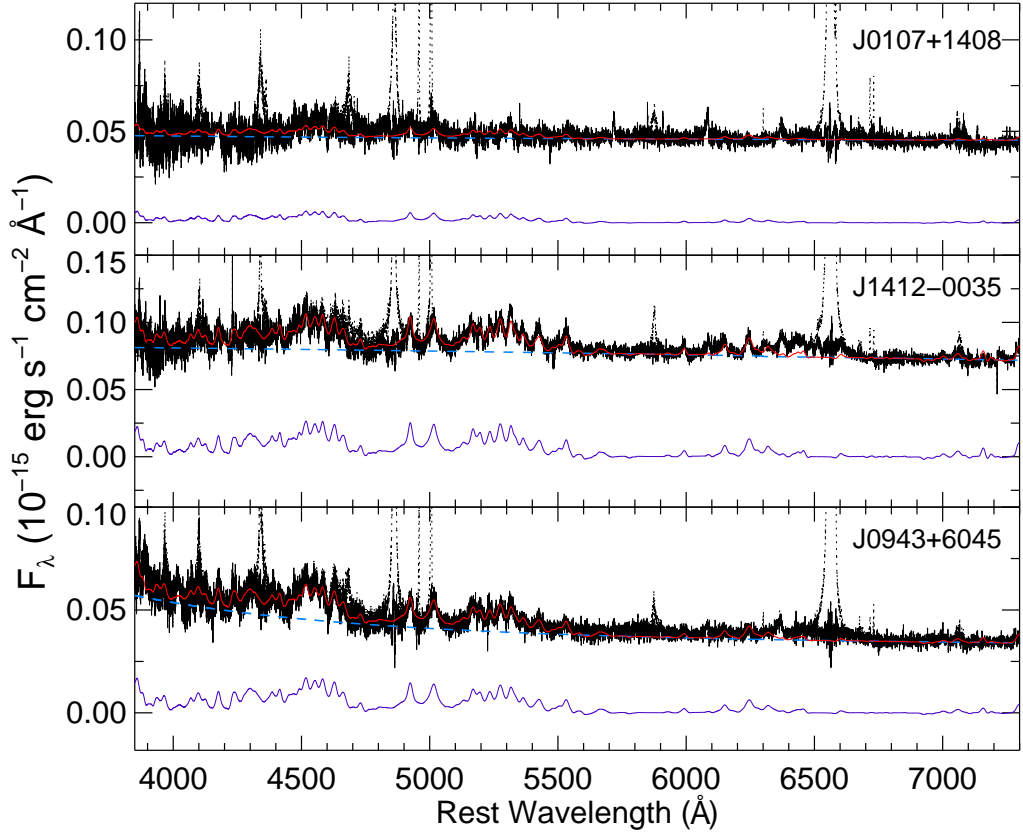}
\caption{\textit{Upper panel.}  Emission line-subtracted spectrum of J0107+1408.  Emission lines are shown with dotted lines, the total continuum fit is shown in red, including the combined reddened power laws as the dashed blue line, and the reddened \ion{Fe}{2} component is shown in purple.  \textit{Middle panel.}  Continuum fit of J1412--0035 with the same convention as above.  \textit{Lower panel.}  Continuum fit of J0943+6045 with same conventions as above.}
\label{AGNfit}
\end{figure}

Our investigation is focused on the properties of the nuclear gas
itself, specifically the emission-line ratios, inferred gas kinematics,
and gas-phase metallicities of the objects at the low end of the BH mass
distribution.  To this end, we want accurate measurements of the
emission-line fluxes and shapes, for which we must remove the
underlying continuum.  At the low nuclear luminosities of most of our
sample, starlight from the galaxy becomes a
significant contributor to the overall continuum, and could bias the
emission-line fluxes if not properly accounted for.  In addition to
the underlying galaxy, a complete model of the AGN continuum
includes a nonstellar power law, typically well-fit by the form
$f_{\lambda} \propto \lambda^{\alpha}$ \citep{2003AJ....126.1131R}.
The optical spectrum also contains emission from numerous broad
\ion{Fe}{2} transitions that form a pseudo-continuum underlying many
other emission lines of interest (see Figure~\ref{AGNfit}).  
Because the components of our fits are coupled, our fitting process
iterates between continuum and emission line fits, as described below.

\begin{figure}[!tp]
\centering
\includegraphics[trim=0mm 0mm 0mm 0mm, clip, scale=0.65,angle=0]{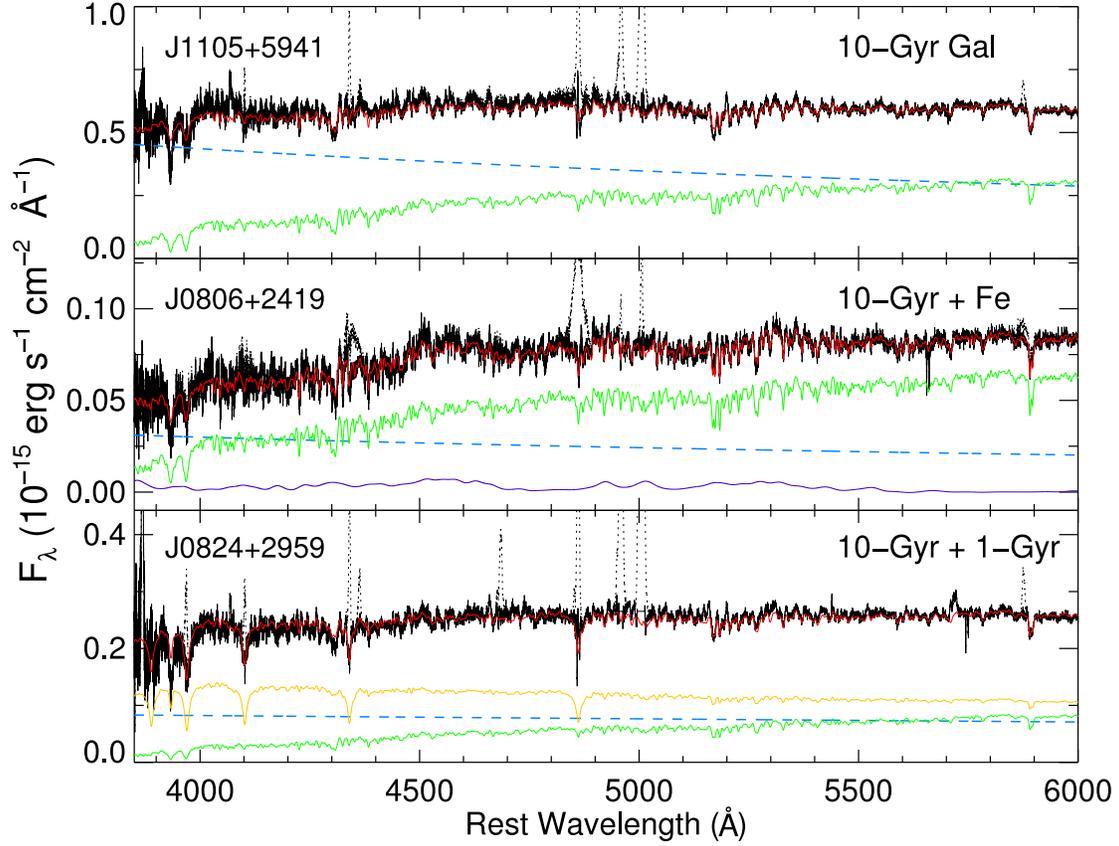}
\caption{\textit{Upper panel.}  Emission line-subtracted spectrum of J1105+5941.  Emission lines are shown with dotted lines, the total continuum fit shown in red, including the reddened power law as the dashed blue line, and the reddened 10-Gyr galaxy component in green.  \textit{Middle panel.}  Spectrum of J0806+2419 with the same convention as above, where the additional reddened \ion{Fe}{2} component is shown in purple.  \textit{Lower panel.}  Spectrum of J0824+2959 with same conventions as above, where the additional reddened 1-Gyr galaxy model is shown in yellow.}
\label{GALfit}
\end{figure}

For computational simplicity, we divide the sample into two classes of
objects -- those where the AGN dominates the continuum and the galaxy
is not generally measurable, and those where the galaxy 
contributes a significant fraction of the continuum (see Figures~\ref{AGNfit} and~\ref{GALfit}).  We treat the fitting of these AGN-dominated and galaxy-dominated spectra separately because there
are strong degeneracies between the power-law shape, the galaxy continuum level,
and extinction.  We use the IRAF routine \texttt{splot} to measure the equivalent width (EW) of the \ion{Ca}{2} K $\lambda 3933$ \AA{} absorption feature in a user-defined window.  We then determine an empirical division between AGN-dominated and galaxy-dominated objects based on this EW.  After inspection of the spectra and comparison to the measured \ion{Ca}{2} EWs, we determine that galaxy-dominated objects typically have \ion{Ca}{2} K
EW~$> 1$ \AA{}, and therefore all such objects are fitted using our galaxy-dominated procedure.\footnote{Two objects with Ca K
  EW$ < 1 $\AA, J1434+0338 and J0931+0635, were moved to the
  galaxy-dominated group after inspection showed \ion{Mg}{1}
  \textit{b} absorption in their spectra.  The Ca K lines were not
  distinctive in these cases because of lower signal-to-noise at the
  blue end of these spectra.}  For both AGN-dominated and galaxy-dominated fitting methods, we start
with an initial continuum fit, then iterate between emission line
fitting and continuum fitting, analogous to the procedure to remove
\ion{Fe}{2} in \cite{2007ApJ...654..754N}.  After three iterations, the H$\alpha$ flux changed by $<0.001$\% for the median objects, but there were two objects where the changes were at the 10\% level at the final iteration.  Still, this 10\% change is within our final errors as described below.  The ratio of H$\beta$ fluxes between iterations also typically converged to within 0.001\% of the final value, but was up to 1\% for two objects.

\subsection{Continuum Fits}

For AGN-dominated objects, the continuum fitting windows included the
following regions: 3900--4330, 4430--4770, 5020--5650, and 7600--7900
\AA.  Examples of our AGN-dominated objects are shown in
Figure~\ref{AGNfit}.  For these objects, we applied a continuum fit
that included two power laws, \ion{Fe}{2} pseudo-continuum, and
reddening.  There are many degeneracies in these fits between internal extinction, Galactic extinction, flux calibration errors, and the intrinsic shapes of the AGN and galaxy continua.  \cite{2005ApJ...619L.151B} and \cite{2011ApJ...739...28X} use a quadratic polynomial in addition to a power-law and galaxy continuum to account for the additional terms.  We have chosen to include a single reddening law instead.  We adopt the reddening prescription in \cite{2000ApJ...533..682C} for star-forming galaxies,
since our objects have blue, disky host galaxies \citep{2008ApJ...688..159G}.  Note that by using the Calzetti et al. reddening law we have implicitly assumed a dust opacity and geometry that may not be applicable to these AGN.  Our original hope was to glean additional physically meaningful information from the fits including extinction.  After comparing the Balmer decrements with the continuum-derived reddening, we determined that due to the degeneracies described above, our continuum-derived extinction terms are not robust.  Therefore, we use the extinction term only to derive satisfactory continuum fits.

Although our spectral region only includes 3800--8500 \AA{}, we find
that two power laws are needed in 38\% of the AGN-dominated objects,
where the power laws are defined as $f_{\lambda} \propto
\lambda^{\alpha}$.  
AGN often show a break in the slope of the 
continuum around 5000 \AA{}, with the blue region steeper than the
red, which is commonly fitted with a broken power law
\citep{2005ApJ...619...41S}.  This upturn in the blue end of the
continuum is not due to the accretion disk emission known as the little
blue bump \citep{1978Natur.272..706S}, which ends around 4000 \AA{},
but represents a real break in the power law continuum.  We cannot tell if this
break is due to increasing galaxy continuum toward the red
\citep{2001AJ....122..549V}, because there are not enough stellar
features to accurately model the galaxy in these AGN-dominated
objects.  Initial parameters for the power laws are based on the H$\alpha$-$L_{5100}$ relation
described in \cite{2005ApJ...630..122G}.  The power-law slopes were free to vary, but were both limited to have indices less than or equal to zero (median index $\alpha=-1.30$) in order to reduce degeneracy with extinction.  For all fits requiring a broken power law, one of the fitted power laws is flat with an index of zero, which dominates the red end of the spectrum, and the other power law dominates the blue end with an average power-law index of $\alpha=-2.26$.  We note again that there is still some degeneracy between the blue power-law slope and the level of extinction.

We fitted the \ion{Fe}{2} pseudo-continuum by scaling, broadening, and
shifting the empirical model from \cite{2002ApJ...565...78B}, which is
based on the prototypical strong \ion{Fe}{2} object I~Zw~1.  Fitting
empirical templates from I~Zw~1 is a well-established method for
dealing with the \ion{Fe}{2} in optical and UV spectra
\citep[e.g.,][]{1992ApJS...80..109B, 2001ApJS..134....1V,2002ApJ...565...78B,
  2006ApJ...644..748L}.  For typical AGN, the I~Zw~1 template fits the
\ion{Fe}{2} pseudo-continuum well \citep{2010ApJS..189...15K}.  While
templates have been made using theoretical \ion{Fe}{2} atomic models
\citep{2000PhDT........10V}, these are limited by the the complexity
of the \ion{Fe}{2} ion and the lack of laboratory data on the atomic
transitions.  We should note that the structure within the
\ion{Fe}{2} pseudo-continuum differs from the structure
of the I~Zw~1 template in several of our galaxies, as seen by
\cite{2010ApJS..189...15K} and \cite{2008ApJ...687...78H}.

For the galaxy-dominated spectra, we applied a continuum fit over the
following wavelength regions to include more stellar absorption lines
than the pure AGN fits: 3900--4300, 5020--5650, 5870--5920,
6150--6530, and 7100--7700 \AA.  Our galaxy-dominated fits included a
single power law, a model galaxy spectrum with an age of 10 Gyr, and
reddening.  The power-law index for these objects was fixed at $\alpha=-1.56$
as in \cite{2001AJ....122..549V} with a starting amplitude based on
the H${\alpha}$-$L_{5100}$ relation \citep{2005ApJ...630..122G}.  This is somewhat steeper than the median index we find for our AGN-dominated objects ($\alpha=-1.30$).  The
galaxy model was made using a \cite{2003MNRAS.344.1000B} single
stellar population model with an age of 10 Gyr, where we assume a
\cite{2003PASP..115..763C} initial mass function, and the metallicity
is assumed to be solar.  Solar metallicity may be an over-estimate for
these galaxies (see Section \S\ref{sec:metal}), but because of the
small aperture of our observations, and the significant dilution from
the AGN continuum, we do not have strong constraints on the stellar
population parameters.  The galaxy model is dominated by emission
from K giants in the optical spectral region.   The continuum fits of  \cite{2005ApJ...619L.151B} and \cite{2011ApJ...739...28X}, which utilize stellar templates rather than single stellar population models, also find continua dominated by K giants.    
The free parameters of this 10-Gyr galaxy
model are a broadening function ($\sigma$), model amplitude, and redshift .  Note that the spectral models have an intrinsic dispersion 
that is larger than the galaxy spectra, and thus 
the fitted $\sigma$ is not physically meaningful. As above, the galaxy
and power law components were multiplied by a reddening function, for
which we varied $A_{\mathrm{V}}$.  

For three of our galaxy-dominated objects, including J1240--0029,
J2156+1103, J0806+2419, individual inspection revealed \ion{Fe}{2}
pseudo-continuum in the spectra, so we fitted them with the previous
galaxy continuum method, but included the \ion{Fe}{2}
pseudo-continuum component as well.  For
four other galaxy-dominated objects (J0824+2959,
J1057+4825, J1143+5500, and J1631+2437) an obvious A star-like
component was necessary in the 3800--4200 \AA{} region to account for
Balmer absorption.  For these objects, we included an additional
galaxy model component, much like the 10-Gyr model, but with an age of
1 Gyr.  We do not find any evidence of systematic differences in host
galaxy morphology \citep{2011ApJ...742...68J} or line ratios for
objects requiring a 1-Gyr galaxy model.  Examples of each fitting
procedure are shown in Figure~\ref{GALfit}.

\subsection{Emission Line Fits}

\begin{figure}[!tp]
\centering
\includegraphics[angle=90,scale=0.65]{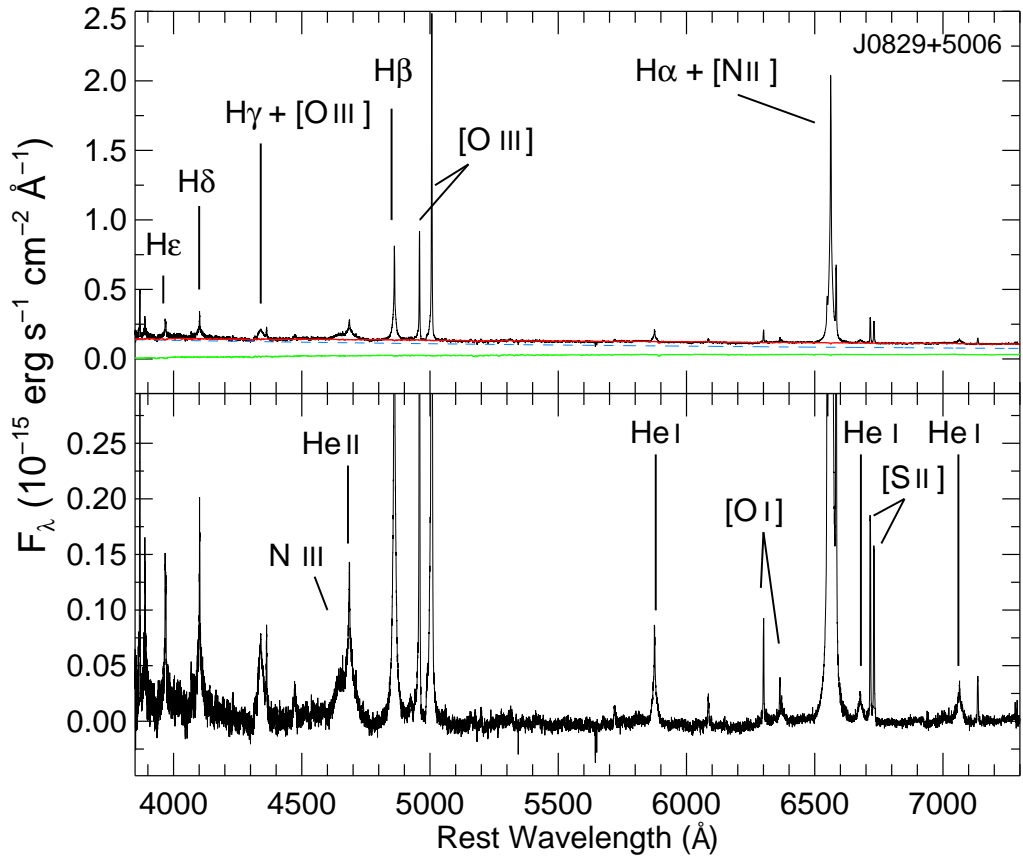}
\caption{\textit{Upper panel.}  Original spectrum of J0829+5006 with total continuum fit shown in red, including a single power law as the dashed-blue line, and 10-Gyr galaxy continuum in green, which have been multiplied by a reddening law.  \textit{Lower panel.}  Zoom in of continuum-subtracted spectrum to enhance the view of weaker emission lines.  All fitted emission lines are labeled.}
\label{linelabel}
\end{figure}

Once the continuum fitting is complete, we have line-only spectra to fit.  The emission line-fitting procedure was the same across all continuum
types.  We fit narrow and broad components to the Balmer lines, from
H${\alpha}$ to H${\epsilon}$, narrow and broad \ion{He}{1} lines
$\lambda$5876, $\lambda$6678, and $\lambda$7065, the [\ion{S}{2}]
$\lambda\lambda$6717, 6730 doublet, the [\ion{N}{2}]
$\lambda\lambda$6548, 6584 doublet, [\ion{O}{1}] lines $\lambda$6300
and $\lambda$6364, [\ion{O}{3}] lines $\lambda\lambda$5007,
4959 and $\lambda$4363, narrow and broad \ion{He}{2} $\lambda$4686, and \ion{N}{3}
$\lambda$4640.  The fitted emission lines are illustrated in
Figure~\ref{linelabel} and are summarized in Table~\ref{tablelines},
where we list the line species, rest wavelengths, and the components
used in the fit.  We present derived quantities of our low-mass AGNs in Table~\ref{tableGHmeas3}, and the emission-line measurements for all of
our objects in Tables~\ref{tableGHmeas1} and~\ref{tableGHmeas2}.  Note that the fluxes presented have only been corrected for Galactic reddening.
Widths of multi-component lines are calculated by constructing the
total fit of the line and measuring the FWHM of this reconstruction.

For each object, we fit the NLR emission lines using sets of Gauss-Hermite functions, much like \cite{2007ApJ...662..131S}.  The expression for a set of Gauss-Hermite functions is as follows:  \\

\begin{equation}
F(x) =A e^{-x^{2}/2} [1+ h_{3}f_{3}(x)+h_{4}f_{4}(x)]\\
\end{equation}
\begin{equation}
f_{3}(x) =\frac{1}{\sqrt{6}}(2\sqrt{2}x^{3}-3\sqrt{2}x)\\
\end{equation}
\begin{equation}
f_{4}(x) =\frac{1}{\sqrt{24}}(4x^{4}-12x^{2}+3)\\
\end{equation}
where \begin{equation}
x=(\lambda-\lambda_{\mathrm{cent}})/\sigma.
\end{equation}

When using a single set of Gauss-Hermite functions, each emission line
is fit with five parameters: line centroid, width ($\sigma$), flux, $h_{3}$,
and $h_{4}$.  The $h_{3}$ parameter quantifies the asymmetry of the
line and $h_{4}$ measures the kurtosis, or boxiness, of the line.  Therefore, we can in principle quantitatively compare NLR line shapes among objects.
This works well for NLR emission with the exception of
[\ion{O}{3}], where 74\% of our objects are not well-described by a
single set of Gauss-Hermite functions, as described below.
In general, we base the NLR line shapes on a fit to the [\ion{S}{2}]
doublet because it is in a relatively clean region of the spectrum, is
not blended with other lines, and is usually strong enough to
constrain the line shape \citep[e.g.][]{1997ApJS..112..391H,
  2004ApJ...610..722G}.  Unless otherwise noted, all NLR lines are
fixed to have the same velocity width, $h_{3}$, and $h_{4}$ as the
[\ion{S}{2}] fit.  The wavelengths for each species are fixed to their
relative laboratory ratios.  The flux ratio of the [\ion{N}{2}]
doublet is fixed to its laboratory value of 2.96.  We also fix the
[\ion{O}{3}] $\lambda$4363 line to have the same shape and number of
components as the [\ion{O}{3}] $\lambda$5007 fit, such that its flux is
the only free parameter.  In two cases, J0806+2419 and 1534+0408, the 
narrow H$\beta$ line was so weak that we imposed a limit that H$\beta$ could not exceed H$\alpha_{\mathrm{NLR}}/3$.

For the [\ion{O}{3}] $\lambda$5007 line, we try an initial fit with
one Gauss-Hermite function for each line, with starting values taken
from the [\ion{S}{2}] fit, but all five parameters free.  As
described above, [\ion{O}{3}] lines are often complex in AGN and can
have strong wavelength shifts and/or asymmetries, so we allow the code
to fit the lines with two summed Gauss-Hermite functions, and again
allow the code to determine if there is a statistically significant
improvement with the additional component.  We find that 74\% of our
spectra require two components for the [\ion{O}{3}] fit, and of those,
85\% had a blueshifted second component.

While we tried fitting the BLR lines with Gauss-Hermite functions as
well, we found that more than one Gauss-Hermite component is required to model the
broad lines.  This is in contrast with
\cite{2007ApJ...662..131S}, where they were able to model H$\alpha$
and H$\beta$ with single sets of Gauss-Hermite functions.  Our observations 
have much higher resolution and S/N than the
SDSS spectra they used, and thus we have robust measurements of two component 
BLRs.  Therefore, we
chose to fit the broad lines with sums of Gaussians, because the
interpretation of summed Gauss-Hermite functions becomes no more
intuitive than sums of Gaussians, and in fact, has more degeneracies
among the fit parameters.

\begin{figure}[!tp]
\centering
\includegraphics[scale=0.85]{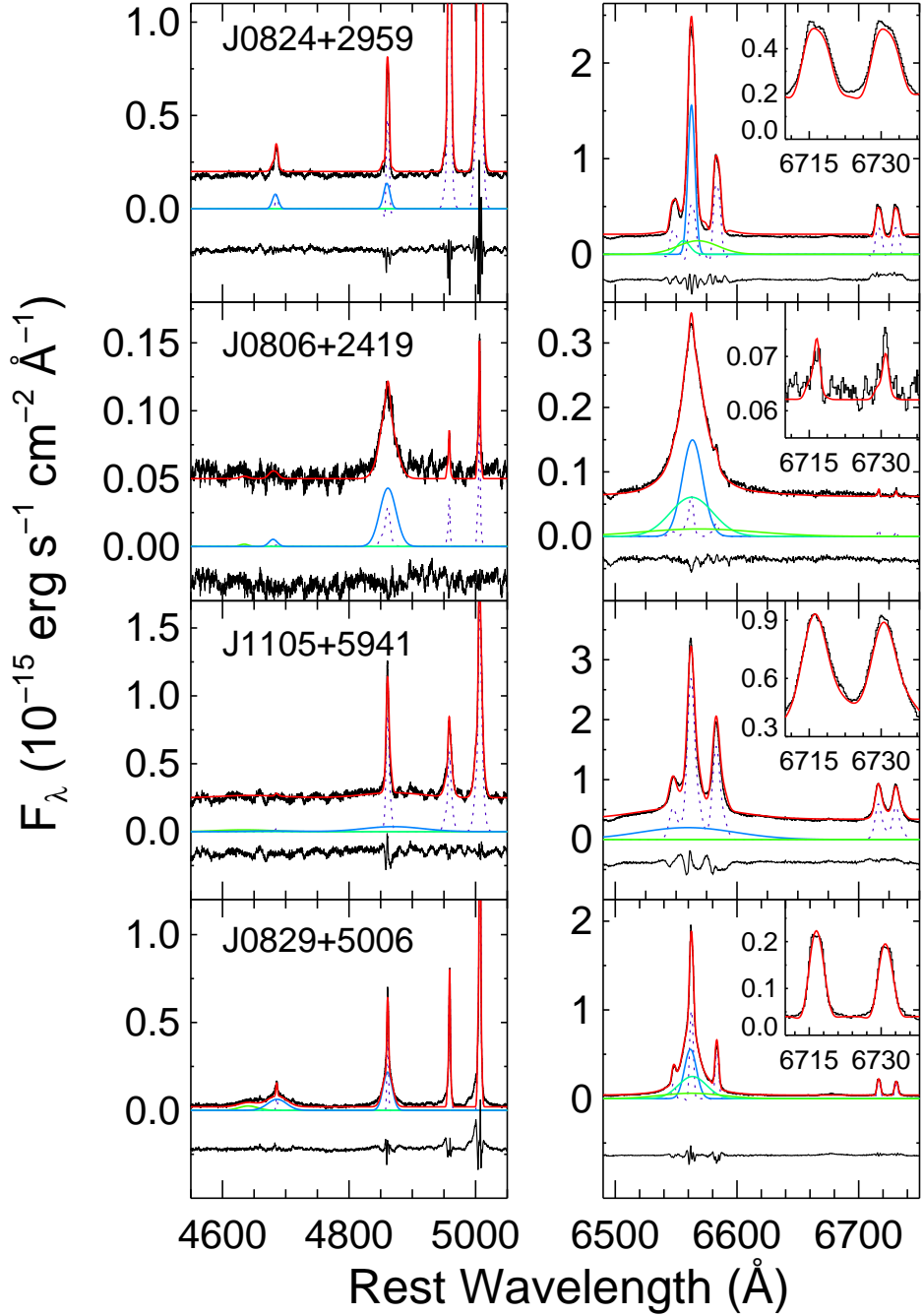}
\caption{\textit{Left column.} H$\beta$ fits, where dotted purple lines show narrow components of H$\beta$, \ion{He}{2} and [\ion{O}{3}].  Blue, cyan, and green lines are broad components of H$\beta$, \ion{He}{2}, and \ion{N}{3}.  \textit{Right column.} H$\alpha$ fits, where dotted purple lines are [\ion{N}{2}] and H$\alpha_{\mathrm {NLR}}$ fits, and blue, cyan, and green show components of H$\alpha_{\mathrm {BLR}}$.  In the insets we zoom in on the [\ion{S}{2}] fit.  For both columns, the panels show (from top to bottom) J0824+2959, J0806+2419, J1105+5941, and J0829+5006.}
\label{linefits}
\end{figure}

For the broad H$\alpha$, we allow the code the freedom to use up to
four Gaussians and to pick the most effective fit for the number of
free parameters included.  Using a criterion inspired by
\cite{2005AJ....129.1783H}, the code determines the benefits of adding
each additional Gaussian to the fit by comparing the number of
additional model parameters to the improvement in $\chi^{2}$.  We apply this
same procedure to the broad H$\beta$, with up to three possible
summed Gaussians.  Examples of our H$\alpha$ and H$\beta$ fits are
shown in Figure~\ref{linefits}.  We then subtract the fit of the
H$\alpha$ complex from the line spectrum, in order to fit the
\ion{He}{1} lines.

\subsection{Error Analysis}
\label{sec:errors}
To estimate uncertainties on our emission line fits, we generate Monte Carlo
simulations of each galaxy spectrum.  To construct the model, we take the measured
values from the final continuum and emission line fits and reconstruct
the spectrum, to which we add Gaussian random noise using the measured error
array.  For each object, we construct 100 such fake spectra and fit
them using our continuum and emission line fitting prescription.  The
final uncertainty is defined by the distribution of measurements of
the artificial spectra.  Specifically, we use the width encapsulating
68\% of the results.  

Of course, we also incur systematic errors in our continuum fitting.  To estimate their magnitude, 
we explore the consequences of removing components from the most complicated fits: those 
requiring a young (1-Gyr old) galaxy component.  We explore fits without the 1-Gyr continuum component and with or without an FeII pseudocontinuum component.  Even in this rather 
extreme case, the flux ratio of the narrow H$\alpha$/H$\beta$ lines typically only varies by 10\% or less across all the different fits, although one object showed a factor of two increase with the addition of the 1-Gyr galaxy component.

We also compare our fitted measurements of the
narrow and broad components of H$\alpha$ to those made by
\cite{2011ApJ...739...28X} on many of the same objects, and find that
93\% of our measured widths are consistent to within 1$\sigma$ and
75\% have fluxes consistent to within 20\% of the measured value.
This provides a valuable additional measure of systematic errors and
is consistent with our measured uncertainties.  

\section{Gas-Phase Metallicity}
\label{sec:metal}

Because AGN are some of the most distant observable objects,
metallicities in AGN are used as a possible tracer of the history
of chemical evolution in the universe \citep{1999ARA&A..37..487H}.
While investigation of high-redshift AGN typically probes the BLR metallicity, here we focus entirely on the metallicity of the NLR,
which should be more representative of the gas-phase metallicity of
the central region of the galaxy.  AGN typically have super-solar
metallicities, as measured with diagnostic emission lines in the NLR.
\cite{2006MNRAS.371.1559G} found that out of a sample of $\sim 23,000$
AGN from SDSS, only 40 showed sub-solar metallicities.  To some
degree, this result is biased by the difficulty of finding low-mass
BHs in late-type, star-forming galaxies with sub-solar gas-phase
metallicities.  Only a handful of low-metallicity, dwarf galaxies which harbor AGN have been detected to date \citep{1999ApJ...520..564K, 2008ApJ...687..133I}.  Thus, we still do not know how intrinsically rare these low-mass BHs
are \citep[e.g.][]{2007ApJ...670...92G}.  We suspect that when
AGN are found in low-mass galaxies, the gas-phase metallicities
will be low and in accord with the mass-metallicity relation of
inactive galaxies \citep[e.g.][]{2004ApJ...613..898T}.  We test that
supposition directly here using the NLR emission lines.

We investigate the gas-phase metallicities of the objects in our
low-mass, low-luminosity sample using several rest-frame optical NLR
emission-line diagnostics.  Because nitrogen is a
secondary element, as the abundances of metals relative to hydrogen
increase, the abundance of nitrogen relative to other metals also
increases.  As a result, the ratio of nitrogen to hydrogen is a good
metallicity diagnostic, and in particular, [\ion{N}{2}]$/$H$\alpha$
has traditionally been used in research on AGN and other nebulae
\citep{1981ARA&A..19...77P, 1986ApJ...309..544E}.  The locus of 
low-metallicity objects is easily seen in the well-known 
Baldwin, Phillips, Terlevich (BPT) diagram \citep{1981PASP...93....5B}.
For a given [\ion{O}{3}]$/$H$\beta$, AGN that have a lower
[\ion{N}{2}]$/$H$\alpha$ are likely at lower relative metallicity.

\begin{figure}[!tp]
\centering
\includegraphics[trim=20mm 25mm 0mm 0mm, clip, scale=0.8]{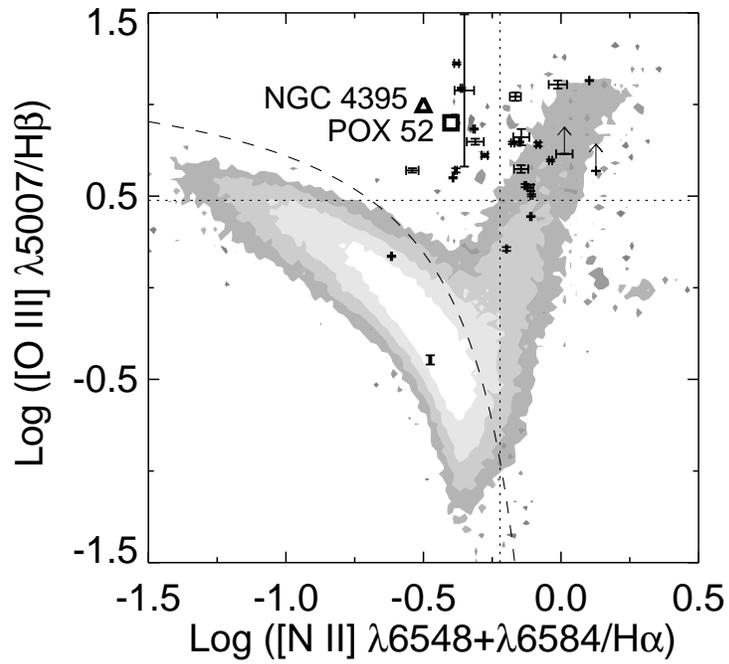}
\caption{BPT diagram of SDSS DR7 galaxy sample in grayscale contours, and our low-mass AGN as crosses, with low-mass AGN NGC 4395 and POX 52 for comparison.  Also plotted is the theoretical \cite{2003MNRAS.346.1055K} divider (dashed) between starburst galaxies (below) and AGN (above), and the dotted \cite{1997ApJS..112..315H} lines, which separate Seyferts (upper right) from LINERs (lower right).}
\label{contour}
\end{figure}

For the sake of comparison to typical AGN, we compare our sample of
low-mass AGN with a subset of the SDSS DR7 galaxy sample at $z \approx 0.1$. 
Spectral measurements for the SDSS subset have been taken from the catalog compiled by the Max-Planck-Institute
for Astrophysics/Johns Hopkins University (MPA/JHU) group\footnote{http://www.mpa-garching.mpg.de/SDSS/DR7/}.  The fitting
process used to construct the catalog is described in
\cite{2004MNRAS.351.1151B} and \cite{2004ApJ...613..898T}.  Note that the catalog selects narrow-line AGN.  We select
only objects whose H$\alpha$, H$\beta$, [\ion{O}{3}], and [\ion{N}{2}]
emission lines had errors $< 10\%$ of the flux of the lines, which
results in a total of 147,816 objects, plotted in grayscale in Figure~\ref{contour}.  Star-forming galaxies lie to the left of the dotted
line at log~[\ion{N}{2}]$/$H$\alpha <-0.2$, and AGN are in the plume
extending to the upper right.  Two of our galaxies, J1143+5500 and J2156+1103, fall into the star-forming region as denoted by the Kauffman et al. dashed line, indicating contamination from \ion{H}{2} regions within the host galaxies.  While the presence of broad emission lines in our objects confirms the presence of an AGN in these two galaxies, they must have substantial contributions from star formation within the host galaxy.  Compared to local AGN in DR7, our sample of
low-mass AGN extend to lower [\ion{N}{2}]$/$H$\alpha$ than
the SDSS AGN, which seems to indicate that the low-mass AGN include objects at relatively
lower metallicities than the bulk of SDSS AGN.  \cite{2011A&A...535A..72H} also see this trend
with the large sample of low-mass SDSS AGN presented in
\cite{2007ApJ...670...92G}.

We want to determine absolute metallicities for our
low-mass AGN to confirm whether they do represent the apparently rare,
sub-solar AGN.  However, it is difficult to define metallicities for
AGN on an absolute scale.  
\cite{2002ApJ...572..753D} and \cite{2004ApJS..153....9G} have developed photoionization models that include the photoionization structure and emission
from dusty, radiation pressure-dominated gas. To this end, they employ MAPPINGS III, a photoionization and shock code whose development is detailed in \cite{1982ApJ...261..183D} and \cite{1993ApJS...88..253S}.  MAPPINGS III considers patches of gas to have the
whole range of ionization structure, thereby producing emission which
contributes to virtually all of the lines observed in an NLR.
\cite{2002ApJ...572..753D} argue that by causing the radiation
pressure to dominate the pressure gradient of the gas, their models
can predict emission line ratios that are consistent with the observed
values seen in real systems \citep{2004ApJS..153...75G}.  Note that MAPPINGS III reports a total metallicity, not a gas-phase metallicity, so the MAPPINGS III metallicities may be systematically higher than other literature values.  For a more
extensive overview of photoionization modeling methods, see also
\citet{2007ASPC..373..511G}.

\begin{figure}[!tp]
\centering
\subfigure{
\includegraphics[trim=10mm 12mm 0mm 10mm, clip, scale=0.45]{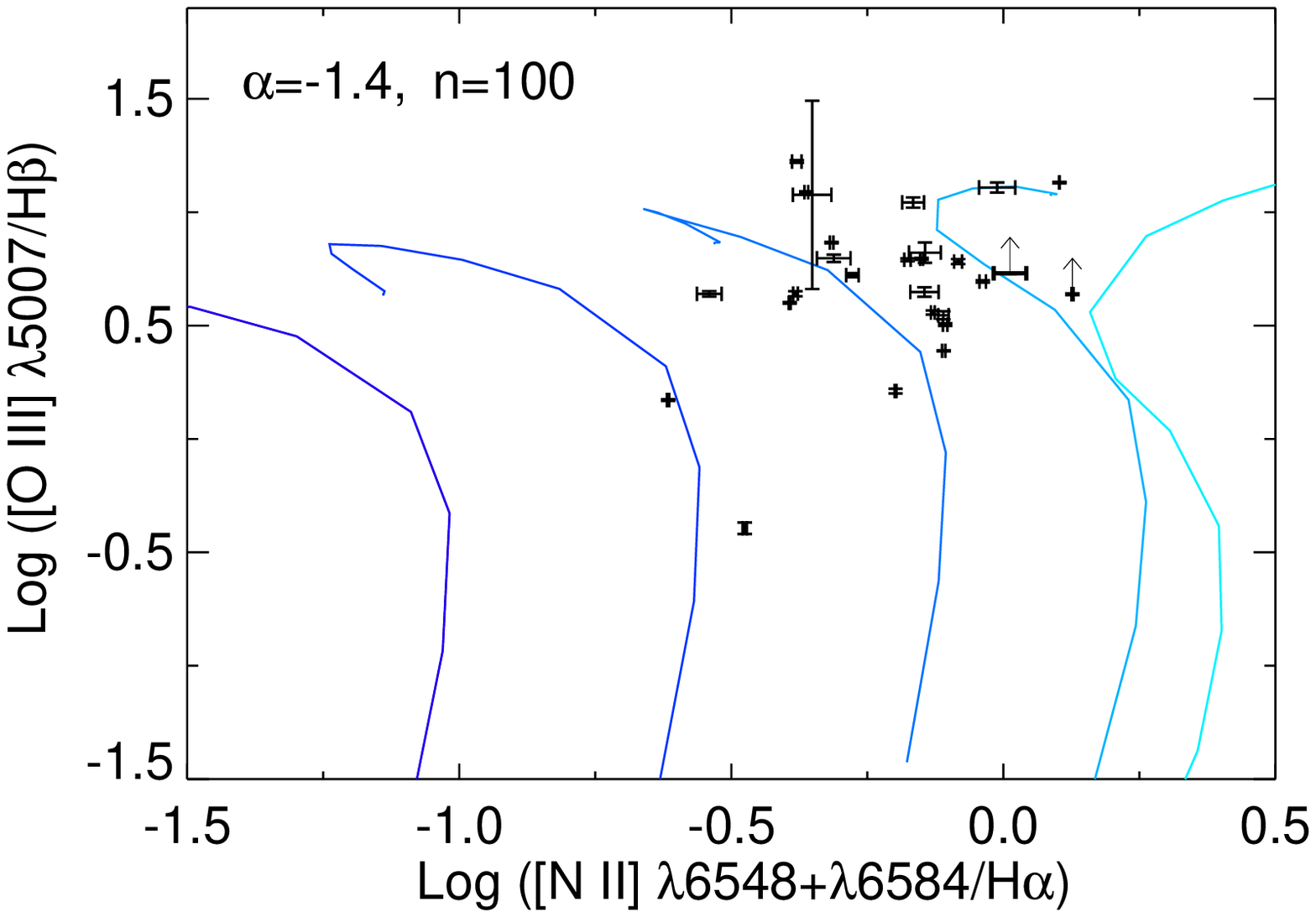}
\label{n100a14}
}
\subfigure{
\includegraphics[trim=20mm 12mm 0mm 10mm, clip, scale=0.45]{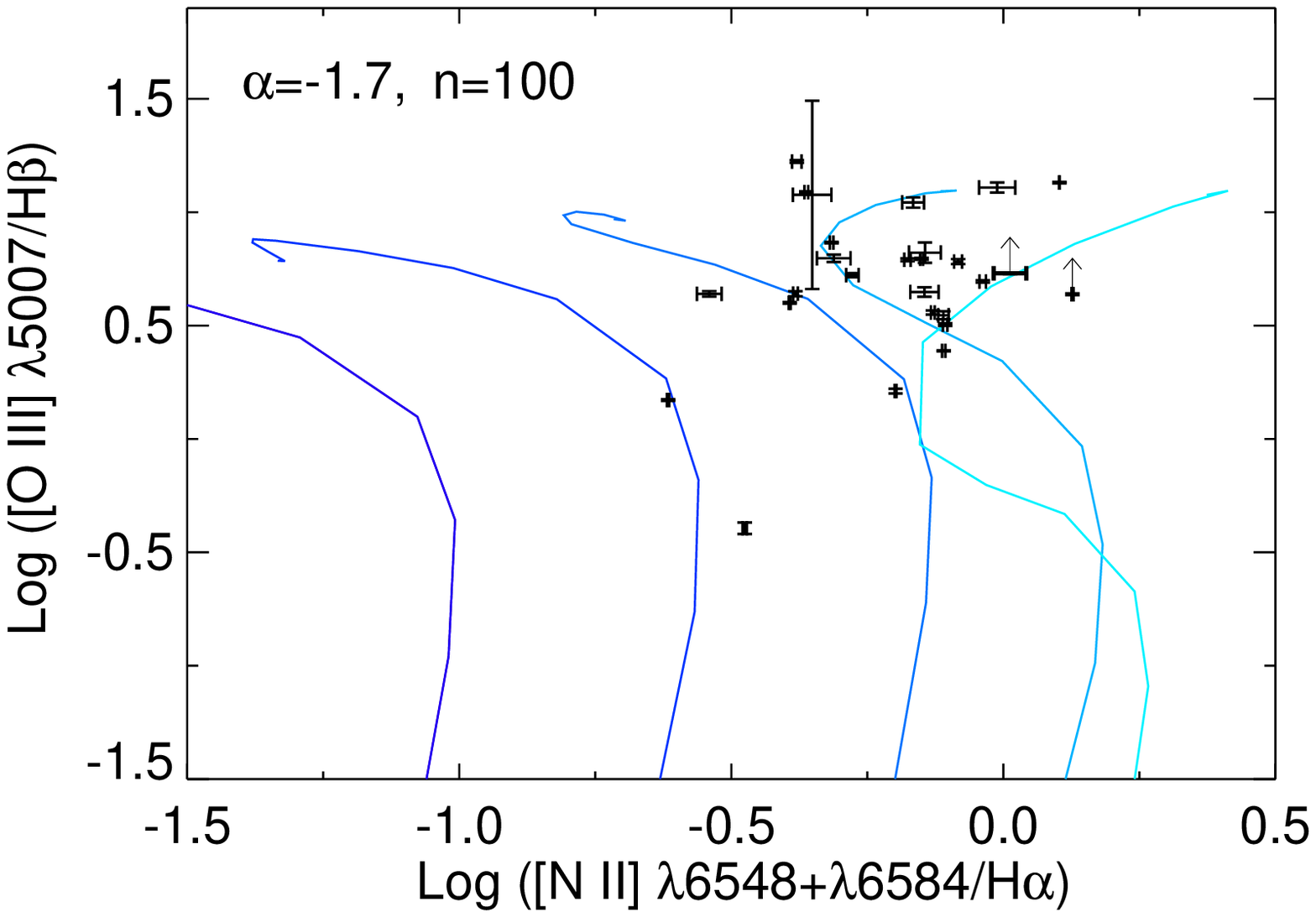}
\label{n100a17}
}
\subfigure{
\includegraphics[trim=10mm 0mm 0mm 10mm, clip, scale=0.45]{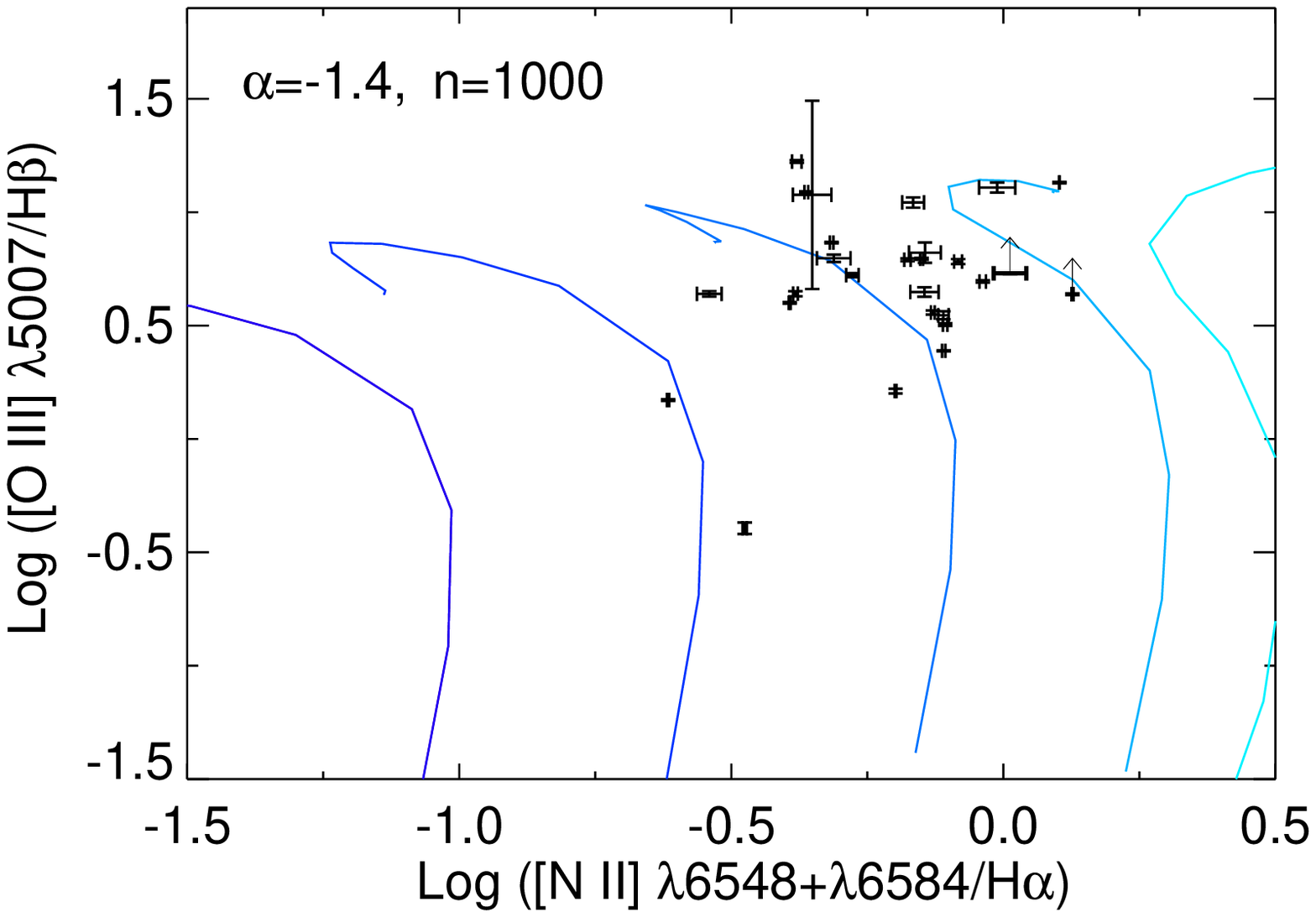}
\label{n1000a14}
}
\subfigure{
\includegraphics[trim=20mm 0mm 0mm 10mm, clip, scale=0.45]{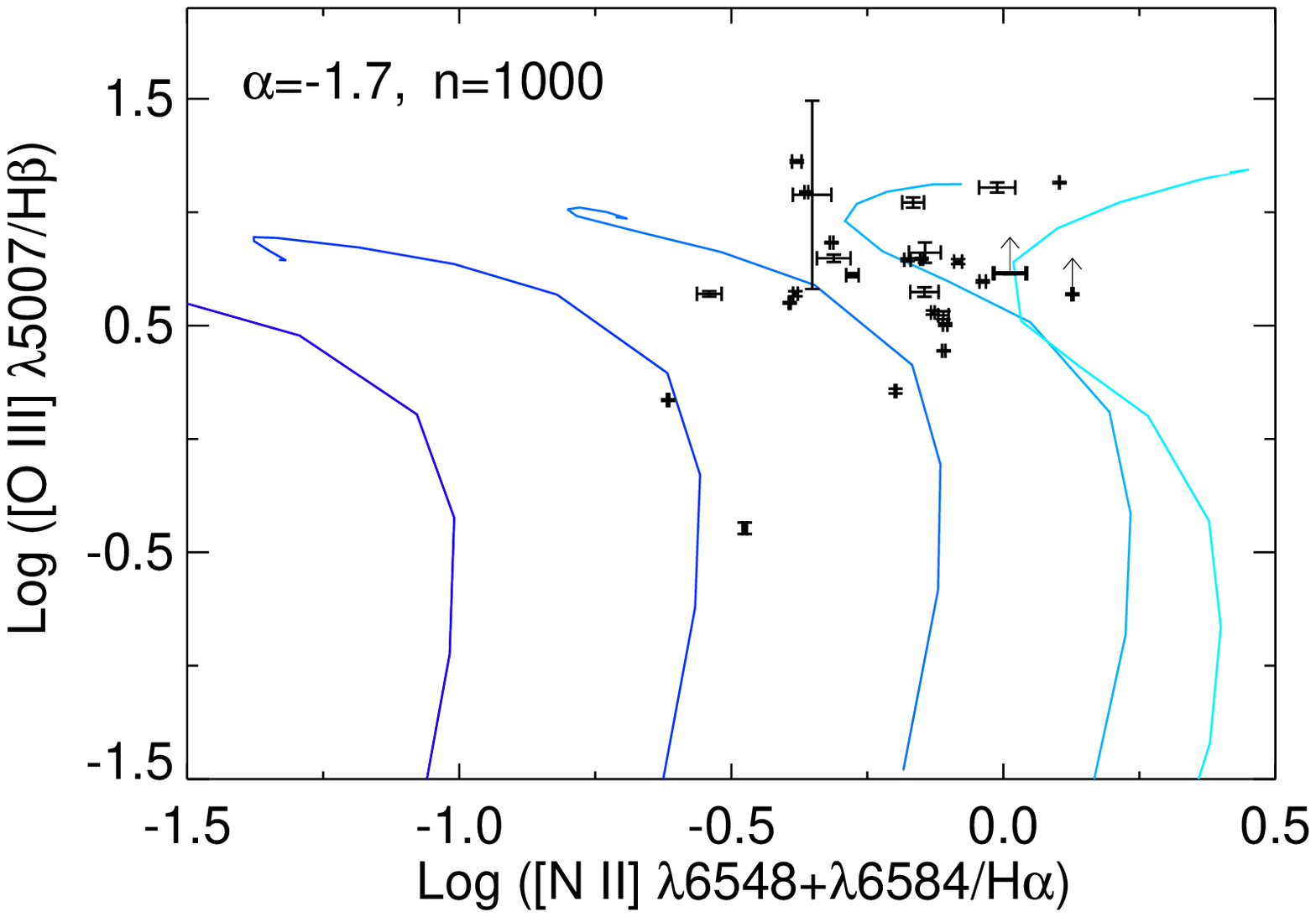}
\label{n1000a17}
}
\caption{MAPPINGS III models used to determine the metallicities in our sample, where $Z=0.25, 0.5, 1, 2,$ and $4~\zsun$, from left to right, and log $U$ ranges from $ -4$ to $0$ from bottom to top (\textit{Top row:} n$=100$~cm\percu, \textit{Bottom row:} n$=1000~$cm\percu, \textit{Left column:} $\alpha = -1.4$, \textit{Right column:} $\alpha = -1.7$).  Using these grids, our sample shows a range of metallicity estimates, with most objects having super-solar metallicities.}
\label{groves}
\end{figure}

We estimate metallicities in our objects by comparing the
[\ion{O}{3}]$/$H$\beta$ vs. [\ion{N}{2}]$/$H$\alpha$ measurements for
our sample of AGN to the predictions from the MAPPINGS III grid of models.
For our sample, we are able to estimate the density in the NLR using
the [\ion{S}{2}] doublet ($100 \lesssim n_{e} \lesssim 1300$, see Table~\ref{tableGHmeas3}) and to constrain the shape of the AGN continuum based on
previous observations of $\alpha$ for AGN ($\alpha= -1.56$; Vanden
Berk et al. 2001).  This allows us to utilize models that are
appropriate for our particular sample.  These dusty,
radiation pressure-dominated models are publicly available within the
library of grids provided in the IDL code \textit{ITERA} \citep{2010NewA...15..614G}.  In Figure~\ref{groves}, we show our
low-mass AGN measurements compared to four grids of models, which
represent the full variety of physical parameters we see in our
objects, including $n_{e}=100$~cm\percu\ and $1000~$cm\percu, and
$\alpha=-1.4$ and $-1.7$. While there are some regions of the grids
that can be described by several models, especially at high log $U$
and steep indices ($\alpha = -1.7$), generally the metallicity
increases with increasing [\ion{N}{2}]$/$H$\alpha$.  The metallicity
estimates derived from the MAPPINGS III code can vary by up to a factor of
three in metallicity, depending on the range of $\alpha$ considered.
MAPPINGS III metallicity estimates for individual objects can range from
sub-solar to several times solar depending on the assumed spectral
shape.  Our metallicity estimates are presented in Table~\ref{tableGHmeas3}.  There are also several objects that lie at the upper edge of the models, which may reflect limitations of the models.

\begin{figure}[!tp]
\centering
\includegraphics[scale=0.8]{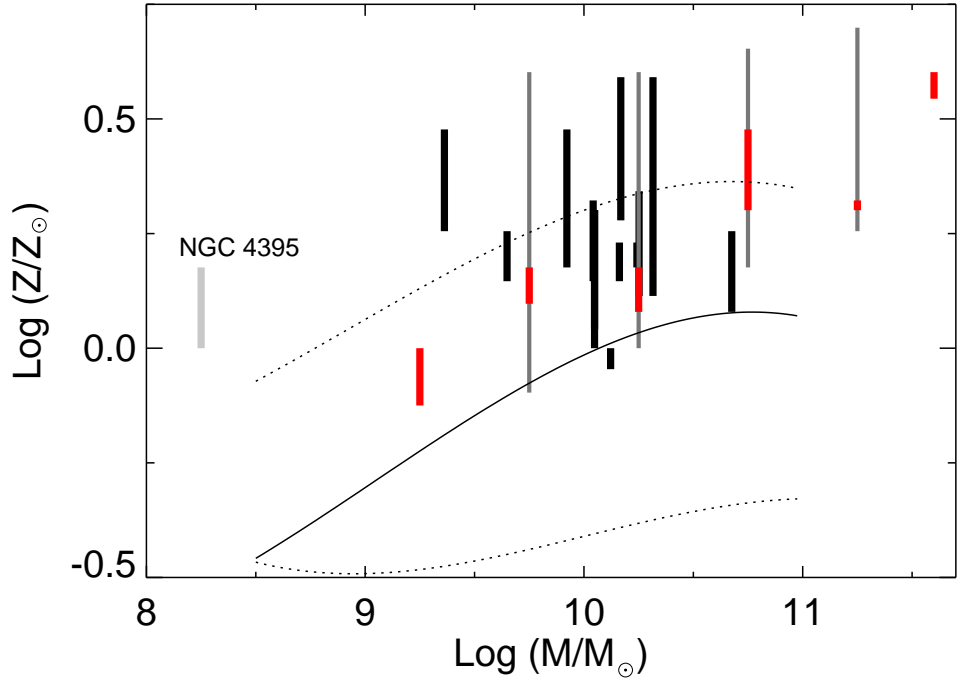}
\caption{Host galaxy stellar mass vs. NLR metallicities for our sample (Fig.~\ref{groves}) as solid black bars.  The range in $Z$ represents the effect of a reasonable range in assumed ionizing slope in the modeling.  Solid red bars are obscured AGN from the SDSS DR7 with metallicities derived in the same way.  The 1$\sigma$ variation over each mass bin is shown in grey.  The solid curve represents the \cite{2008ApJ...681.1183K} stellar mass-metallicity relation based on [\ion{O}{3}] and [\ion{N}{2}], which is bracketed by dotted lines showing the most extreme relations they present (KK04 is uppermost and P05 is lowermost).}
\label{MZ}
\end{figure}

To do a relative comparison, we investigate trends between our
low-mass AGN and the SDSS sample of AGN using the MAPPINGS III
models.   To derive the metallicity for each of our galaxies, 
we consider where an object lies on a particular MAPPINGS III
grid, and interpolate between the two nearest metallicity tracks to
determine a metallicity ($Z$) for that object (see Figure~\ref{groves}).  For each object, we estimate the metallicity from grids
with the nearest density, $n_{e}=100$~cm\percu\ or $n_{e}=1000$~cm\percu.  The
average change in metallicity between models with $n_{e}=100$~cm\percu\
and $n_{e}=1000$~cm\percu\ at fixed $\alpha=-1.4$ is only $0.02~\zsun$, so
density differences are negligible.  However, the difference in metallicity for 
$\alpha=-1.4$ versus $\alpha=-1.7$ is $0.3~\zsun$, and since the
errors in our slopes are large, we incorporate this uncertainty into
our results.  We estimate the metallicity using both $\alpha=-1.4$ and
$\alpha=-1.7$, but exclude estimates for the $\alpha=-1.7$,
$Z=4~\zsun$ model, because it is degenerate with the other models
in the region of interest (see Figure~\ref{groves}).  This gives a
range of $Z$ for each object presented as black bars on Figure~\ref{MZ}.

To highlight the difficulties in deriving absolute metallicities, we 
consider the particular case of NGC 4395, a well-studied, low-mass
AGN.  We have an ESI spectrum of NGC 4395, for which
extensive measurements and Cloudy modeling are published
\citep{1999ApJ...520..564K}, and we find that our observed emission-
line ratios agree with the previous measurements
[log~([\ion{N}{2}]$/$H$\alpha)~\sim~-0.5$,
log~([\ion{O}{3}]$/$H$\beta)~\sim~1.0$].  \citet{1999ApJ...520..564K}
estimate a metallicity of $Z\sim~$0.25~$\zsun$, which is consistent
with estimates made from the \ion{H}{2} regions.  Since then, the
definition of solar metallicity has been revised because of new oxygen
measurements \citep{2004A&A...417..751A}, such that solar metallicity
is now defined to be a factor of two lower than previously thought.
Therefore the estimate of metallicity from \cite{1999ApJ...520..564K}
actually supports a metallicity for NGC 4395 of 0.5~$\zsun$.  To
estimate a metallicity from the MAPPINGS III grids, we need to
determine the density and power-law index for NGC 4395.  
Using the [\ion{S}{2}] $\lambda 6717,
6731$ doublet, we estimate that the density in the NLR of NGC 4395 is
about 1300 cm\percu, 
and \citet{1999ApJ...520..564K} also constrains
$\alpha_{\mathrm ox} \sim -1.7$ for NGC 4395.  Using these
values, we estimate the metallicity of NGC 4395 to be $Z\sim
1.5~\zsun$ from the MAPPINGS III models that have n$=1000~$cm\percu and
$\alpha=-1.7$ (Figure~\ref{n1000a17}).  There is a significant zero-point
offset between the two sets of photoionization models, as has been
well described in other contexts.  Again, we note that the absolute scale of the gas-phase metallicities remain highly uncertain.  For instance, one reason the MAPPINGS III values appear high is that the depleted metals are counted (i.e., we calculate the total, rather than gas-phase metallicity).  Other factors, such as the photoionization modeling methodology, are likely at play as well.  In the following, we use metallicity estimates from the MAPPINGS III models so that we can compare directly 
with \cite{2006MNRAS.371.1559G}, but focus only on relative trends 
between objects rather than absolute metallicities.  Our approach is 
analogous to that advocated by \citet{2008ApJ...681.1183K} for 
star-forming galaxies.

For comparison to AGN with more massive BHs, we selected AGN from the SDSS DR7
galaxy sample described above using the cuts from
\cite{1997ApJS..112..315H} for Seyferts ([\ion{S}{2}]$/$H$\alpha >
0.4$ and [\ion{O}{3}]$/$H$\beta > 3$).  This results in 3747
galaxies, which are binned according to the stellar mass reported by
the MPA/JHU group in increments of 0.5 dex between
$9.0<~$log~$M_{*}/\msun<12.0$.  For each stellar mass bin, we estimate $Z/\zsun$ from the MAPPINGS III grids using the median [\ion{N}{2}]/H$\alpha$ and [\ion{O}{3}]/H$\beta$ value for that bin (red bars in Figure~\ref{MZ} denote the range of metallicities that result from the range of $\alpha$ as above).  We also estimate the range of $Z$
for the central 1$\sigma$ of the objects in each bin in the same way (grey bars
in Figure~\ref{MZ}), except for the highest- and lowest-mass bins which had fewer than 40 objects.  Because these SDSS AGN cover the same
range of densities as our low-mass AGN (300-1500 cm\percu), we
estimate their $Z$ from the same MAPPINGS III grids as described above.
 
We present our relative metallicity comparison in the context of the
$M_{*}$-$Z$ relation in Figure~\ref{MZ}.  The $M_{*}$-$Z$ relation is
only defined for star-forming galaxies and has never been measured in
AGN host galaxies before to our knowledge.  For several of our low-mass AGN, we
derive a stellar mass from the \textit{Hubble Space Telescope} 
\citep{2008ApJ...688..159G} \textit{I}-band luminosities and a
mass-to-light ratio based on the SDSS \textit{g-r} color following
\cite{2003ApJS..149..289B}. These stellar masses and our metallicity estimates are presented in Table~\ref{tableGHmeas3}.\footnote{We also calculate stellar masses using a
fixed mass-to-light ratio of 0.6 in the $I$-band 
\citep[motivated by][]{2008ApJ...688..159G} to estimate the lower limit of the
stellar masses.  With such a low mass-to-light ratio the masses would
decrease by 0.4 dex.}  To expand the low-mass end of the sample, we
include NGC 4395 for comparison, with metallicities estimated from the
MAPPINGS III grids.  We also plot the \citet{2008ApJ...681.1183K}
$M_{*}$-$Z$ relation for star-forming galaxies that is based on
[\ion{O}{3}] and [\ion{N}{2}] (referred to as PP04 O3N2), which is the
closest analog to our method but was derived for a star-forming
spectral energy distribution.  For those low-mass AGN with
$10.0<$~log~$M_{*}/\msun \le 10.5$, which includes most of our
objects, we find that the mean $\langle Z/\zsun \rangle=0.11$, which is
somewhat higher than the $Z$ predicted by the $M_{*}$-$Z$ relation for
star-forming galaxies shown by the solid curve.  In
comparison, the metallicities of our objects agree with the SDSS AGNs at the same
mass.

Even without modeling, our low-mass AGN include objects that have a lower ratio of [\ion{N}{2}]$/$H$\alpha$ relative to the SDSS AGN (Figure~\ref{contour}), which suggests lower metallicities.  This confirms our suspicion that since our host galaxies 
extend to lower mass than the SDSS galaxies, the low-mass AGN include lower-$Z$ objects.  
There appears to be a systematic difference between the active and inactive galaxies, in the
sense that the active ones appear to have higher average metallicities
at fixed stellar mass (Figure~\ref{MZ}).  Either we are seeing systematic differences
due to methodology (i.e. different ionizing spectra) or it is a real
effect.  Since the NLR is compact compared to the interstellar medium
of the galaxy as a whole at these luminosities
\citep[e.g.][]{2003ApJ...597..768S}, we are probing the
high-metallicity central region of the galaxy.  Given well-known
metallicity gradients in spiral galaxies
\citep[e.g.][]{1999PASP..111..919H}, we may be slightly overestimating
the overall gas-phase metallicity in the active galaxies.

Our investigation shows that determining absolute metallicities for
AGN is a very complex problem, and as suggested in other contexts, we
can only trust relative metallicity information derived using the same
methodology across samples. While the AGN in our sample have similar
NLR metallicities as the SDSS AGN at fixed stellar mass, our low-mass AGN include objects at
somewhat lower metallicities than SDSS AGN, at least in part because
they occupy lower-mass host galaxies.  We estimate metallicities using
[\ion{N}{2}]$/$H$\alpha$ and photoionization
models. \cite{2008AJ....136.1179B} present the Seyfert 2 counterparts
to our low-mass Type 1 AGN, which have overlapping
[\ion{N}{2}]$/$H$\alpha$ and [\ion{O}{3}]$/$H$\beta$ values and
therefore similarly likely include low-$Z$ objects as well.  Because of the
uncertainties from the models we cannot say for sure whether our
objects are the rare, sub-solar metallicity objects addressed by
\cite{2006MNRAS.371.1559G}.  Conversely, when directly compared to the
metallicity predictions from the $M_{*}$-$Z$ relation for star-forming
galaxies, AGN are consistent with obeying a comparable
mass-metallicity relation as inactive galaxies, although the scatter
is large.

\subsection{Metallicity Systematics}
\label{sec:metalsys}
As we will see in \S\ref{sec:SED}, there is tentative evidence for a steeper far-UV continuum slope in these systems of $\alpha=-2.0$.  If we assume a steeper slope of $\alpha=-2.0$ for the photoionization models, this would only minimally impact the metallicity estimate of objects with estimates $Z<1~\zsun$, causing $Z$ to increase by $\sim0.2$.  A steeper slope of $\alpha=-2.0$ could potentially increase the measured metallicities by up to a factor of two for objects with $Z>1~\zsun$, but this does not change our previous conclusions that some of these low-mass AGN potentially represent the rare, low-metallicity AGN sought in \cite{2006MNRAS.371.1559G}.

Line emission from star formation in the host galaxy could also potentially affect our metallicity estimates.
To estimate the magnitude of the changes, we take two representative AGN with metallicities of 
solar and twice solar.  We then add varying levels of star formation, assuming gas at the same 
metallicity and deriving the appropriate line ratios from \cite{2004MNRAS.348L..59P}.  For instance, an 
AGN with a solar gas phase metallicity will have [\ion{O}{3}]/H$\beta\approx6$, while the star forming regions will have [\ion{O}{3}]/H$\beta\approx 0.2$.  In the extreme case that 75\% of the Balmer emission comes from star formation, the observed line ratio would be [\ion{O}{3}]/H$\beta=1.7$.  For solar metallicity, we underestimate the metallicity estimates only by 10--30\% for a 25--75\% contribution from star formation to the Balmer line flux.  At twice solar, 50--75\% contribution from star formation leads to a factor of 2--3 underestimate in metallicity.  However, only two sources,  J1143+5500 and J2156+1103, lie far enough from the AGN locus to be consistent with this level of contamination (Fig.~\ref{contour}).  Thus, we conclude that contributions from star formation do not change our metallicity distribution significantly.

\section{Spectral Energy Distribution}
\label{sec:SED}

In theory, the lower masses of these BHs should correspond to a hotter
accretion disk temperature and thus measurable differences in the
spectral energy distributions (SEDs) of our sample from those of more
typical AGN.  Testing this expectation will help to bridge the
gap in BH behavior and accretion theory from stellar mass BHs to the
largest SMBHs.  In addition, the nature of the accretion disk
surrounding these low-mass AGN can inform our understanding of the
primordial intermediate-mass BHs that formed within the first galaxies and
participated in the reionization of the Universe
\citep{2009ApJ...696L.146M}.  We know that the SEDs of accretion disks
around stellar mass BHs are harder and flatter than those of quasars
\citep{2002apa..book.....F}. In fact, the peak blackbody temperature
for disks in AGN with low-mass BHs may also move into the X-ray band
\citep{2012MNRAS.420.1848D}.  We have seen hints of SED changes in
this sample in previous work, both in $\alpha_{\mathrm ox}$
\citep{2009ApJ...698.1515D} and in EW$_{H\beta}$
\citep{2002MNRAS.337..275C, 2005ApJ...630..122G}.  While most of our
objects have been studied at radio \citep{2006ApJ...636...56G} and
X-ray \citep{2007ApJ...656...84G, 2009ApJ...698.1515D,
  2009MNRAS.394..443M} wavelengths, we have no direct constraints on
the far-UV continuum.  We can gain an indirect handle on the accretion
disk shape, however, since those photons are responsible for
photoionizing the emission line gas that we observe in the optical.

Specifically, we can infer the slope of the continuum in the far UV by
contrasting the \ion{He}{2} and narrow H$\beta$ components.  As
described in \cite{1978MNRAS.183..479P}, since both H$\beta$ and
\ion{He}{2} are recombination lines, the ratio of their relative
strengths should yield a measure of the relative intensity of the
continuum at 912 \AA{} and 228 \AA{}, the ionization edge for
\ion{He}{2}.  Assuming the ionizing source is well-described by a
power law of the form $I_{\nu} \propto \nu^{\alpha_{\rm UV}}$, the intensities
should behave as follows:

\begin{equation}
\frac{I_{4686}}{I_{H\beta}}=1.99 \times 4^{\alpha_{\rm UV}}
\end{equation}

\begin{figure}[!tp]
\centering
\includegraphics[scale=0.6]{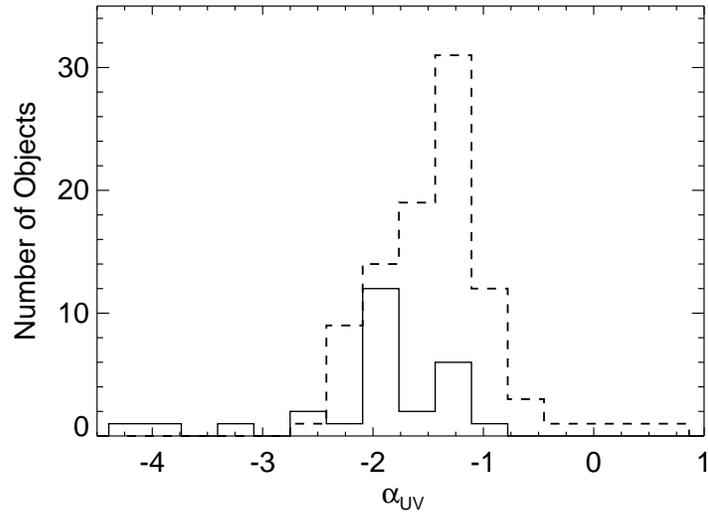}
\caption{Distribution of derived UV slope, from \ion{He}{2} to H$\beta$ ratio, for the GH objects (solid) and the Ho \& Kim sample (dotted).}
\label{alpha}
\end{figure}

As the BH mass decreases at fixed Eddington ratio, we would expect to move closer to the peak of the big blue bump, and thus to observe shallower $\alpha_{\rm UV}$ as a function of mass. Of course, many other factors may be at play, including differing Eddington ratio distributions or different reddening \citep[e.g.][]{2007ApJ...659..211B}.  The deduced UV power-law slopes for our objects range from
$-4<\alpha_{\rm UV}<-0.4$, with a median value of $-1.9$.  To compare
to a sample of typical AGN, we consider the low-redshift AGN presented
by \cite{2009ApJS..184..398H}, which have
$\mbh\sim10^{6}-10^{8}~M_{\odot}$ and a median absolute magnitude of
$M_{B}\sim-19$ derived from their H$\beta$ luminosities.  Because their observations were taken on the Magellan 6.5~m Clay Telescope, they have data of similar quality to ours.  They
measured many emission lines, including \ion{He}{2} and H$\beta$.  We
use their measured line ratios to calculate $\alpha_{\rm UV}$ for
their 94 objects, and present the distribution of the two samples in
Figure~\ref{alpha}.  The comparison sample has a range from $-2.5 <
\alpha_{\rm UV}<0.3$, with a median slope of $-1.4$.  A KS test of the
two distributions results in $P<0.001$, meaning that there is less than 0.1\% probability that
they are drawn from the same parent sample.  We find that the distributions are statistically different whether we include the three low-mass AGN with upper limits in our distribution or not.

Given this difference in distributions, and the significant shift in the median, we also look for a correlation between \mbh\ and $\alpha_{\rm UV}$.   We find a mild correlation, with a correlation coefficient $r=0.18$ for the range of BH masses included here, from $5.5 <$~log~$M_{\mathrm{BH}}/\msun < 10$.  A least-squares fit to the data yields the following relationship:  log~$\mbh\propto0.10~\alpha_{\rm UV}$.  
Previous work \citep{2005ApJ...619...41S, 2007ApJ...668..682D} investigate possible correlations between the UV continuum slope and \mbh\ and do not find compelling evidence for a trend between UV slope and \mbh.  We probe a bluer region of the UV continuum and extend to lower \mbh~than those authors, and we see a tantalizing hint for a correlation between \mbh\ and $\alpha_{\rm UV}$.  On the other hand, the sense of the trend is counter to our naive expectations for lower-mass BHs.  It would be useful to compile a sample including our low-mass BHs with matched Eddington ratios and examine both the X-ray and UV spectral slopes at the same time \citep[e.g.][]{2009ApJ...698.1515D}.

\section{Composite Spectra}
\label{sec:comps}

An effective way to highlight general spectral characteristics is to
make composite spectra.  Other authors in AGN research have used
composite spectra to investigate trends in AGN across a wide range of
redshifts \citep{2001AJ....122..549V}, to probe weak emission lines in
high luminosity AGN \citep{1991ApJ...373..465F}, and weak emission and
absorption lines in $z=2-3$ star-forming galaxies that house AGN
\citep{2011ApJ...733...31H}.  One advantage of a composite spectrum
with our high spectral resolution is that it gives us the opportunity
to investigate low-contrast features that would otherwise be difficult
to detect, including faint emission lines, line wings, and
intermediate-line regions.  Some of our objects exhibit
high-ionization, forbidden Fe lines, such as [\ion{Fe}{7}]~$\lambda
6087$, [\ion{Fe}{10}]~$\lambda 6374$, and [\ion{Fe}{11}]~$\lambda
7892$.  These coronal lines are thought to originate in the inner
edge of the NLR, where densities are low enough to allow forbidden
line emission, but the ionization parameter is high enough to enable
ions like Fe$^{+9}$ to exist.  \cite{2009MNRAS.394L..16M} compare the
line shapes and shifts of the centroid of these lines to [\ion{O}{3}].
They find that the high-ionization lines are at similar offsets, and
therefore similar velocities, to the blue wing of [\ion{O}{3}], which
may originate in outflowing material \citep{1984ApJ...286..171D, 2002ApJ...568..627C,
  2009MNRAS.397..172G}.  With the luxury of very high spectral
resolution and S/N, we can both investigate the trends highlighted in
previous works in more detail and determine whether they hold for
these low-mass, low-luminosity systems.

\subsection{Composite of Entire Sample}
\label{sec:compall}
\begin{figure}[!tp]
\centering
\includegraphics[trim=7mm 26mm 0mm 5mm, clip, scale=0.8]{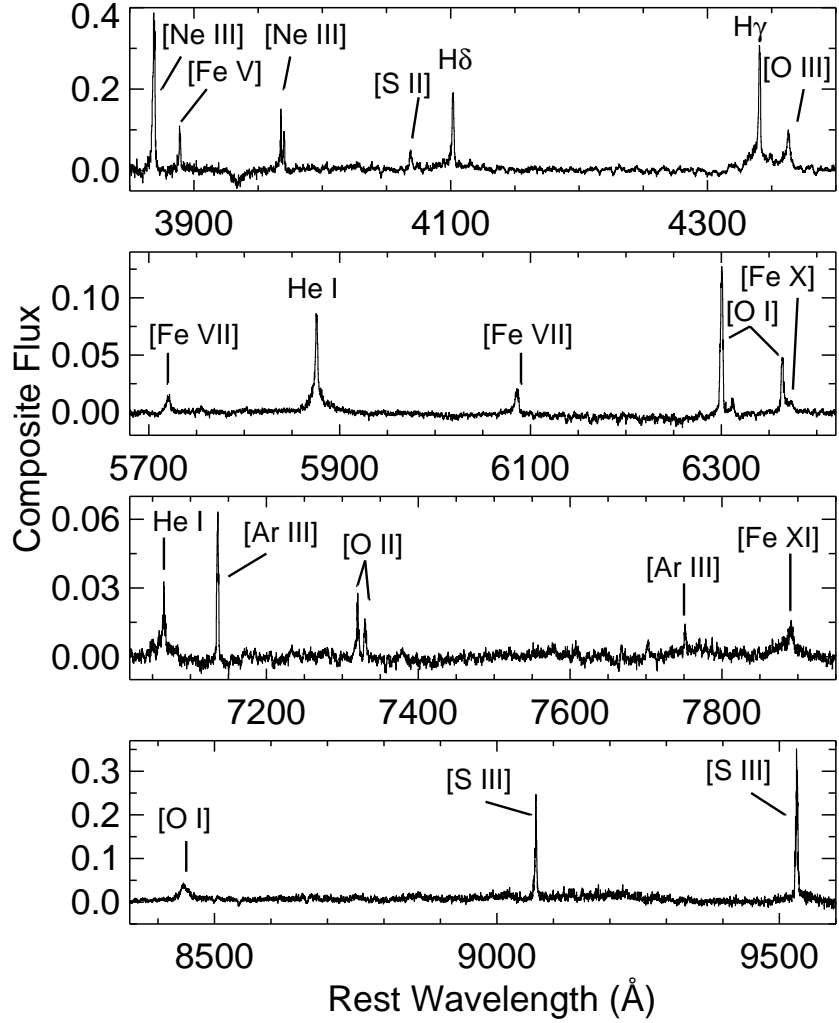}
\caption{Continuum-subtracted composite spectrum of low-mass AGN showing weak emission line features listed in Table~\ref{tablecomp}.}
\label{compGH}
\end{figure}

We construct our composite spectra by interpolating all the spectra to the
same rest-wavelength grid using [\ion{S}{2}], normalizing them to the
rest-frame AGN continuum at 5600 \AA{}, and then taking the median
flux value of contributing objects at each wavelength.  This method is
similar to the median composite construction presented in
\cite{2001AJ....122..549V}.  To highlight the weak features, we
construct our composites from the continuum-subtracted spectra.  In
Figure~\ref{compGH}, we present the composite spectrum of the original
spectra including all 27 low-mass AGN, which highlights the many weak
NLR emission that we observe, and in Table 5 we present our
measurements of the many NLR emission lines in this composite.

In our continuum-subtracted composite, which highlights emission line
features, we first investigate broad H$\alpha$ and H$\beta$, to
compare the BLRs in our low-mass AGN to those of more massive AGN.
\cite{2005ApJ...630..122G} investigate the relationship between the
FWHM of H$\alpha$ and H$\beta$ for a sample of over 200 AGN from the
SDSS, which were selected to have high S/N and low galaxy
contamination.  They find that H$\beta$ is commonly broader than
H$\alpha$, and fit a relation between the two.  Within our composite
spectrum, we also find that the wings of the H$\beta$ line are broader
than those of H$\alpha$.  This trend can be seen in our individual
fits as well, with a median FWHM of H$\alpha$ of 840 km s$^{-1}$ and a
median FWHM of H$\beta$ of 1195 km s$^{-1}$.  In the accepted model of
the BLR, where the emission lines are primarily broadened by their
Keplerian velocities, H$\beta$, being broader, is emitted from gas
interior to the region emitting H$\alpha$ \citep[][and references
therein]{2006agna.book.....O}.

\begin{figure}[!tp]
\centering
\subfigure{
\includegraphics[scale=0.45,angle=0]{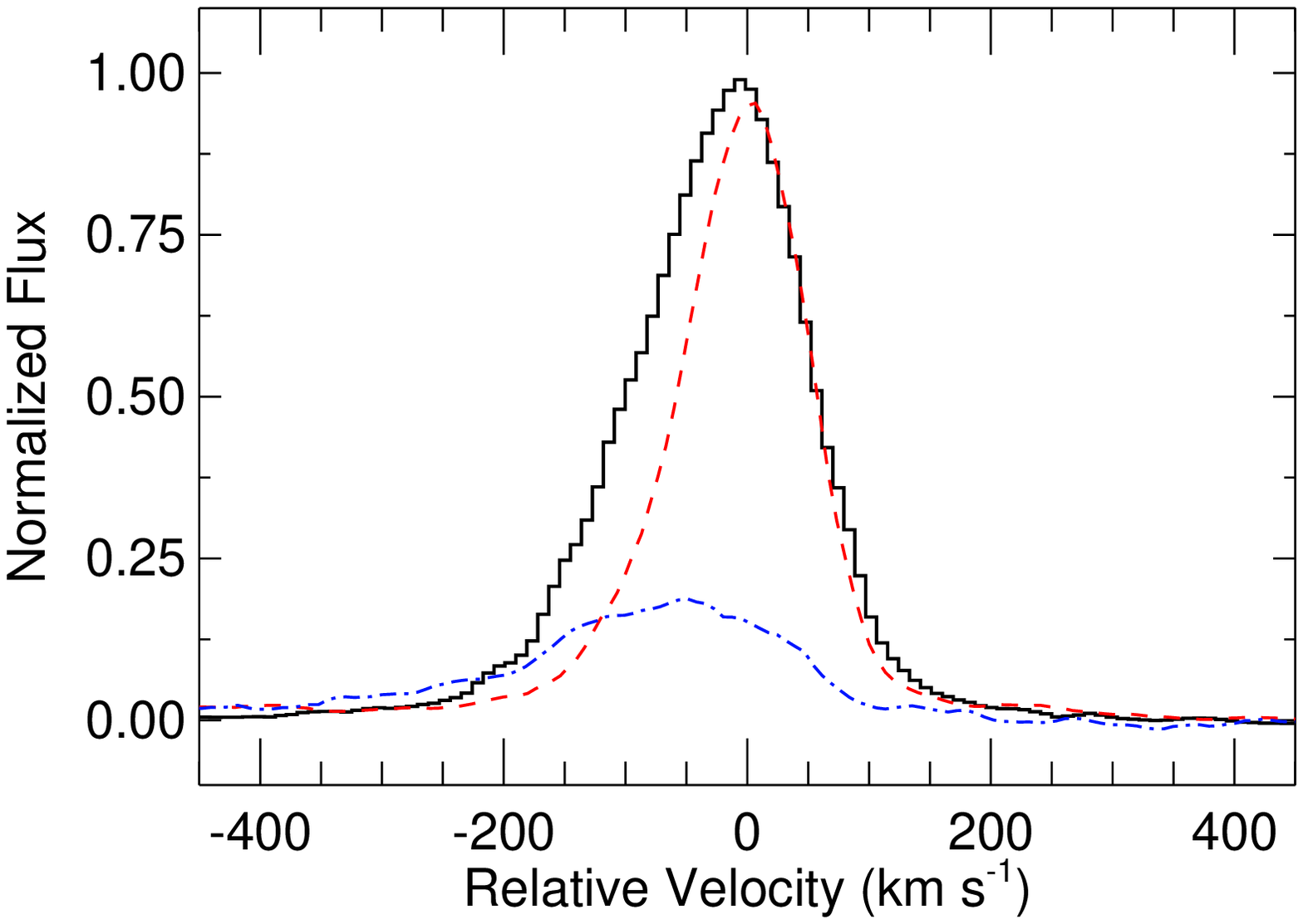}
\label{velplotGH}
}
\subfigure{
\includegraphics[trim=15mm 0mm 0mm 0mm, clip, scale=0.45,angle=0]{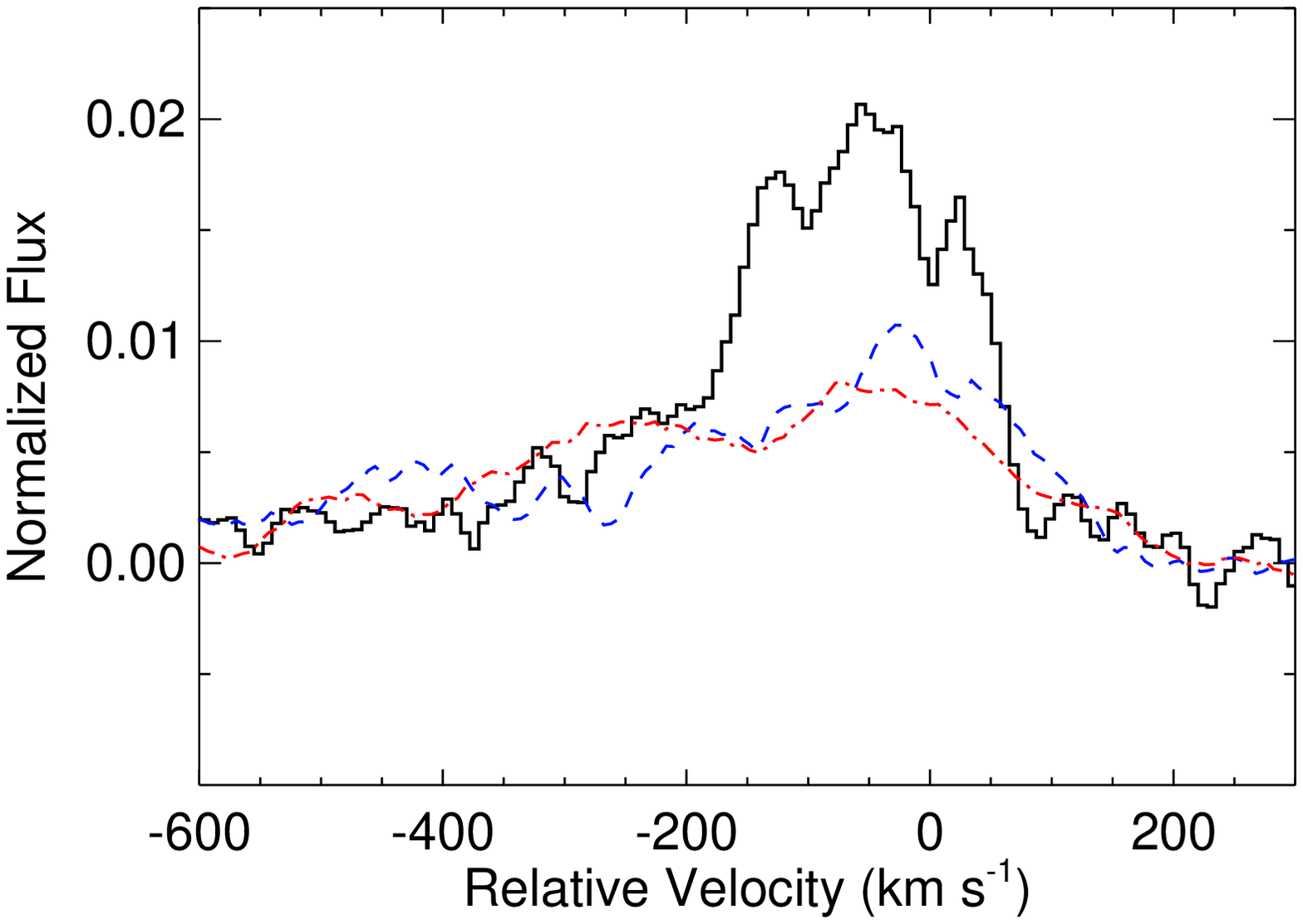}
\label{velplotfe7}
}
\caption{\textit{left panel:} Line profiles of [\ion{O}{3}] in black (solid), [\ion{S}{2}] $\lambda 6731$ in red (dashed), and [\ion{Fe}{7}] $\lambda 6087$ in blue (dot-dashed) in velocity space.  The line fluxes have been scaled for easier visibility. \textit{right panel:} Line profiles of [\ion{Fe}{7}] $\lambda 6087$ in black (solid), [\ion{Fe}{10}] $\lambda 6375$ in red (dot-dashed), and [\ion{Fe}{11}] $\lambda 7892$ in blue (dashed) in velocity space.  In both panels, $v=0$ is with respect to [\ion{S}{2}].}
\label{velplots}
\end{figure}

Turning to the NLR, we investigate the line shapes of some of the
strongest NLR emission lines, including [\ion{O}{3}], [\ion{S}{2}] $\lambda
6731$, and the high-ionization [\ion{Fe}{7}], to see if the results
from \cite{2009MNRAS.394L..16M} hold for AGN of lower mass (see
Figure~\ref{velplotGH}).  While [\ion{S}{2}] and [\ion{O}{3}] have
very similar profiles on the red side, [\ion{O}{3}] shows substantial
excess flux on the blue side.  It turns out that [\ion{Fe}{7}], much
like the profiles in \cite{2009MNRAS.394L..16M}, lies mostly under the
blue wing of [\ion{O}{3}].  The common velocity structure of the 
blue wing and the high ionization lines supports the idea that the
high-ionization Fe lines are emitted in an outflow from the inner
face of the dusty torus, and appears to hold true even for our smaller
scale low-mass AGN.

We also present multiple high-ionization Fe lines in
Figure~\ref{velplotfe7}, which all have ionization
potential~$>~100$~eV.  In the composite of all 27 low-mass AGN, the
[\ion{Fe}{7}] line is blueshifted by $\sim75$~km~s$^{-1}$ and the
[\ion{Fe}{10}] line blueshifted by $\sim125$~km~s$^{-1}$ relative to [\ion{S}{2}].  The
[\ion{Fe}{10}] line is also broader than the [\ion{Fe}{7}] line.
Unfortunately, the composite shows only a marginal detection in the
[\ion{Fe}{11}] line, but the [\ion{Fe}{7}] and [\ion{Fe}{10}] lines
suggest that as the ionization level of these Fe lines increases, they
get broader and more blueshifted.  This pattern has been seen in
high-ionization, forbidden lines in AGN by many authors
\citep{1978ApJ...221..501G, 1976MNRAS.177P.121C, 1984MNRAS.208..347P}.  In comparison to \cite{2009MNRAS.397..172G}, who investigate 63 AGN from SDSS which are selected to have high-ionization forbidden lines, the increasing FWHM and blueshift of high-ionization Fe lines in our low-mass AGN most closely resemble their NSL1 and Seyfert 1.5 subsets.  
Increasing FWHM and blueshift at high ionization potential have also
been identified using near-infrared, high-ionization, forbidden lines
of various elements, including S, Si, Fe, Ca, and Al, by \cite{2011ApJ...743..100R}.  One would expect this behavior if, as the
high-ionization, forbidden lines increase in ionization potential,
they are emitted from closer to the nucleus than other NLR lines, and
are nearer the launching point of an outflow
\citep{1981ApJ...247..403H, 1984MNRAS.207..867W}.  A possible
alternative explanation is that the high-ionization emission results
from inflowing material on the far side of the central engine, which
is still blueshifted from our point of view; however, the higher
ionization lines would still be spatially closer to the nucleus with
their stronger blueshifts.

Because the high-ionization Fe lines suggest a possible relationship
between ionization potential and line widths and shifts, we
investigate several lower-ionization, narrow lines to look for trends
with ionization potential, as discussed in \cite{1986ARA&A..24..171O}
and references therein. We fit the composite spectra with our 
emission-line fitting procedure as described above.  In addition, we measure
line widths and shifts of weaker lines presented here using a single
Gaussian fit to each line in the continuum-subtracted composite
spectrum, because there is not enough signal in the high-ionization
lines to make a more complicated fit informative.  We
do this analysis on the composite spectrum rather than individual
spectra because the high-ionization lines are weak enough that they
are difficult to measure in individual spectra and benefit from the
combined signal in the composite.  In our individual spectra, only
about 20\% of our objects have detectable [\ion{Fe}{10}] or
[\ion{Fe}{11}], despite 16 of 27 showing [\ion{Fe}{7}].

\begin{figure}[!tp]
\centering
\subfigure{
\includegraphics[trim=0mm 0mm 0mm 0mm, clip, scale=0.6]{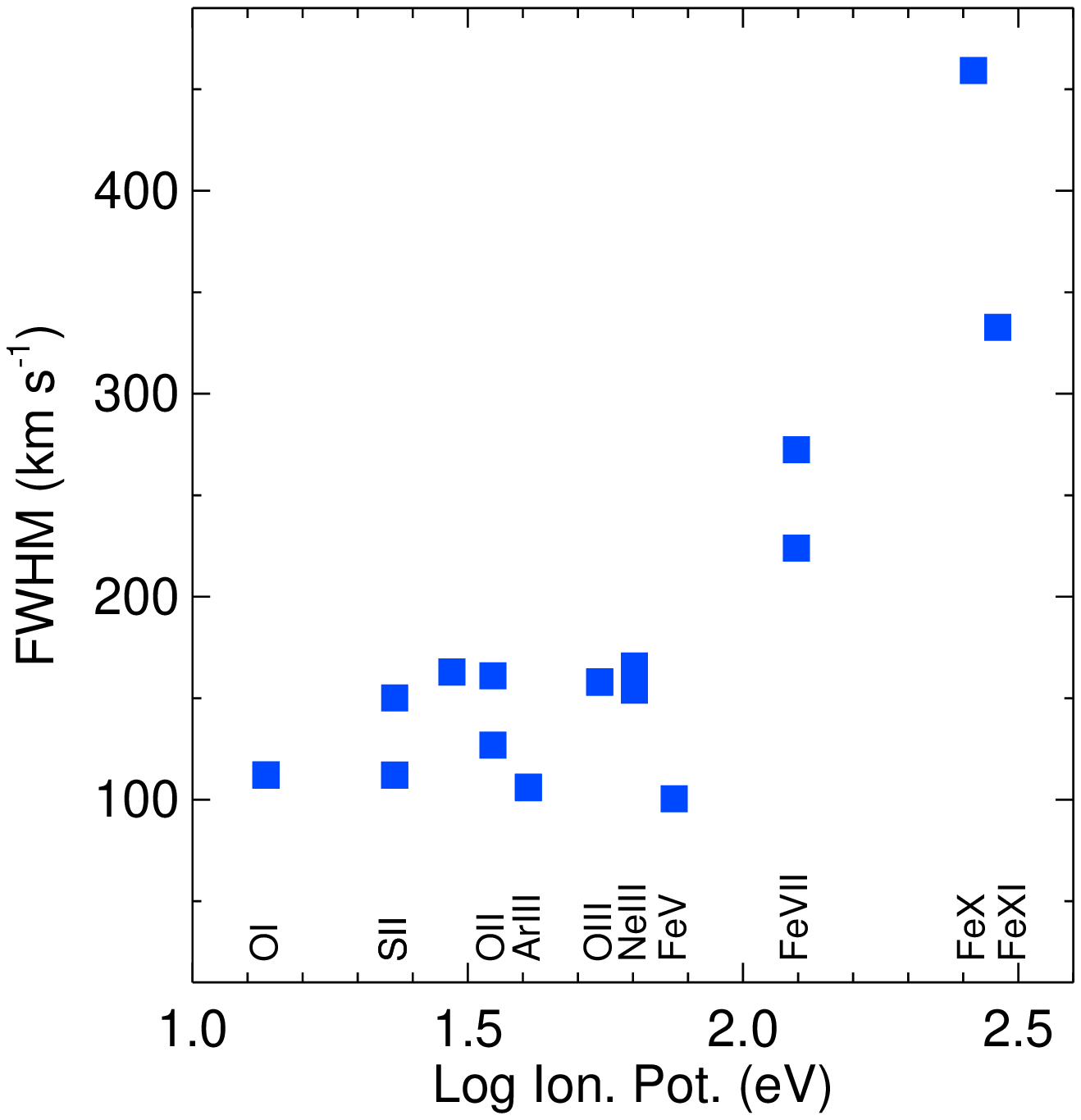}
\label{GHwidth}
}
\subfigure{
\includegraphics[trim=0mm 0mm 0mm 0mm, clip, scale=0.6]{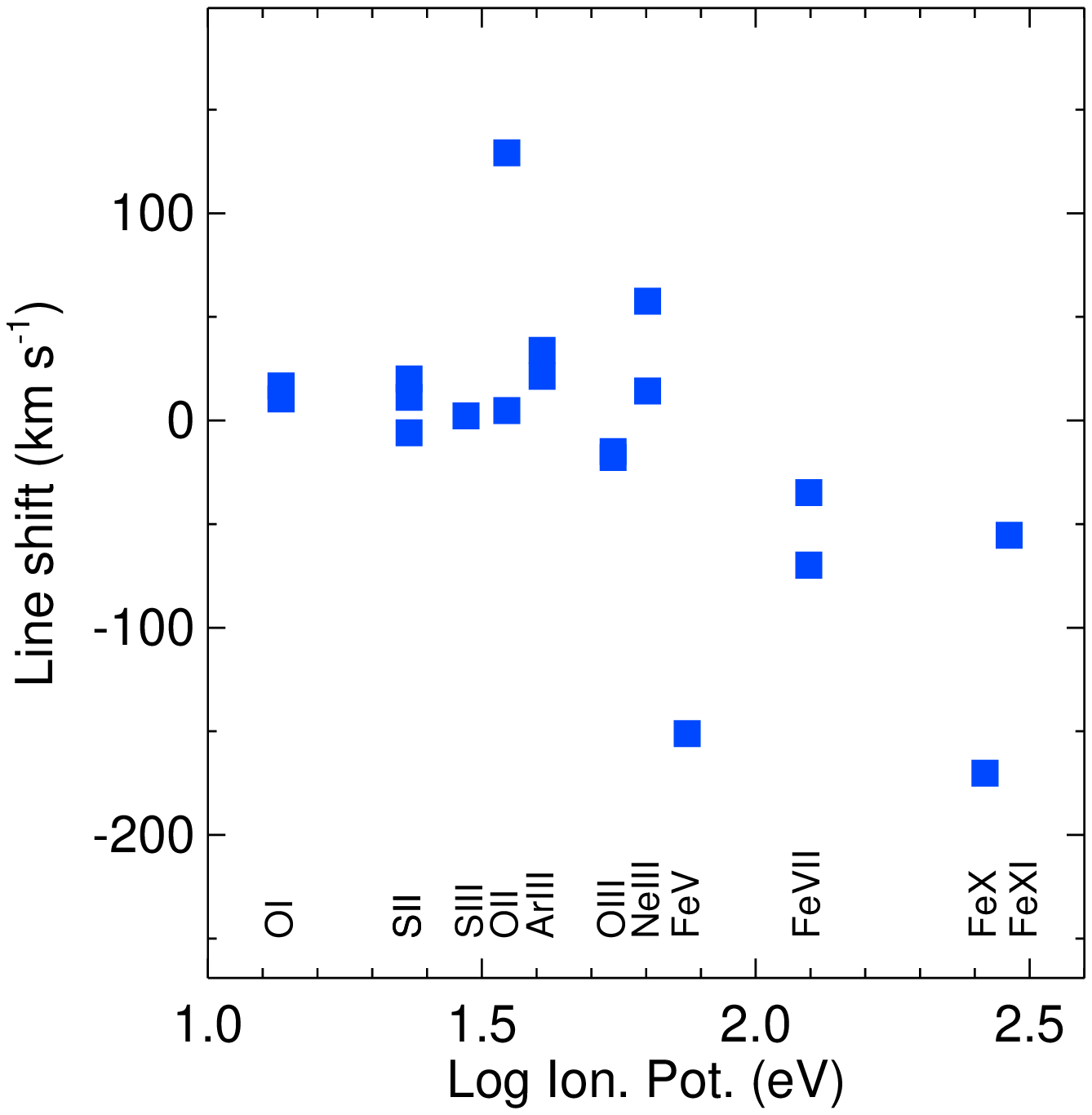}
\label{GHshift}
}
\caption{\textit{Top:}  Line widths (FWHM) plotted for NLR forbidden lines from the continuum-subtracted composite of low-mass AGN. \textit{Bottom:}  Line shifts (relative to the centroid predicted from [\ion{S}{2}]) for narrow lines from the continuum-subtracted composite of low-mass AGN. Negative shifts are blueward.}
\end{figure}

In Figures~\ref{GHwidth} and~\ref{GHshift}, we show the measured line
widths and shifts for a selection of NLR emission lines of varying
ionization states in the composite spectrum.  In accordance with the
trend we see in Figure~\ref{velplotfe7}, as the ionization potential
of the lines increases, the lines become broader (correlation
coefficient of $r = 0.88$) and more blueshifted ($r=-0.56$).  Again,
this is consistent with an ionization structure of the NLR in which
the lines with higher ionization potential are emitted closer to the
central engine and thus are broader.  We postulate that the blueshift
indicates an outflowing component on these small scales
\citep{2008ApJ...680..926K, 2011ApJ...739...69M}.  Note that nuclear outflows are not typically aligned with the host galaxy in any systematic way.  In larger samples, no correlation is seen between the presence of a blue wing and the inclination of the host galaxy \citep[e.g. ][]{2005ApJ...627..721G}, and radio jets, which indicate the direction of nuclear outflows, seem to be randomly oriented with respect to their host galaxies \citep[e.g.][]{2000ApJ...537..152K}.  Another commonly
discussed correlation is that between FWHM and critical density.  We also find a
weak trend between critical density and FWHM, as discussed by
\cite{1984ApJ...285..458F} and others.  However, with a correlation
coefficient of $r=0.54$, this correlation is weaker than that with
ionization potential, at least in this sample.

\subsection{Division by Physical Properties}
\label{sec:compsubsets}
While the previous composites are useful for evaluating
characteristics of our entire sample, we can also investigate how our
spectra change in relation to specific physical or spectral
characteristics.  To this end, we divide our sample into several
subsets.  We investigate many characteristics in this
manner, including luminosity, Eddington ratio, $\mbh$, FWHM of
H$\alpha$, presence or absence of a blue wing in [\ion{O}{3}], shift
of the centroid of [\ion{O}{3}], $h_{3}$ and $h_{4}$ of the NLR fits,
and NLR density determined from the [\ion{S}{2}] line ratio.  In most
cases no interesting trends were seen, or they were redundant with those
that we show here based on Eddington ratio, luminosity, and the
presence of a blue wing in [\ion{O}{3}].

For each characteristic, we divide the entire sample of 27 objects
into two subsets as described below and construct a
continuum-subtracted composite spectrum of that subset.  For the
luminosity subsets, we derive luminosities from $L_{\mathrm H \alpha}$
using the formalism in \cite{2005ApJ...630..122G} to calculate
$L_{\mathrm {bol}}$.  We then divide the sample in half at the median
luminosity (log $L_{\mathrm {bol}}=40.5$~erg~s$^{-1}$, with a full
range of $39.3~$erg~s$^{-1}~<~$log~$L_{\mathrm
  {bol}}~<~41.5$~erg~s$^{-1}$).  To investigate Eddington ratio, we
bifurcate the sample about the median Eddington ratio (log $L/\ledd=-0.4$, with a full
range of $-1.8<~$log~$L/\ledd~<~0.3$).  To evaluate spectra with or
without a blue wing in [\ion{O}{3}], we visually inspect the
individual continuum-subtracted spectra to determine the presence or
absence of a blue wing in [\ion{O}{3}] in each object and then form a
composite of objects with a blue wing and another composite for those
without.  Dividing by visual inspection is necessary because of the
difficulty of parameterizing blue asymmetry with measurements from the
two component [\ion{O}{3}] fits.  For each of these six composite
spectra, we investigate emission line behavior among the NLR emission
lines shown in Figure~\ref{GHwidth}.  
We present the velocity dispersion in the NLR ($\sigma$)
for each subset in Table~\ref{subsets}, where we subtracted the
instrumental resolution ($\sigma_{\mathrm {inst}}=22~$km~s$^{-1}$) in
quadrature.  These values are averages over all of the NLR emission
lines shown in Figure~\ref{GHwidth} or averages of the high-ionization
Fe lines, where the ionization potential~$>100$~eV.

\begin{figure}[!tp]
\centering
\subfigure{
\includegraphics[scale=0.45,angle=0]{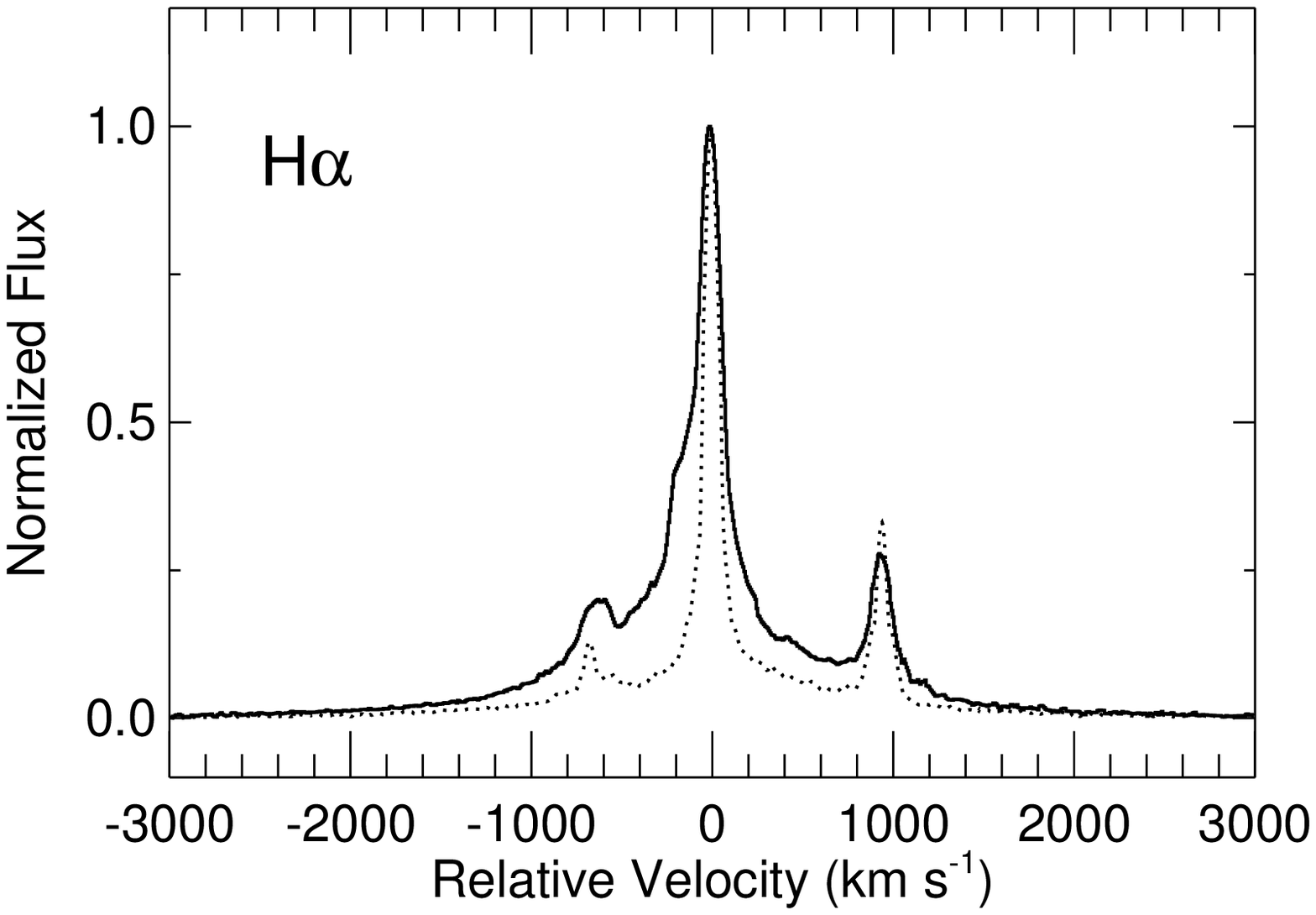}
\label{HalineLedd}
}
\subfigure{
\includegraphics[trim=20mm 0mm 0mm 0mm, clip,scale=0.45,angle=0]{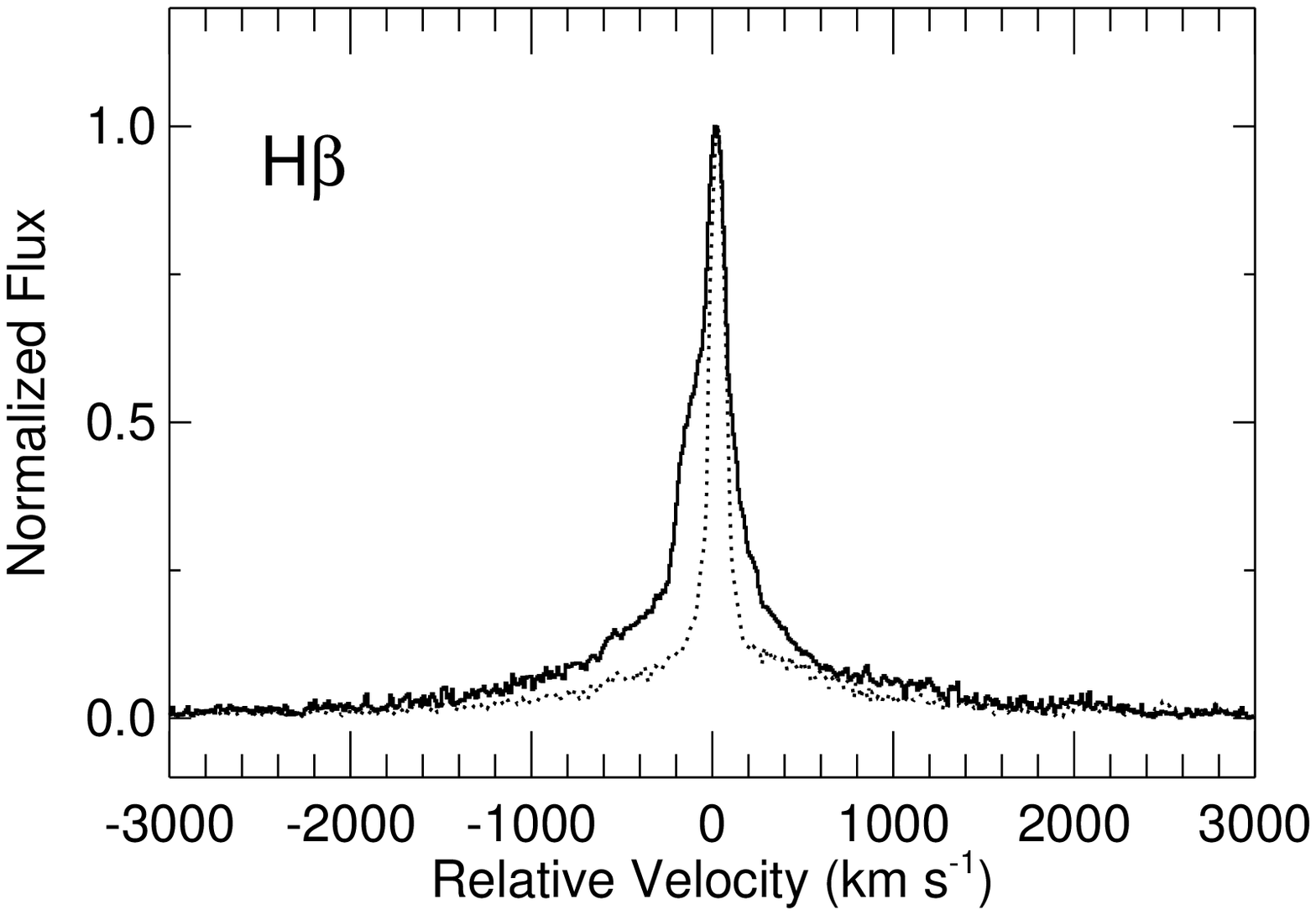}
\label{HblineLedd}
}
\caption{Composites of the H$\alpha$ (\textit{left panel}) and H$\beta$ regions (\textit{right panel}) for high $L/\ledd$ (solid) and low $L/\ledd$ (dotted) objects.  Notice the excess ILR in the high $L/\ledd$ composite for both lines.}
\label{LeddHa}
\end{figure}

Investigation of these subsets yields interesting results for both the BLR and the NLR.  We find that the BLR emission lines differ significantly between the high-$L/\ledd$ and the
low-$L/\ledd$ composites.  In Figure~\ref{LeddHa}, we show
the H$\alpha$ and H$\beta$ regions for both composites.  While the
wings of the lines ($v>1200$~km s$^{-1}$) are similar, the
high-$L/\ledd$ composite shows excess flux at intermediate
velocities in both the broad H$\alpha$ and H$\beta$ emission lines.  The
difference in intermediate BLR component is unique to the $L/\ledd$
subset.  The intermediate-width component in the high-$L/\ledd$ composite
has a FWHM$=840$~km~s$^{-1}$, compared to the width of the broad
component at FWHM$=2,400$~km~s$^{-1}$.  The intermediate line
component is still broader than the average NLR width, shown in Table
6, and also broader than the high-ionization Fe lines.  Our discovery of an 
intermediate line region supports
the conclusion of \cite{2008ApJ...683L.115H} that changes in
Eddington ratio among objects can account for differences in an 
intermediate component of the Balmer lines.  Increased intermediate-line emission 
in high-$L/\ledd$ AGN could bias their derived BH masses, causing 
underestimates of the true \mbh~\citep{2006A&A...456...75C, 2009ApJ...696..741W}.

We also find that NLR emission-line properties are dependent on luminosity, 
$L/\ledd$, and the presence of a blue wing in [\ion{O}{3}].  When investigating 
the effects of luminosity, we find that the
high-$L_{\mathrm {bol}}$ composite has broader NLR emission lines than
both the low-$L_{\mathrm {bol}}$ composite and the general composite
of our whole sample. This is in keeping with past findings
\citep{1983ApJ...266..485P, 1985MNRAS.213...33W, 1992ApJS...79...49W},
where higher luminosities result in broader line widths, commonly
thought to be due to the inferred larger gravitational potential of
the bulge of the galaxy \citep{2003ApJ...583..159H}.  The
high-$L/\ledd$ subset also possesses similarly broad NLR emission
lines compared to the low-$L/\ledd$ subset.  Given the small dynamic
range in $\mbh$ represented here, it is not surprising that we see
similar trends in $L_{\mathrm {bol}}$ and $L/\ledd$.

There are well-known correlations between the line widths and line shifts of 
[\ion{O}{3}] as a function of Eddington ratio, such that high-$L/\ledd$ objects 
tend to have broader [\ion{O}{3}] with a more blueshifted peak 
\citep{2005AJ....130..381B}.  This is often interpreted in the context of 
disk winds or outflows that are more prevalent at high $L/\ledd$ 
\citep{1984ApJ...286..171D}.  Blueshifted [\ion{O}{3}] clouds at high velocities (up to 3200 km s\per) are seen in NGC 1068, and are associated with outflows induced by the jet and/or radiation pressure from the AGN itself \citep[e.g.][]{2002ApJ...568..627C, 2012ApJ...746...86G}.  Such radial motion might be expected to reduce the covering factor of NLR gas and lower the EW of lines like [\ion{O}{3}].
The well-known Eigenvector 1 (EV1) ties these trends together \citep{1992ApJS...80..109B}.  At the high 
$L/\ledd$ end of EV1, the ratio of [\ion{O}{3}]$/$H$\beta$ drops and 
the incidence of blue asymmetries in H$\beta$ increases.  The [\ion{O}{3}] EW may drop in high EV1 objects as well, but the 
data remain inconclusive on this point 
\citep{2005AJ....130..381B, 2009ApJ...706..995L}.  In a related trend, 
$\sigma_{\mathrm{NLR}}/\sigma_{*}$ may increase with increasing $L/\ledd$ due to increasing nonvirial motions caused by outflows
\citep{2005ApJ...627..721G, 2009ApJ...699..638H}.  At the highest luminosities, 
the correlation between $\sigma_{\mathrm{NLR}}$ and $\sigma_{*}$ 
seems to disappear altogether \citep{2009ApJ...702..441G}.

Examining our low-mass AGN alone, we do not find a consistent story connecting the NLR to Eddington ratio.  Turning first to the ratio of $\sigma_{\mathrm{NLR}}/\sigma_{*}$, there does not appear to be any correlation with $L/\ledd$ \citep{2011ApJ...739...28X}.  Unfortunately, the stellar velocity dispersions are often overestimated due to contamination from the galaxy disk \citep[see also][]{2011ApJ...742...68J}, which complicates our interpretation.  Secondly, we investigate the presence of a 
blue wing in [\ion{O}{3}] and find no correlation with $L/\ledd$ ($r=0.06$).  
If, as we suggest above, blue wings are indicative of an outflowing component, 
then we do not see evidence for stronger outflows at higher Eddington ratios, 
as we might expect \citep{2000ApJ...543..686P}.   Furthermore, if $L/\ledd$ 
were tied to outflows, we might expect to find lower EW narrow-lines at 
higher $L/\ledd$. Instead, we see higher EW NLR lines in the higher $L/\ledd$ composite.  Perhaps our dynamic range in $L/\ledd$ (or $\mbh$ or luminosity) is too narrow to truly evaluate the interdependence of outflows and accretion rate.

\begin{figure}[!tp]
\centering
\includegraphics[scale=0.6]{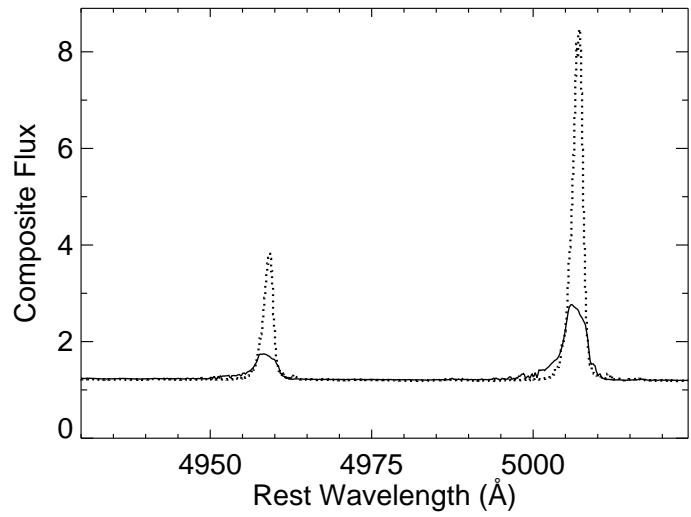}
\caption{Composite spectra with a blue wing in [\ion{O}{3}] (solid) and without (dotted).  In order to compare EWs, we plot each continuum-subtracted composite plus the average power-law AGN continuum of objects in that composite, where the power law is normalized to 5600~\AA.  Note that symmetric [\ion{O}{3}] lines have much higher EW.}
\label{o3wingplot}
\end{figure}

\begin{figure}[!tp]
\centering
\includegraphics[trim=0mm 0mm 0mm 0mm,clip,scale=0.8, angle=0]{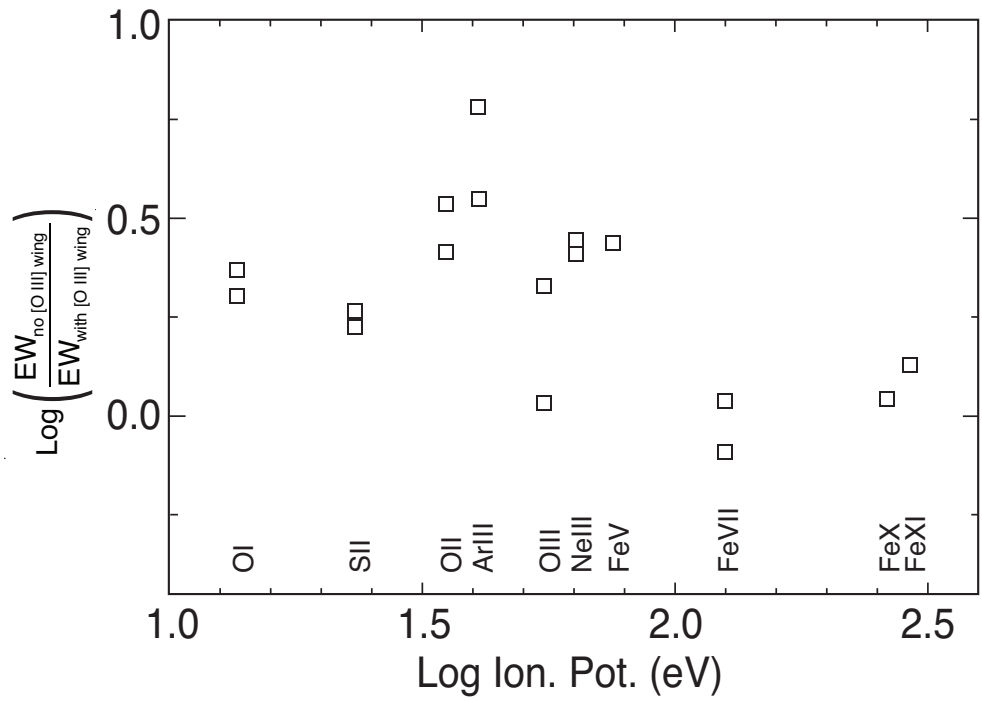}
\caption{Logarithmic NLR emission line EW ratios vs. ionization potentials.  We compare the EWs in the composites with and without a blue wing.}
\label{o3wingIP}
\end{figure}

On the other hand, we do see evidence that blue wings are associated with 
outflows.  We might expect that in objects with strong outflows, the gas 
would be more disturbed, leading to a lower covering factor
and thus a lower EW \citep{2005MNRAS.358.1043B, 2009ApJ...706..995L}.  
We do find that the EW of [\ion{O}{3}] in our objects is lower, relative to the AGN power-law continuum, when there is a blue wing in [\ion{O}{3}] (Figure~\ref{o3wingplot}). The EWs of other 
low-ionization NLR emission lines are also lower (Figure~\ref{o3wingIP}).  
Furthermore, if we divide the sample into those with and without blue asymmetry in the low-ionization lines (e.g., the $h_{3}$ measurement of [\ion{S}{2}]) then we also see low 
EW narrow-lines in the blue-asymmetric subsample.  Thus, we do see indirect 
evidence that blue wings indicate outflowing material (as we suggested above 
based on the velocity of the high ionization lines, \S\ref{sec:compall}).  
Interestingly enough, \textit{all} the low-ionization narrow lines, 
not just [\ion{O}{3}], appear to behave similarly, both in terms of EWs and 
asymmetric lines.

From our sample alone, we see evidence for varying levels of
disturbance in the NLR, but we see no direct tie with $L/\ledd$ or
luminosity.  Given the limited dynamic range in luminosity, and the large uncertainties on Eddington ratio, it is hard to draw strong conclusions from these data alone.  Perhaps the
absolute luminosities of the sample are too low
\citep{2005ApJ...630..122G} or the uncertainties on our $L/\ledd$
estimates are too large.

\section{Conclusions}
\label{sec:concl}

We present observations of a sample of 27 low-mass AGN
($10^{4}~M_{\odot}<\mbh<2~\times~10^6~M_{\odot}$), observed with the
Echellette Spectrograph and Imager on the Keck Telescope.  Large
samples of low-mass AGN have not existed until recently
\citep{2004ApJ...610..722G}, and so we compare their emission-line
properties to those of well-studied higher-mass AGN.

We investigate the NLR metallicities of these objects using emission
line ratios, particularly [\ion{N}{2}]$/$H$\alpha$ and
[\ion{O}{3}]$/$H$\beta$.  Our sample includes objects with weaker
[\ion{N}{2}]$/$H$\alpha$ than higher-mass AGN for a given
[\ion{O}{3}]$/$H$\beta$, which implies that those objects have lower metallicities.  Thus we see some galaxies with similar metallicities to the rare AGN with sub-solar metallicity sought by \cite{2006MNRAS.371.1559G}, yet we cannot determine their metallicities on an
absolute scale \citep[e.g.][]{2008ApJ...681.1183K}.  Additionally, we do
see weak evidence for a correlation between galaxy stellar mass and
gas-phase metallicity in these systems.  Most likely, the lack of
low-metallicity AGN highlighted by \cite{2006MNRAS.371.1559G} is
attributable to the difficulties of finding AGN in low-mass galaxies
\citep[e.g.][]{2007ApJ...670...92G} rather than extra enrichment of
the NLR by the AGN itself.

We examine the continuum properties of the accretion disk for our
low-mass AGN.  By using the emission from the recombination lines
[\ion{He}{2}] and H$\beta$, we infer the slope of the far-UV continuum from 200--900\AA{}, near the peak of the big blue bump.  We find that the
far-UV slopes in our low-mass AGN are steeper than those of low-redshift, 
higher-mass AGN.  While we know
that our low-mass AGN are very radio-quiet \citep{2006ApJ...636...56G}
and have flat X-ray slopes \citep{2009ApJ...698.1515D} compared to typical AGN, they 
have quite similar optical properties and unexpectedly, somewhat steeper UV slopes.  
These steeper inferred UV continua imply that changing continuum shape could explain the inverse Baldwin effect seen between H$\beta$ FWHM and luminosity \citep{2002MNRAS.337..275C,
  2005ApJ...630..122G}.

Despite tentative evidence for different SEDs, the NLR structure is also similar between low-mass and typical AGN.
Using composite spectra, we are able to measure the widths and
velocities of high-ionization Fe lines. As found by
\cite{2009MNRAS.394L..16M}, the [\ion{Fe}{7}] line exhibits a similar
width and blue shift as the blue wing of [\ion{O}{3}], pointing to a
common physical origin for both transitions in a radially flowing
component.  As seen in previous work, we find that the width of NLR
lines and their blueshift correlates with the ionization potential of
the line.  We see no evidence for a dramatic change in the NLR
structure in this mass and luminosity regime.

Making composite spectra in subsets of luminosity, Eddington ratio,
and presence or absence of a blue wing in [\ion{O}{3}], we find first
that the high-luminosity composite has much broader NLR emission lines than
the low-luminosity counterpart within our sample, as has been seen in more-massive AGN.  We also find that the subset showing the presence of a blue wing in [\ion{O}{3}] exhibits weak NLR emission.  We posit that in the objects with a blue wing, the outflow drives gas away from the central region, leading to a lower covering fraction \cite[e.g.][]{2009ApJ...706..995L}.  However, contrary to our expectations from typical AGN, the presence of a blue wing does not correlate with high Eddington ratio, possibly due to our small dynamic range in this sample.  Lastly, we find that the high Eddington ratio composite has excess emission at intermediate velocities in the Balmer lines.  This
corresponds to the ILR identified by \cite{2008ApJ...683L.115H} and
\cite{2009ApJ...700.1173Z}.  This ILR emission could potentially bias
$\mbh$ estimates for AGN at high $L/\ledd$ by decreasing the measured
FWHM of the Balmer lines, thus underestimating $\mbh$, as pointed out by 
\cite{2006A&A...456...75C} and \citet{2009ApJ...700.1173Z}.  

Overall, the low-mass AGN studied here seem to behave in much the same
way as more massive AGN, as traced by optical emission lines.  They
are likely to have relatively low gas-phase metallicities in the NLR
than their larger brethren, and may have steeper far-UV continuum slopes, but the structure and organization of their emission line regions seems to be largely similar.  

\section{Acknowledgements}
\label{sec:ack}
We would like to thank Edward L. Robinson and Greg Shields for helpful conversations during the course of this project, and Brent Groves for discussions and help with the MAPPINGS III models.  We would also like to thank the referee, for a thorough and insightful report that improved this paper a great deal.  Research by A.J.B. is supported by NSF grant AST-1108835.  The data
presented herein were obtained at the W.M. Keck Observatory, which is
operated as a scientific partnership among the California Institute of
Technology, the University of California and the National Aeronautics
and Space Administration. The Observatory was made possible by the
generous financial support of the W.M. Keck Foundation.  The authors
wish to recognize and acknowledge the very significant cultural role
and reverence that the summit of Mauna Kea has always had within the
indigenous Hawaiian community.  We are most fortunate to have the
opportunity to conduct observations from this mountain.

\clearpage

\begin{deluxetable}{l c l r c}
\tablenum{1}
\tablecolumns{6}
\tabletypesize{\footnotesize}
\tablewidth{0pc}
\tablecaption{Low-Mass AGN Observations \label{tableobjs}}
\tablehead{\colhead{Object} & \colhead{Flag} & \colhead{$z$} & \colhead{S/N} & \colhead{Fit Class}}
\startdata
SDSS J010712.03+140844.9 & GH01  &     0.0770 &     17 & AGN \\ 
SDSS J024912.86-081525.6 & GH02  &     0.0296 &     18 & AGN \\ 
SDSS J032515.59+003408.4 & GH03  &     0.1021  &    12 & AGN \\ 
SDSS J080629.80+241955.6 &	  &     0.0415 &     31 & GAL \\ 
SDSS J080907.58+441641.4 &	  &     0.0540 &     24 & GAL \\ 
SDSS J081550.23+250640.9 &	  &     0.0726 &     20 & GAL \\ 
SDSS J082443.28+295923.5 &	  &     0.0254 &     54 & GAL \\ 
SDSS J082912.67+500652.3 & GH04  &     0.0435 &     37 & GAL \\ 
SDSS J094310.12+604559.1 & GH05  &     0.0745 &     25 & AGN \\ 
SDSS J101108.40+002908.7 & GH06  &     0.1001  &     12 & AGN \\ 
SDSS J105755.66+482502.0 &	  &     0.0732 &     26 & GAL \\ 
SDSS J110501.97+594103.6 &	  &     0.0336 &      57 & GAL \\ 
SDSS J114343.76+550019.3 &	  &     0.0271 &     26 & GAL \\ 
SDSS J122342.81+581446.1 &	  &     0.0146 &     34 & GAL \\ 
SDSS J124035.81-002919.4 & GH10  &     0.0810 &     31 & GAL \\ 
SDSS J125055.28-015556.6 & GH11  &     0.0815 &     20 & GAL \\ 
SDSS J131310.12+051942.1 &	  &     0.0488 &     24 & GAL \\ 
SDSS J141234.67-003500.0 & GH13  &     0.1271  &     17 & AGN \\ 
SDSS J143450.62+033842.5 & GH14  &     0.0283 &     20 & GAL \\ 
SDSS J153425.59+040806.7 &	  &     0.0395 &     9 & AGN \\ 
SDSS J162636.40+350242.0 &	  &     0.0341  &     38 & GAL \\ 
SDSS J163159.59+243740.2 &	  &     0.0436 &     43 & GAL \\ 
SDSS J170246.09+602818.9 & GH16  &     0.0691  &     20 & GAL \\ 
SDSS J172759.15+542147.0 & GH17  &     0.0995 &     14 & AGN \\ 
SDSS J215658.30+110343.1 &        &      0.1080 &     39 & GAL \\ 
SDSS J232159.06+000738.8 & GH18  &      0.1838 &     12 & GAL \\ 
SDSS J233837.10-002810.3 & GH19  &     0.0356 &     29 & GAL \\ 
\enddata
\tablecomments{~Low Mass AGN Sample Observations.  Col. (1): SDSS Designation.  Col (2):  \citet{2004ApJ...610..722G} object designations.  Col (3):  Redshift.  Col. (4): Median S/N per pixel at 5500~\AA{}.  Col (5):  Fit classification, as described in Section \S~\ref{sec:method}.}
\label{observed}
\end{deluxetable}
\clearpage

\clearpage

\clearpage

\clearpage

\begin{deluxetable}{l c c}
\tablenum{2}
\tablecolumns{4}
\tablewidth{0pc}
\tablecaption{Fitted Emission Lines}
\tablehead{\colhead{Emission Line} & \colhead{Wavelength (\AA{})} & \colhead{Possible components}}
\startdata
\ion{He}{1}, narrow 	& 7065, 6678, 5876           	& 1 G-H Poly each\\
\ion{He}{1}, broad   	& 7065, 6678, 5876		& 1 Gaussian each\\
$[$\ion{S}{2}]         & 6730, 6717					& 1 G-H Poly each\\
$[$\ion{N}{2}]		& 6584, 6548				& 1 G-H Poly each\\
H$\alpha$ -- H$\epsilon$, narrow 	& 6563, 4861, 4341, 4102, 3970		& 1 G-H Poly each\\
H$\alpha$, broad   	& 6563 							&   $\leq$ 4 Gaussians \\
$[$\ion{O}{1}]		& 6364, 6300				& 1 G-H Poly each\\
$[$\ion{O}{3}]	& 5007, 4959, 4363	&  $\leq$ 2 G-H Polys each\\
H$\beta$, broad   	& 4861 			&  $\leq$ 3 Gaussians  \\
\ion{He}{2}, narrow 	& 4686			& 1 G-H Poly\\
\ion{He}{2}, broad   	& 4686			&  $\leq$ 2 Gaussians\\
\ion{N}{3}			& 4640			& 1 Gaussian\\
H$\gamma$ -- H$\epsilon$, broad   & 4341, 4102, 3970	&  1 Gaussian each\\
\enddata
\tablecomments{Col. (1): Emission Line Species.  Col (2):  Rest wavelength~(\AA{}).  Col (3):  Possible components in fit, where G-H Poly is a set of Gauss-Hermite functions.}
\label{tablelines}
\end{deluxetable}
\clearpage

\begin{deluxetable}{l c r c c c}
\tablenum{3}
\tablecolumns{6}
\tabletypesize{\scriptsize}
\tablewidth{0pc}
\tablecaption{Derived Measurements}
\tablehead{
\colhead{Object} & \colhead{Log~$\mbh/\msun$} & \colhead{Log~$L/\ledd$} & 
\colhead{Log~$M_{\mathrm{gal}}/\msun$} & \colhead{$n_e$ (cm$\percu$)} & \colhead{$Z/\zsun$} }
\startdata
J0107+1408 &  $
    6.03$ & $    -0.40$ & \nodata & $      360$ & $     2.0 \pm 
     0.1
$\\
J0249-0815 &  $
    5.32$ & $    -0.66$ & \nodata & $     1010$ & $     2.0 \pm 
     0.3
$\\
J0325+0034 &  $
    5.98$ & $    -0.57$ & $    10.05$ & $      230$ & $     1.5 \pm 
     0.5
$\\
J0806+2419 &  $
    5.95$ & $    -0.71$ & $    10.24$ & $      100$ & $     1.6 \pm 
     0.2
$\\
J0809+4416 &  $
    5.94$ & $    -0.61$ & $    10.04$ & $      140$ & $     1.8 \pm 
     0.4
$\\
J0815+2506 &  $
    5.55$ & $    -0.50$ & $    10.24$ & $      170$ & $     1.8 \pm 
     0.5
$\\
J0824+2959 &  $
    5.33$ & $    -0.93$ & $    10.16$ & $      590$ & $     1.5 \pm 
     0.2
$\\
J0829+5006 &  $
    5.76$ & $    -0.12$ & \nodata & $      270$ & $     2.3 \pm 
     0.7
$\\
J0943+6045 &  $
    5.76$ & $     0.02$ & \nodata & $      190$ & $     1.9 \pm 
     0.2
$\\
J1011+0029 &  $
    6.18$ & $    -0.56$ & \nodata & $      210$ & $     0.8 \pm 
     0.2
$\\
J1057+4825 &  $
    5.68$ & $    -0.30$ & $    10.12$ & $       60$ & $     0.9 \pm 
     0.1
$\\
J1105+5941 &  $
    6.89$ & $    -1.81$ & $    10.68$ & $      440$ & $     1.5 \pm 
     0.3
$\\
J1143+5500 &  $
    5.97$ & $    -1.25$ & \nodata & $      500$ & $     0.5 \pm 
     0.1
$\\
J1223+5814 &  $
    6.04$ & $    -0.93$ & $     9.65$ & $      850$ & $     1.6 \pm 
     0.2
$\\
J1240-0029 &  $
    5.93$ & $    -0.09$ & \nodata & $      400$ & $     1.6 \pm 
     0.4
$\\
J1250-0155 &  $
    6.80$ & $    -1.25$ & \nodata & $      160$ & $     0.9 \pm 
     0.2
$\\
J1313+0519 &  $
    5.32$ & $    -0.40$ & $    10.17$ & $      340$ & $     2.9 \pm 
     1.0
$\\
J1412-0035 &  $
    6.15$ & $     0.11$ & \nodata & $     1030$ & $     1.8 \pm 
     0.2
$\\
J1434+0338 &  $
    5.65$ & $    -0.68$ & \nodata & $      240$ & $     2.5 \pm 
     0.6
$\\
J1534+0408 &  $
    4.79$ & $    -0.21$ & $     9.36$ & $      360$ & $     2.4 \pm 
     0.6
$\\
J1626+3501 &  $
    5.54$ & $    -0.39$ & $    10.32$ & $      190$ & $     2.6 \pm 
     1.4
$\\
J1631+2437 &  $
    5.33$ & $     0.02$ & $     9.92$ & $      160$ & $     2.2 \pm 
     0.8
$\\
J1702+6028 &  $
    6.16$ & $    -0.31$ & \nodata & $       10$ & $     0.9 \pm 
     0.1
$\\
J1727+5421 &  $
    5.68$ & $    -0.14$ & \nodata & $      290$ & $     1.6 \pm 
     0.6
$\\
J2156+1103 &  $
    6.45$ & $    -0.65$ & \nodata & $      230$ & $     0.6 \pm 
     0.1
$\\
J2321+0007 &  $
    6.56$ & $    -0.78$ & \nodata & $      290$ & $     2.2 \pm 
     0.8
$\\
J2338-0028 &  $
    5.92$ & $    -1.14$ & \nodata & $      540$ & $     2.0 \pm 
     0.1
$\\
\enddata
\tablecomments{Col. (1): Object name.  Col. (2):  Log $\mbh/\msun$, from Xiao et al. (2011).  Col. (3):  Log $L_{bol}/\ledd$ calculated from our H$\alpha_{BLR}$ measurements.   Col. (4):  Log $M_{*}/\msun$ for  objects with reported $I$-band luminosities from  \cite{2008ApJ...688..159G} (see \S\ref{sec:metal}).  Col. (5):  Density, calculated from the [\ion{S}{2}] doublet.  Col. (6):  $Z/\zsun$, as determined in \S\ref{sec:metal}.}
\label{tableGHmeas3}
\end{deluxetable}

\begin{deluxetable}{l r c r r l r l r c r l r l r c r }
\tablenum{4}
\tablecolumns{17}
\tabletypesize{\scriptsize}
\tablewidth{0pc}
\rotate
\tablecaption{Emission Line Measurements: H$\alpha$ Region}
\tablehead{\colhead{} & 
\multicolumn{4}{c}{[\ion{S}{2}]~$\lambda6717$} & \colhead{} & 
\colhead{[\ion{S}{2}]~$\lambda6731$} & \colhead{} & 
\multicolumn{3}{c}{H$\alpha_{NLR}$} & \colhead{} & 
\colhead{[\ion{N}{2}]~$\lambda6584$} & \colhead{} & 
\multicolumn{3}{c}{H$\alpha_{BLR}$}
\\
\cline{2-5} \cline{7-7} \cline{9-11} \cline{13-13} \cline{15-17}
\\
\colhead{Object} &
\colhead{$f/f$(H$\beta$)} &
\colhead{FWHM} &
\colhead{$h_{3}$} &
\colhead{$h_{4}$} &
\colhead{} &
\colhead{$f/f$(H$\beta$)} &
\colhead{} &
\colhead{$f/f$(H$\beta$)} &
\colhead{FWHM} &
\colhead{$\Delta$v} &
\colhead{} &
\colhead{$f/f$(H$\beta$)} &
\colhead{} &
\colhead{$f/f$(H$\beta$)} &
\colhead{FWHM} &
\colhead{$\Delta$v} }
\startdata
J0107+1408 &  $
    0.63 \pm     0.04$ & $      90 \pm       20$ & $   -0.03$ & $   -0.03
$ & & $
    0.57 \pm     0.04
$ & & $
     2.6 \pm      0.3$ & $     120 \pm       40$ & $       0
$ & & $
    0.63 \pm     0.04
$ & & $
    50.6 \pm      2.6$ & $     960 \pm      400$ & $     -10
$\\
J0249-0815 &  $
    0.66 \pm     0.04$ & $     110 \pm       40$ & $    0.01$ & $    0.05
$ & & $
    0.75 \pm     0.05
$ & & $
     3.8 \pm      0.3$ & $     130 \pm       30$ & $      23
$ & & $
    0.66 \pm     0.04
$ & & $
    18.3 \pm      1.0$ & $     790 \pm      250$ & $     -68
$\\
J0325+0034 &  $
    0.36 \pm     0.03$ & $     120 \pm       30$ & $   -0.08$ & $    0.01
$ & & $
    0.30 \pm     0.03
$ & & $
     2.9 \pm      0.3$ & $     150 \pm       20$ & $       0
$ & & $
    0.36 \pm     0.03
$ & & $
    21.5 \pm      0.9$ & $    1070 \pm      190$ & $    -229
$\\
J0806+2419 &  $
    0.36 \pm     0.35$ & $      70 \pm      120$ & $   -0.13$ & $    0.10
$ & & $
    0.29 \pm     0.28
$ & & $
     6.0 \pm      5.8$ & $     220 \pm      140$ & $      -2
$ & & $
    0.68 \pm     0.65
$ & & $
   104.7 \pm    100.3$ & $    1120 \pm      210$ & $      31
$\\
J0809+4416 &  $
    0.67 \pm     0.02$ & $     160 \pm       10$ & $   -0.01$ & $   -0.04
$ & & $
    0.53 \pm     0.02
$ & & $
     3.4 \pm      0.1$ & $     180 \pm       20$ & $      12
$ & & $
    0.67 \pm     0.02
$ & & $
     9.7 \pm      0.2$ & $    1020 \pm       80$ & $      81
$\\
J0815+2506 &  $
    0.56 \pm     0.03$ & $     130 \pm       40$ & $   -0.10$ & $    0.04
$ & & $
    0.45 \pm     0.03
$ & & $
     2.9 \pm      0.2$ & $     150 \pm       40$ & $       0
$ & & $
    0.56 \pm     0.03
$ & & $
    15.9 \pm      0.7$ & $     840 \pm      110$ & $     -22
$\\
J0824+2959 &  $
    0.64 \pm     0.02$ & $     270 \pm       10$ & $    0.05$ & $   -0.13
$ & & $
    0.64 \pm     0.02
$ & & $
     6.1 \pm      0.2$ & $     310 \pm       10$ & $      24
$ & & $
    0.64 \pm     0.02
$ & & $
     3.3 \pm      0.1$ & $    1240 \pm       20$ & $    -236
$\\
J0829+5006 &  $
    0.48 \pm     0.01$ & $     150 \pm       10$ & $    0.02$ & $   -0.05
$ & & $
    0.41 \pm     0.01
$ & & $
     2.7 \pm      0.1$ & $     160 \pm       10$ & $       0
$ & & $
    0.48 \pm     0.01
$ & & $
    13.5 \pm      0.2$ & $     690 \pm       30$ & $     -33
$\\
J0943+6045 &  $
    >0.29 \pm     0.10$ & $     140 \pm       30$ & $   -0.08$ & $   -0.02
$ & & $
    >0.23 \pm     0.08
$ & & $
     >1.4 \pm      0.5$ & $     150 \pm       40$ & $      59
$ & & $
    >0.45 \pm     0.16
$ & & $
    >48.1 \pm     16.1$ & $     660 \pm      420$ & $     -22
$\\
J1011+0029 &  $
    0.20 \pm     0.02$ & $     100 \pm       20$ & $   -0.01$ & $    0.04
$ & & $
    0.16 \pm     0.02
$ & & $
     2.7 \pm      0.2$ & $     210 \pm       20$ & $      81
$ & & $
    0.20 \pm     0.02
$ & & $
    28.8 \pm      0.8$ & $    1220 \pm      250$ & $      46
$\\
J1057+4825 &  $
    0.60 \pm     0.02$ & $      80 \pm       10$ & $   -0.04$ & $    0.06
$ & & $
    0.44 \pm     0.02
$ & & $
     3.8 \pm      0.1$ & $      70 \pm       10$ & $      23
$ & & $
    0.60 \pm     0.02
$ & & $
     3.6 \pm      0.3$ & $     640 \pm      140$ & $     -45
$\\
J1105+5941 &  $
    0.80 \pm     0.02$ & $     300 \pm       20$ & $    0.07$ & $    0.06
$ & & $
    0.74 \pm     0.02
$ & & $
     4.1 \pm      0.1$ & $     350 \pm       20$ & $       0
$ & & $
    0.80 \pm     0.02
$ & & $
     3.4 \pm      0.1$ & $    4240 \pm      780$ & $    -125
$\\
J1143+5500 &  $
    0.61 \pm     0.01$ & $      70 \pm        10$ & $   -0.01$ & $    0.04
$ & & $
    0.43 \pm     0.01
$ & & $
     3.3 \pm      0.1$ & $      70 \pm        10$ & $      23
$ & & $
    0.21 \pm     0.01
$ & & $
     1.1 \pm      0.1$ & $    1290 \pm       70$ & $    -171
$\\
J1223+5814 &  $
    0.51 \pm     0.01$ & $     100 \pm       10$ & $   -0.04$ & $    0.00
$ & & $
    0.55 \pm     0.01
$ & & $
     4.6 \pm      0.1$ & $     140 \pm       20$ & $      12
$ & & $
    0.51 \pm     0.01
$ & & $
    12.2 \pm      0.2$ & $    1150 \pm      150$ & $     -22
$\\
J1240-0029 &  $
    0.51 \pm     0.01$ & $     140 \pm       10$ & $   -0.01$ & $    0.03
$ & & $
    0.47 \pm     0.01
$ & & $
     4.2 \pm      0.1$ & $     160 \pm       10$ & $       0
$ & & $
    0.51 \pm     0.01
$ & & $
    10.1 \pm      0.1$ & $     840 \pm       80$ & $    -183
$\\
J1250-0155 &  $
    0.71 \pm     0.01$ & $     110 \pm       10$ & $   -0.14$ & $    0.07
$ & & $
    0.56 \pm     0.01
$ & & $
     3.5 \pm      0.1$ & $     110 \pm       10$ & $      -2
$ & & $
    0.36 \pm     0.01
$ & & $
     1.9 \pm      0.1$ & $    2260 \pm      160$ & $      11
$\\
J1313+0519 &  $
    0.80 \pm     0.02$ & $     150 \pm       10$ & $   -0.01$ & $    0.03
$ & & $
    0.71 \pm     0.02
$ & & $
     3.5 \pm      0.1$ & $     180 \pm       10$ & $       0
$ & & $
    0.80 \pm     0.02
$ & & $
     6.6 \pm      0.2$ & $     720 \pm      180$ & $     -56
$\\
J1412-0035 &  $
    >0.30 \pm     0.11$ & $     150 \pm       30$ & $   -0.15$ & $    0.04
$ & & $
    >0.35 \pm     0.12
$ & & $
     >2.1 \pm      0.8$ & $     180 \pm       40$ & $    -215
$ & & $
    >0.53 \pm     0.18
$ & & $
    >57.5 \pm     19.2$ & $     670 \pm      370$ & $     -23
$\\
J1434+0338 &  $
    0.64 \pm     0.03$ & $     120 \pm       30$ & $   -0.05$ & $    0.04
$ & & $
    0.53 \pm     0.03
$ & & $
     3.4 \pm      0.2$ & $     120 \pm       20$ & $      23
$ & & $
    0.71 \pm     0.02
$ & & $
    14.4 \pm      0.4$ & $     860 \pm      190$ & $     -68
$\\
J1534+0408 &  $
    0.55 \pm     0.07$ & $      70 \pm       20$ & $    0.01$ & $    0.00
$ & & $
    0.49 \pm     0.07
$ & & $
     3.0 \pm      0.4$ & $     100 \pm       10$ & $      10
$ & & $
    0.55 \pm     0.07
$ & & $
    34.4 \pm      3.5$ & $     480 \pm      130$ & $      24
$\\
J1626+3501 &  $
    0.40 \pm     0.02$ & $      70 \pm       20$ & $   -0.02$ & $   -0.03
$ & & $
    0.32 \pm     0.01
$ & & $
     3.6 \pm      0.1$ & $      90 \pm       10$ & $      23
$ & & $
    0.68 \pm     0.02
$ & & $
    24.2 \pm      0.5$ & $     750 \pm       40$ & $     -45
$\\
J1631+2437 &  $
    0.56 \pm     0.02$ & $     180 \pm       20$ & $    0.02$ & $    0.07
$ & & $
    0.45 \pm     0.02
$ & & $
     3.4 \pm      0.1$ & $     210 \pm       20$ & $       0
$ & & $
    0.56 \pm     0.02
$ & & $
    10.5 \pm      0.2$ & $     600 \pm      140$ & $    -332
$\\
J1702+6028 &  $
    0.47 \pm     0.02$ & $     180 \pm       30$ & $   -0.06$ & $   -0.15
$ & & $
    0.33 \pm     0.02
$ & & $
     4.5 \pm      0.2$ & $     260 \pm       20$ & $     -22
$ & & $
    0.48 \pm     0.02
$ & & $
    31.5 \pm      0.9$ & $    1040 \pm      160$ & $    -321
$\\
J1727+5421 &  $
    0.41 \pm     0.02$ & $     150 \pm       60$ & $   -0.08$ & $   -0.01
$ & & $
    0.35 \pm     0.01
$ & & $
     3.1 \pm      0.2$ & $     200 \pm       70$ & $       0
$ & & $
    0.41 \pm     0.02
$ & & $
    12.3 \pm      0.3$ & $     810 \pm      690$ & $      12
$\\
J2156+1103 &  $
    0.18 \pm     0.01$ & $     160 \pm       30$ & $   -0.08$ & $   -0.01
$ & & $
    0.15 \pm     0.01
$ & & $
     2.8 \pm      0.1$ & $     190 \pm       20$ & $      11
$ & & $
    0.24 \pm     0.01
$ & & $
     6.7 \pm      0.1$ & $    1540 \pm      230$ & $     120
$\\
J2321+0007 &  $
    0.65 \pm     0.06$ & $     180 \pm      110$ & $   -0.05$ & $   -0.01
$ & & $
    0.56 \pm     0.07
$ & & $
     3.6 \pm      0.3$ & $     200 \pm       80$ & $      12
$ & & $
    0.65 \pm     0.05
$ & & $
    27.5 \pm      1.4$ & $    1670 \pm     1020$ & $    -114
$\\
J2338-0028 &  $
    1.22 \pm     0.01$ & $     120 \pm       10$ & $    0.01$ & $    0.07
$ & & $
    1.19 \pm     0.01
$ & & $
     3.8 \pm      0.1$ & $     130 \pm       10$ & $      12
$ & & $
    1.22 \pm     0.01
$ & & $
     2.6 \pm      0.1$ & $    1490 \pm       80$ & $     -33
$\\
\enddata
\tablecomments{Col. (1): Object name.  Cols. (2-6): Fitting parameters for [\ion{S}{2}]~$\lambda 6717$.  Col. (7): Flux of [\ion{S}{2}]~$\lambda 6731$, where all other parameters are fixed to [\ion{S}{2}]~$\lambda 6717$.  Cols. (8-10):  Fitting parameters for H$\alpha_{NLR}$, where $h_{\mathrm{3}}$ and $h_{\mathrm{4}}$ are fixed to [\ion{S}{2}]~$\lambda 6717$, and $\Delta$v is the centroid shift compared to [\ion{S}{2}].  Col. (11): Flux of [\ion{N}{2}], where all other parameters are fixed.  Cols. (12-14):  Parametric measurements for total H$\alpha_{\mathrm{BLR}}$ fit.  FWHMs and $\Delta$v are in km~s$^{-1}$, and fluxes are relative to H$\beta_{NLR}$, which can be found in Table~\ref{tableGHmeas2}.  Fluxes have been corrected for Galactic extinction using values from \cite{1998ApJ...500..525S} and the reddening law of \cite{1994ApJ...422..158O}.  Upper and lower limits are three sigma estimates.}
\label{tableGHmeas1}
\end{deluxetable}
\clearpage

\begin{deluxetable}{l c c r l r c r l r c r l c c r }
\tablenum{5}
\tablecolumns{16}
\tabletypesize{\scriptsize}
\tablewidth{0pc}
\rotate
\tablecaption{Emission Line Measurements: H$\beta$ Region}
\tablehead{\colhead{} & 
\multicolumn{3}{c}{[\ion{0}{3}]~$\lambda5007$} & \colhead{} & 
\multicolumn{3}{c}{H$\beta_{NLR}$} & \colhead{} & 
\multicolumn{3}{c}{\ion{He}{2}~$\lambda4686_{NLR}$} & \colhead{} & 
\multicolumn{3}{c}{H$\beta_{BLR}$}
\\
\cline{2-4} \cline{6-8} \cline{10-12} \cline{14-16}
\\
\colhead{Object} &
\colhead{$f/f$(H$\beta$)} &
\colhead{FWHM} &
\colhead{$\Delta$v} &
\colhead{} &
\colhead{$F$(H$\beta$)} &
\colhead{FWHM} &
\colhead{$\Delta$v} &
\colhead{} &
\colhead{$f/f$(H$\beta$)} &
\colhead{FWHM} &
\colhead{$\Delta$v} &
\colhead{} &
\colhead{$f/f$(H$\beta$)} &
\colhead{FWHM} &
\colhead{$\Delta$v} }
\startdata
J0107+1408 & $
    12.9 \pm      0.7$ & $     150 \pm       10$ & $       7
$ & & $
    0.14 \pm     0.01$ & $     160 \pm       30$ & $      30
$ & & $
    0.29 \pm     0.05$ & $     150 \pm       30$ & $     1.2
$ & & $
    14.3 \pm      0.8$ & $    1380 \pm      330$ & $      18
$\\
J0249-0815 & $
    11.1 \pm      0.6$ & $     110 \pm       10$ & $      30
$ & & $
    0.15 \pm     0.01$ & $     160 \pm       30$ & $      41
$ & & $
    0.33 \pm     0.05$ & $     150 \pm       30$ & $     1.3
$ & & $
     4.7 \pm      0.3$ & $    1190 \pm      380$ & $     -15
$\\
J0325+0034 & $
     6.3 \pm      0.3$ & $     100 \pm       10$ & $       7
$ & & $
    0.13 \pm     0.01$ & $     400 \pm       20$ & $      -4
$ & & $
    0.16 \pm     0.04$ & $     380 \pm       20$ & $    12.7
$ & & $
     3.8 \pm      0.2$ & $    3120 \pm      260$ & $     375
$\\
J0806+2419 & $
    12.0 \pm     11.5$ & $     180 \pm       20$ & $     -24
$ & & $
    0.08 \pm     0.06$ & $     230 \pm     1250$ & $      24
$ & & $
    0.13 \pm     0.29$ & $     220 \pm     1250$ & $    11.5
$ & & $
    38.4 \pm     36.9$ & $    1530 \pm      670$ & $      14
$\\
J0809+4416 & $
     3.2 \pm      0.1$ & $     160 \pm       10$ & $      -4
$ & & $
    0.56 \pm     0.01$ & $     190 \pm       20$ & $      41
$ & & $
    0.10 \pm     0.02$ & $     180 \pm       20$ & $     1.2
$ & & $
     2.8 \pm      0.1$ & $    1550 \pm      280$ & $     179
$\\
J0815+2506 & $
     3.6 \pm      0.2$ & $     130 \pm       20$ & $      18
$ & & $
    0.11 \pm     0.01$ & $     180 \pm       40$ & $      30
$ & & $
    0.14 \pm     0.03$ & $     170 \pm       40$ & $     1.3
$ & & $
     4.4 \pm      0.2$ & $    1410 \pm      440$ & $      30
$\\
J0824+2959 & $
    16.8 \pm      0.3$ & $     360 \pm        10$ & $      38
$ & & $
    2.83 \pm     0.03$ & $     340 \pm       10$ & $      32
$ & & $
    0.15 \pm     0.01$ & $     320 \pm       10$ & $    -0.2
$ & & $
     0.8 \pm      0.1$ & $     730 \pm       90$ & $     -32
$\\
J0829+5006 & $
     6.3 \pm      0.1$ & $     180 \pm        10$ & $      18
$ & & $
    1.56 \pm     0.01$ & $     200 \pm       10$ & $      18
$ & & $
    0.17 \pm     0.01$ & $     190 \pm       10$ & $     1.3
$ & & $
     3.0 \pm      0.1$ & $    1040 \pm       40$ & $      18
$\\
J0943+6045 & $
     >4.4 \pm      1.5$ & $     190 \pm       20$ & $     -19
$ & & $
    <0.27 \pm     0.08$ & $     160 \pm       40$ & $      47
$ & & $
    >0.53 \pm     0.18$ & $     600 \pm       40$ & $    26.9
$ & & $
    >14.6 \pm      4.9$ & $     760 \pm      240$ & $      47
$\\
J1011+0029 & $
     4.4 \pm      0.2$ & $     160 \pm       10$ & $      30
$ & & $
    0.19 \pm     0.01$ & $     360 \pm       30$ & $     110
$ & & $
    0.43 \pm     0.03$ & $     340 \pm       30$ & $     1.3
$ & & $
     7.3 \pm      0.3$ & $    1690 \pm      290$ & $      87
$\\
J1057+4825 & $
     1.7 \pm      0.1$ & $     300 \pm       50$ & $     -96
$ & & $
    0.34 \pm     0.01$ & $      80 \pm       20$ & $      41
$ & & $
    0.01 \pm     0.01$ & $      80 \pm       20$ & $    12.9
$ & & $
     0.6 \pm      1.5$ & $     720 \pm     1090$ & $      -4
$\\
J1105+5941 & $
     2.5 \pm      0.1$ & $     400 \pm       30$ & $      -4
$ & & $
    6.10 \pm     0.03$ & $     370 \pm       20$ & $      30
$ & & $
    0.03 \pm     0.01$ & $     350 \pm       20$ & $     1.2
$ & & $
     0.8 \pm      0.1$ & $    6740 \pm     1980$ & $     594
$\\
J1143+5500 & $
     1.5 \pm      0.1$ & $      70 \pm        10$ & $      30
$ & & $
    1.07 \pm     0.01$ & $      80 \pm        10$ & $      53
$ & & $
    0.01 \pm     0.01$ & $      70 \pm        10$ & $     1.4
$ & & $
     0.1 \pm      0.1$ & $     460 \pm       10$ & $     317
$\\
J1223+5814 & $
    12.3 \pm      0.2$ & $     100 \pm        10$ & $      18
$ & & $
    1.40 \pm     0.01$ & $     160 \pm       10$ & $      30
$ & & $
    0.31 \pm     0.01$ & $     150 \pm       10$ & $     1.3
$ & & $
     1.9 \pm      0.1$ & $    1730 \pm      400$ & $      41
$\\
J1240-0029 & $
     7.4 \pm      0.1$ & $     150 \pm       10$ & $      18
$ & & $
    1.99 \pm     0.01$ & $     170 \pm       10$ & $      30
$ & & $
    0.14 \pm     0.01$ & $     160 \pm       10$ & $     1.2
$ & & $
     2.4 \pm      0.1$ & $    1130 \pm      100$ & $    -130
$\\
J1250-0155 & $
     4.0 \pm      0.1$ & $     150 \pm       10$ & $       5
$ & & $
    1.79 \pm     0.01$ & $     120 \pm       10$ & $      26
$ & & $
    0.07 \pm     0.01$ & $     110 \pm       10$ & $     4.8
$ & & $
     0.6 \pm      0.1$ & $   10170 \pm     6670$ & $     563
$\\
J1313+0519 & $
     5.0 \pm      0.1$ & $     160 \pm       10$ & $      -4
$ & & $
    0.48 \pm     0.01$ & $     200 \pm       10$ & $      30
$ & & $
    0.14 \pm     0.01$ & $     190 \pm       10$ & $     1.3
$ & & $
     1.3 \pm      0.1$ & $    1110 \pm      260$ & $      -4
$\\
J1412-0035 & $
     >5.4 \pm      1.9$ & $     190 \pm       10$ & $     -21
$ & & $
    <0.16 \pm     0.04$ & $     120 \pm       30$ & $    -288
$ & & $
    >0.07 \pm     0.05$ & $     110 \pm       30$ & $     4.8
$ & & $
    >19.4 \pm      6.5$ & $     950 \pm      400$ & $       8
$\\
J1434+0338 & $
     6.1 \pm      0.2$ & $     140 \pm       10$ & $      -4
$ & & $
    0.31 \pm     0.01$ & $     150 \pm       20$ & $      41
$ & & $
    0.16 \pm     0.03$ & $     140 \pm       20$ & $     1.2
$ & & $
     2.8 \pm      0.1$ & $    1220 \pm      270$ & $      64
$\\
J1534+0408 & $
     6.7 \pm      0.7$ & $      80 \pm       10$ & $      18
$ & & $
    0.05 \pm     0.01$ & $     200 \pm       20$ & $      40
$ & & $
    0.13 \pm     0.08$ & $     190 \pm       10$ & $     8.6
$ & & $
     9.5 \pm      1.0$ & $     500 \pm      210$ & $      49
$\\
J1626+3501 & $
     3.7 \pm      0.1$ & $      90 \pm      130$ & $      18
$ & & $
    0.36 \pm     0.01$ & $     100 \pm       10$ & $      53
$ & & $
    0.12 \pm     0.02$ & $     100 \pm       10$ & $     1.3
$ & & $
     3.9 \pm      0.2$ & $     910 \pm      150$ & $       7
$\\
J1631+2437 & $
     6.2 \pm      0.1$ & $     210 \pm       10$ & $      18
$ & & $
    2.13 \pm     0.03$ & $     220 \pm       20$ & $      30
$ & & $
    0.13 \pm     0.02$ & $     210 \pm       20$ & $     1.2
$ & & $
     3.2 \pm      0.1$ & $     400 \pm       50$ & $    -326
$\\
J1702+6028 & $
     4.4 \pm      0.2$ & $     200 \pm       10$ & $     -27
$ & & $
    0.76 \pm     0.02$ & $     420 \pm       30$ & $     -39
$ & & $
    0.13 \pm     0.03$ & $     400 \pm       30$ & $    24.2
$ & & $
     5.6 \pm      0.2$ & $    3890 \pm      940$ & $     -15
$\\
J1727+5421 & $
     5.3 \pm      0.2$ & $     170 \pm       20$ & $      18
$ & & $
    0.64 \pm     0.01$ & $     250 \pm       40$ & $      30
$ & & $
    0.24 \pm     0.02$ & $     240 \pm       40$ & $     1.4
$ & & $
     2.8 \pm      0.1$ & $    1300 \pm      690$ & $      53
$\\
J2156+1103 & $
     0.5 \pm      0.1$ & $     440 \pm      170$ & $    -201
$ & & $
    1.82 \pm     0.01$ & $     220 \pm       20$ & $      45
$ & & $
    <0.31 \pm     0.11$ & $      50 \pm       20$ & \nodata
 & & $
     1.0 \pm      0.2$ & $    4260 \pm      470$ & $    -408
$\\
J2321+0007 & $
     4.5 \pm      0.3$ & $     210 \pm       40$ & $     -15
$ & & $
    0.08 \pm     0.01$ & $     250 \pm       70$ & $      30
$ & & $
    0.21 \pm     0.04$ & $     230 \pm       70$ & $     1.2
$ & & $
     8.3 \pm      0.5$ & $    3930 \pm     1200$ & $      76
$\\
J2338-0028 & $
    13.6 \pm      0.1$ & $     120 \pm        10$ & $      30
$ & & $
    1.29 \pm     0.01$ & $     140 \pm       10$ & $      41
$ & & $
    0.37 \pm     0.01$ & $     130 \pm       10$ & $     1.2
$ & & $
     0.5 \pm      0.1$ & $    1200 \pm      470$ & $       7
$\\
\enddata
\tablecomments{Col. (1): Object name.  Cols. (2-4): Parametric measurements for total [\ion{O}{3}]~$\lambda {5007}$ fit.  Cols. (5-7):  Fitting parameters for H$\beta_{NLR}$.  Note that $F$(H$\beta$) is in units of $10^{-15}~$ergs~s\per~cm\persq~\AA{}\per, and $h_{\mathrm{3}}$ and $h_{\mathrm{4}}$ are fixed to [\ion{S}{2}]~$\lambda 6717$.  Cols. (8-10): Fitting parameters for [\ion{He}{2}]$_{NLR}$, $h_{\mathrm{3}}$ and $h_{\mathrm{4}}$ are fixed to [\ion{S}{2}]~$\lambda 6717$.  Cols. (11-13):  Parametric measurements for total H$\beta_{BLR}$ fit.  FWHMs and $\Delta$v are in km~s$^{-1}$, and fluxes are relative to H$\beta_{NLR}$.  Fluxes have been corrected for Galactic extinction using values from \cite{1998ApJ...500..525S} and the reddening law of \cite{1994ApJ...422..158O}.  Upper and lower limits are three sigma estimates.}
\label{tableGHmeas2}
\end{deluxetable}
\clearpage

\clearpage

\clearpage

\clearpage

\clearpage

\clearpage

\clearpage

\begin{deluxetable}{l c c c c}
\tablenum{6}
\tablecolumns{4}
\tabletypesize{\footnotesize}
\tablewidth{0pc}
\tablecaption{Composite NLR Emission Lines}
\tablehead{\colhead{Line} & \colhead {$\lambda$}&
\colhead{FWHM} & 
\colhead{$f/f_{\mathrm H\beta}$} & 
\colhead{Ion. Pot.}}
\startdata
$[$\ion{Ne}{3}] & 
  3868.7 & 165 &     0.42 &     63.5
 \\
$[$\ion{Fe}{5}] & 
  3889.0 & 100 &     0.07 &     75.0
 \\
$[$\ion{Ne}{3}] & 
  3967.2 & 153 &     0.03 &     63.5
 \\
H$\epsilon$ & 
  3970.2 & 112 &     0.06 &     13.6
 \\
$[$\ion{S}{2}] & 
  4068.9 & 150 &     0.04 &     23.3
 \\
H$\delta$ & 
  4101.9 & 112 &     0.16 &     13.6
 \\
H$\gamma$ & 
  4340.6 & 112 &     0.26 &     13.6
 \\
$[$\ion{O}{3}] & 
  4362.9 & 158 &     0.13 &     54.9
 \\
\ion{He}{2} & 
  4685.9 & 199 &     0.19 &     54.4
 \\
H$\beta$ & 
  4861.1 & 208 &     1.00 &     13.6
 \\
$[$\ion{O}{3}] & 
  5006.7 & 157 &     4.35 &     54.9
 \\
$[$\ion{Fe}{7}] & 
  5720.3 & 272 &     0.03 &    125.0
 \\
\ion{He}{1} & 
  5875.8 & 115 &     0.09 &     24.5
 \\
$[$\ion{Fe}{7}] & 
  6085.6 & 223 &     0.05 &    125.0
 \\
$[$\ion{O}{1}] & 
  6300.3 & 112 &     0.19 &     13.6
 \\
$[$\ion{O}{1}] & 
  6364.1 & 112 &     0.07 &     13.6
 \\
$[$\ion{Fe}{10}] & 
  6371.4 & 459 &     0.05 &    262.1
 \\
$[$\ion{N}{2}] & 
  6548.1 & 162 &     0.38 &    29.6
 \\
H$\alpha$ & 
  6563.0 & 111 &     2.02 &     13.6
 \\
\ion{He}{1} & 
  6678.2 & 115 &     0.02 &     24.5
 \\
$[$\ion{S}{2}] & 
  6716.7 & 112 &     0.48 &     23.3
 \\
$[$\ion{S}{2}] & 
  6731.1 & 112 &     0.44 &     23.3
 \\
\ion{He}{1} & 
  7065.3 & 115 &     0.03 &     24.5
 \\
$[$\ion{Ar}{3}] & 
  7135.8 & 105 &     0.09 &     40.7
 \\
$[$\ion{Ar}{3}] & 
  7751.6 & 106 &     0.01 &     40.7
 \\
$[$\ion{O}{2}] & 
  7320.1 & 160 &     0.04 &     35.1
 \\
$[$\ion{O}{2}] & 
  7330.1 & 126 &     0.02 &     35.1
 \\
$[$\ion{Fe}{11}] & 
  7890.5 & 332 &     0.05 &    290.2
 \\
$[$\ion{O}{1}] & 
  8445.5 & 255 &     0.06 &     13.6
 \\
$[$\ion{S}{3}] & 
  9069.1 & 104 &     0.32 &     34.8
 \\
$[$\ion{S}{3}] & 
  9531.0 & 125 &     0.63 &     34.8
 \\
 \\
\enddata
\tablecomments{Narrow emission-line measurements of continuum-subtracted composite spectrum of low-mass AGN.  Col. (1):  Line ID.  Col. (2):  Measured centroid in \AA{}.  Col. (3):  FWHM in km~s\per.  Col. (4): Flux relative to narrow H$\beta$.  Col. (5):  Ionization potential in eV.}
\label{tablecomp}
\end{deluxetable}
\clearpage

\clearpage

\clearpage

\begin{deluxetable}{l c c c}
\tablenum{7}
\tablecolumns{4}
\tabletypesize{\footnotesize}
\tablewidth{0pc}
\tablecaption{Subset Properties}
\tablehead{\colhead{} & \colhead {}&
\multicolumn{2}{c}{Ave. $\sigma$ (km s$^{-1}$)} \\ \cline{3-4}
\colhead{Composite} & \colhead{No. of objects}& 
 \colhead{All~NLR} & \colhead{High-ion~Fe}}
\startdata
Whole sample & 27 & 
      70 &      135
 \\
\hline
High $L_{bol}$ & 14 & 
      98 &      188
 \\
Low $L_{bol}$ & 13 & 
      53 &       98
 \\
\hline
High $L_{bol}/L_{Edd}$ & 14 & 
      90 &      175
 \\
Low $L_{bol}/L_{Edd}$ & 13 & 
      55 &      102
 \\
\hline
Includes [O III] Wing & 17 & 
      85 &      136
 \\
No [O III] Wing & 10 &  
      66 &      131
 \\
\enddata
\tablecomments{NLR emission-line properties for subset composites.  Averages are taken over all measured NLR emission lines, or over high-ionization Fe lines (ionization potential~$>100$~eV) as noted.}
\label{subsets}
\end{deluxetable}
\clearpage

\clearpage

\clearpage

\clearpage


\begin{thebibliography}{105}
\expandafter\ifx\csname natexlab\endcsname\relax\def\natexlab#1{#1}\fi

\bibitem[{{Adelman-McCarthy} {et~al.}(2006){Adelman-McCarthy}, {Ag{\"u}eros},
  {Allam}, {et~al.}}]{2006ApJS..162...38A}
{Adelman-McCarthy}, J.~K., {Ag{\"u}eros}, M.~A., {Allam}, S.~S., {et~al.} 2006,
  \apjs, 162, 38

\bibitem[{{Asplund} {et~al.}(2004){Asplund}, {Grevesse}, {Sauval}, {Allende
  Prieto}, \& {Kiselman}}]{2004A&A...417..751A}
{Asplund}, M., {Grevesse}, N., {Sauval}, A.~J., {Allende Prieto}, C., \&
  {Kiselman}, D. 2004, \aap, 417, 751

\bibitem[{{Baldwin} {et~al.}(1981){Baldwin}, {Phillips}, \&
  {Terlevich}}]{1981PASP...93....5B}
{Baldwin}, J.~A., {Phillips}, M.~M., \& {Terlevich}, R. 1981, \pasp, 93, 5

\bibitem[{{Barth} {et~al.}(2005){Barth}, {Greene}, \&
  {Ho}}]{2005ApJ...619L.151B}
{Barth}, A.~J., {Greene}, J.~E., \& {Ho}, L.~C. 2005, \apjl, 619, L151

\bibitem[{{Barth} {et~al.}(2008){Barth}, {Greene}, \&
  {Ho}}]{2008AJ....136.1179B}
---. 2008, \aj, 136, 1179

\bibitem[{{Barth} {et~al.}(2004){Barth}, {Ho}, {Rutledge}, \&
  {Sargent}}]{2004ApJ...607...90B}
{Barth}, A.~J., {Ho}, L.~C., {Rutledge}, R.~E., \& {Sargent}, W.~L.~W. 2004,
  \apj, 607, 90

\bibitem[{{Baskin} \& {Laor}(2005)}]{2005MNRAS.358.1043B}
{Baskin}, A. \& {Laor}, A. 2005, \mnras, 358, 1043

\bibitem[{{Bell} {et~al.}(2003){Bell}, {McIntosh}, {Katz}, \&
  {Weinberg}}]{2003ApJS..149..289B}
{Bell}, E.~F., {McIntosh}, D.~H., {Katz}, N., \& {Weinberg}, M.~D. 2003, \apjs,
  149, 289

\bibitem[{{Bentz} {et~al.}(2006){Bentz}, {Peterson}, {Pogge}, {Vestergaard}, \&
  {Onken}}]{2006ApJ...644..133B}
{Bentz}, M.~C., {Peterson}, B.~M., {Pogge}, R.~W., {Vestergaard}, M., \&
  {Onken}, C.~A. 2006, \apj, 644, 133

\bibitem[{{Bonning} {et~al.}(2007){Bonning}, {Cheng}, {Shields}, {Salviander},
  \& {Gebhardt}}]{2007ApJ...659..211B}
{Bonning}, E.~W., {Cheng}, L., {Shields}, G.~A., {Salviander}, S., \&
  {Gebhardt}, K. 2007, \apj, 659, 211
  
\bibitem[{{Boroson}(2005)}]{2005AJ....130..381B}
{Boroson}, T. 2005, \aj, 130, 381

\bibitem[{{Boroson}(2002)}]{2002ApJ...565...78B}
{Boroson}, T.~A. 2002, \apj, 565, 78

\bibitem[{{Boroson} \& {Green}(1992)}]{1992ApJS...80..109B}
{Boroson}, T.~A. \& {Green}, R.~F. 1992, \apjs, 80, 109

\bibitem[{{Brinchmann} {et~al.}(2004){Brinchmann}, {Charlot}, {White},
  {et~al.}}]{2004MNRAS.351.1151B}
{Brinchmann}, J., {Charlot}, S., {White}, S.~D.~M., {et~al.} 2004, \mnras, 351,
  1151

\bibitem[{{Bromm} \& {Yoshida}(2011)}]{2011ARA&A..49..373B}
{Bromm}, V. \& {Yoshida}, N. 2011, \araa, 49, 373

\bibitem[{{Bruzual} \& {Charlot}(2003)}]{2003MNRAS.344.1000B}
{Bruzual}, G. \& {Charlot}, S. 2003, \mnras, 344, 1000

\bibitem[{{Calzetti} {et~al.}(2000){Calzetti}, {Armus}, {Bohlin},
  {et~al.}}]{2000ApJ...533..682C}
{Calzetti}, D., {Armus}, L., {Bohlin}, R.~C., {et~al.} 2000, \apj, 533, 682

\bibitem[{{Cecil} {et~al.}(2002){Cecil}, {Dopita}, {Groves},
  {et~al.}}]{2002ApJ...568..627C}
{Cecil}, G., {Dopita}, M.~A., {Groves}, B., {et~al.} 2002, \apj, 568, 627

\bibitem[{{Chabrier}(2003)}]{2003PASP..115..763C}
{Chabrier}, G. 2003, \pasp, 115, 763

\bibitem[{{Collin} {et~al.}(2006){Collin}, {Kawaguchi}, {Peterson}, \&
  {Vestergaard}}]{2006A&A...456...75C}
{Collin}, S., {Kawaguchi}, T., {Peterson}, B.~M., \& {Vestergaard}, M. 2006,
  \aap, 456, 75

\bibitem[{{Cooke} {et~al.}(1976){Cooke}, {Elvis}, {Ward},
  {et~al.}}]{1976MNRAS.177P.121C}
{Cooke}, B.~A., {Elvis}, M., {Ward}, M.~J., {et~al.} 1976, \mnras, 177, 121P

\bibitem[{{Croom} {et~al.}(2002){Croom}, {Rhook}, {Corbett},
  {et~al.}}]{2002MNRAS.337..275C}
{Croom}, S.~M., {Rhook}, K., {Corbett}, E.~A., {et~al.} 2002, \mnras, 337, 275

\bibitem[{{Davis} {et~al.}(2007){Davis}, {Woo}, \&
  {Blaes}}]{2007ApJ...668..682D}
{Davis}, S.~W., {Woo}, J.-H., \& {Blaes}, O.~M. 2007, \apj, 668, 682

\bibitem[{{De Robertis} \& {Osterbrock}(1984)}]{1984ApJ...286..171D}
{De Robertis}, M.~M. \& {Osterbrock}, D.~E. 1984, \apj, 286, 171

\bibitem[{{Desroches} {et~al.}(2009){Desroches}, {Greene}, \&
  {Ho}}]{2009ApJ...698.1515D}
{Desroches}, L., {Greene}, J.~E., \& {Ho}, L.~C. 2009, \apj, 698, 1515

\bibitem[Done et al.(2012)]{2012MNRAS.420.1848D} Done, C., Davis, S.~W., 
Jin, C., Blaes, O., \& Ward, M.\ 2012, \mnras, 420, 1848 

\bibitem[Dong et al.(2012)]{2012arXiv1206.3843D} Dong, X.-B., Ho, L.~C., 
Yuan, W., et al.\ 2012, \apj, in press (arXiv:1206.3843)

\bibitem[{{Dopita} {et~al.}(1982){Dopita}, {Binette}, \&
  {Schwartz}}]{1982ApJ...261..183D}
{Dopita}, M.~A., {Binette}, L., \& {Schwartz}, R.~D. 1982, \apj, 261, 183

\bibitem[{{Dopita} {et~al.}(2002){Dopita}, {Groves}, {Sutherland}, {Binette},
  \& {Cecil}}]{2002ApJ...572..753D}
{Dopita}, M.~A., {Groves}, B.~A., {Sutherland}, R.~S., {Binette}, L., \&
  {Cecil}, G. 2002, \apj, 572, 753

\bibitem[{{Evans}(1986)}]{1986ApJ...309..544E}
{Evans}, I.~N. 1986, \apj, 309, 544

\bibitem[{{Filippenko} \& {Halpern}(1984)}]{1984ApJ...285..458F}
{Filippenko}, A.~V. \& {Halpern}, J.~P. 1984, \apj, 285, 458

\bibitem[{{Filippenko} \& {Ho}(2003)}]{2003ApJ...588L..13F}
{Filippenko}, A.~V. \& {Ho}, L.~C. 2003, \apjl, 588, L13

\bibitem[{{Francis} {et~al.}(1991){Francis}, {Hewett}, {Foltz},
  {et~al.}}]{1991ApJ...373..465F}
{Francis}, P.~J., {Hewett}, P.~C., {Foltz}, C.~B., {et~al.} 1991, \apj, 373,
  465

\bibitem[{{Frank} {et~al.}(2002){Frank}, {King}, \&
  {Raine}}]{2002apa..book.....F}
{Frank}, J., {King}, A., \& {Raine}, D.~J. 2002, {Accretion Power in
  Astrophysics: Third Edition}, ed. {Frank, J., King, A., \& Raine, D.~J.}

\bibitem[{{Gelbord} {et~al.}(2009){Gelbord}, {Mullaney}, \&
  {Ward}}]{2009MNRAS.397..172G}
{Gelbord}, J.~M., {Mullaney}, J.~R., \& {Ward}, M.~J. 2009, \mnras, 397, 172

\bibitem[{{Grandi}(1978)}]{1978ApJ...221..501G}
{Grandi}, S.~A. 1978, \apj, 221, 501

\bibitem[{{Greene} \& {Ho}(2004)}]{2004ApJ...610..722G}
{Greene}, J.~E. \& {Ho}, L.~C. 2004, \apj, 610, 722

\bibitem[{{Greene} \& {Ho}(2005{\natexlab{a}})}]{2005ApJ...627..721G}
---. 2005{\natexlab{a}}, \apj, 627, 721

\bibitem[{{Greene} \& {Ho}(2005{\natexlab{b}})}]{2005ApJ...630..122G}
---. 2005{\natexlab{b}}, \apj, 630, 122

\bibitem[{{Greene} \& {Ho}(2007{\natexlab{a}})}]{2007ApJ...670...92G}
---. 2007{\natexlab{a}}, \apj, 670, 92

\bibitem[{{Greene} \& {Ho}(2007{\natexlab{b}})}]{2007ApJ...656...84G}
---. 2007{\natexlab{b}}, \apj, 656, 84

\bibitem[{{Greene} {et~al.}(2008){Greene}, {Ho}, \&
  {Barth}}]{2008ApJ...688..159G}
{Greene}, J.~E., {Ho}, L.~C., \& {Barth}, A.~J. 2008, \apj, 688, 159

\bibitem[{{Greene} {et~al.}(2006){Greene}, {Ho}, \&
  {Ulvestad}}]{2006ApJ...636...56G}
{Greene}, J.~E., {Ho}, L.~C., \& {Ulvestad}, J.~S. 2006, \apj, 636, 56

\bibitem[{{Greene} {et~al.}(2009){Greene}, {Zakamska}, {Liu}, {Barth}, \&
  {Ho}}]{2009ApJ...702..441G}
{Greene}, J.~E., {Zakamska}, N.~L., {Liu}, X., {Barth}, A.~J., \& {Ho}, L.~C.
  2009, \apj, 702, 441

\bibitem[{{Greene} {et~al.}(2012){Greene}, {Zakamska}, \&
  {Smith}}]{2012ApJ...746...86G}
{Greene}, J.~E., {Zakamska}, N.~L., \& {Smith}, P.~S. 2012, \apj, 746, 86

\bibitem[{{Groves}(2007)}]{2007ASPC..373..511G}
{Groves}, B. 2007, in Astronomical Society of the Pacific Conference Series,
  Vol. 373, The Central Engine of Active Galactic Nuclei, ed. {L.~C.~Ho \&
  J.-W.~Wang}, 511--+

\bibitem[{{Groves} \& {Allen}(2010)}]{2010NewA...15..614G}
{Groves}, B.~A. \& {Allen}, M.~G. 2010, \na, 15, 614

\bibitem[{{Groves} {et~al.}(2004{\natexlab{a}}){Groves}, {Dopita}, \&
  {Sutherland}}]{2004ApJS..153....9G}
{Groves}, B.~A., {Dopita}, M.~A., \& {Sutherland}, R.~S. 2004{\natexlab{a}},
  \apjs, 153, 9

\bibitem[{{Groves} {et~al.}(2004{\natexlab{b}}){Groves}, {Dopita}, \&
  {Sutherland}}]{2004ApJS..153...75G}
---. 2004{\natexlab{b}}, \apjs, 153, 75

\bibitem[{{Groves} {et~al.}(2006){Groves}, {Heckman}, \&
  {Kauffmann}}]{2006MNRAS.371.1559G}
{Groves}, B.~A., {Heckman}, T.~M., \& {Kauffmann}, G. 2006, \mnras, 371, 1559

\bibitem[{{G{\"u}ltekin} {et~al.}(2009){G{\"u}ltekin}, {Richstone}, {Gebhardt},
  {et~al.}}]{2009ApJ...698..198G}
{G{\"u}ltekin}, K., {Richstone}, D.~O., {Gebhardt}, K., {et~al.} 2009, \apj,
  698, 198

\bibitem[{{Hainline} {et~al.}(2011){Hainline}, {Shapley}, {Greene}, \&
  {Steidel}}]{2011ApJ...733...31H}
{Hainline}, K.~N., {Shapley}, A.~E., {Greene}, J.~E., \& {Steidel}, C.~C. 2011,
  \apj, 733, 31

\bibitem[{{Hamann} \& {Ferland}(1999)}]{1999ARA&A..37..487H}
{Hamann}, F. \& {Ferland}, G. 1999, \araa, 37, 487

\bibitem[{{Hao} {et~al.}(2005){Hao}, {Strauss}, {Tremonti},
  {et~al.}}]{2005AJ....129.1783H}
{Hao}, L., {Strauss}, M.~A., {Tremonti}, C.~A., {et~al.} 2005, \aj, 129, 1783

\bibitem[{{Heckman} {et~al.}(1981){Heckman}, {Miley}, {van Breugel}, \&
  {Butcher}}]{1981ApJ...247..403H}
{Heckman}, T.~M., {Miley}, G.~K., {van Breugel}, W.~J.~M., \& {Butcher}, H.~R.
  1981, \apj, 247, 403

\bibitem[{{Henry} \& {Worthey}(1999)}]{1999PASP..111..919H}
{Henry}, R.~B.~C. \& {Worthey}, G. 1999, \pasp, 111, 919

\bibitem[{{Ho}(2009)}]{2009ApJ...699..638H}
{Ho}, L.~C. 2009, \apj, 699, 638

\bibitem[{{Ho} {et~al.}(1997{\natexlab{a}}){Ho}, {Filippenko}, \&
  {Sargent}}]{1997ApJS..112..315H}
{Ho}, L.~C., {Filippenko}, A.~V., \& {Sargent}, W.~L.~W. 1997{\natexlab{a}},
  \apjs, 112, 315

\bibitem[{{Ho} {et~al.}(2003){Ho}, {Filippenko}, \&
  {Sargent}}]{2003ApJ...583..159H}
---. 2003, \apj, 583, 159

\bibitem[{{Ho} {et~al.}(1997{\natexlab{b}}){Ho}, {Filippenko}, {Sargent}, \&
  {Peng}}]{1997ApJS..112..391H}
{Ho}, L.~C., {Filippenko}, A.~V., {Sargent}, W.~L.~W., \& {Peng}, C.~Y.
  1997{\natexlab{b}}, \apjs, 112, 391

\bibitem[{{Ho} \& {Kim}(2009)}]{2009ApJS..184..398H}
{Ho}, L.~C. \& {Kim}, M. 2009, \apjs, 184, 398

\bibitem[{{Hu} {et~al.}(2008{\natexlab{a}}){Hu}, {Wang}, {Ho},
  {et~al.}}]{2008ApJ...683L.115H}
{Hu}, C., {Wang}, J., {Ho}, L.~C., {et~al.} 2008{\natexlab{a}}, \apjl, 683,
  L115

\bibitem[{{Hu} {et~al.}(2008{\natexlab{b}}){Hu}, {Wang}, {Ho},
  {et~al.}}]{2008ApJ...687...78H}
{Hu}, C., {Wang}, J.-M., {Ho}, L.~C., {et~al.} 2008{\natexlab{b}}, \apj, 687,
  78

\bibitem[{{Husemann} {et~al.}(2011){Husemann}, {Wisotzki}, {Jahnke}, \&
  {S{\'a}nchez}}]{2011A&A...535A..72H}
{Husemann}, B., {Wisotzki}, L., {Jahnke}, K., \& {S{\'a}nchez}, S.~F. 2011,
  \aap, 535, A72

\bibitem[Izotov 
\& Thuan(2008)]{2008ApJ...687..133I} Izotov, Y.~I., \& Thuan, T.~X.\ 2008, \apj, 687, 133 

\bibitem[{{Jiang} {et~al.}(2011){Jiang}, {Greene}, {Ho}, {Xiao}, \&
  {Barth}}]{2011ApJ...742...68J}
{Jiang}, Y.-F., {Greene}, J.~E., {Ho}, L.~C., {Xiao}, T., \& {Barth}, A.~J.
  2011, \apj, 742, 68

\bibitem[{{Kauffmann} {et~al.}(2003){Kauffmann}, {Heckman}, {Tremonti},
  {et~al.}}]{2003MNRAS.346.1055K}
{Kauffmann}, G., {Heckman}, T.~M., {Tremonti}, C., {et~al.} 2003, \mnras, 346,
  1055

\bibitem[{{Kewley} \& {Ellison}(2008)}]{2008ApJ...681.1183K}
{Kewley}, L.~J. \& {Ellison}, S.~L. 2008, \apj, 681, 1183

\bibitem[{{Kinney} {et~al.}(2000){Kinney}, {Schmitt}, {Clarke},
  {et~al.}}]{2000ApJ...537..152K}
{Kinney}, A.~L., {Schmitt}, H.~R., {Clarke}, C.~J., {et~al.} 2000, \apj, 537,
  152

\bibitem[{{Komossa} {et~al.}(2008){Komossa}, {Xu}, {Zhou}, {Storchi-Bergmann},
  \& {Binette}}]{2008ApJ...680..926K}
{Komossa}, S., {Xu}, D., {Zhou}, H., {Storchi-Bergmann}, T., \& {Binette}, L.
  2008, \apj, 680, 926

\bibitem[{{Kova{\v c}evi{\'c}} {et~al.}(2010){Kova{\v c}evi{\'c}},
  {Popovi{\'c}}, \& {Dimitrijevi{\'c}}}]{2010ApJS..189...15K}
{Kova{\v c}evi{\'c}}, J., {Popovi{\'c}}, L.~{\v C}., \& {Dimitrijevi{\'c}},
  M.~S. 2010, \apjs, 189, 15

\bibitem[{{Kraemer} {et~al.}(1999){Kraemer}, {Ho}, {Crenshaw}, {Shields}, \&
  {Filippenko}}]{1999ApJ...520..564K}
{Kraemer}, S.~B., {Ho}, L.~C., {Crenshaw}, D.~M., {Shields}, J.~C., \&
  {Filippenko}, A.~V. 1999, \apj, 520, 564

\bibitem[{{Leighly} \& {Moore}(2006)}]{2006ApJ...644..748L}
{Leighly}, K.~M. \& {Moore}, J.~R. 2006, \apj, 644, 748

\bibitem[{{Ludwig} {et~al.}(2009){Ludwig}, {Wills}, {Greene}, \&
  {Robinson}}]{2009ApJ...706..995L}
{Ludwig}, R.~R., {Wills}, B., {Greene}, J.~E., \& {Robinson}, E.~L. 2009, \apj,
  706, 995

\bibitem[{{Milosavljevi{\'c}} {et~al.}(2009){Milosavljevi{\'c}}, {Couch}, \&
  {Bromm}}]{2009ApJ...696L.146M}
{Milosavljevi{\'c}}, M., {Couch}, S.~M., \& {Bromm}, V. 2009, \apjl, 696, L146

\bibitem[{{Miniutti} {et~al.}(2009){Miniutti}, {Ponti}, {Greene},
  {et~al.}}]{2009MNRAS.394..443M}
{Miniutti}, G., {Ponti}, G., {Greene}, J.~E., {et~al.} 2009, \mnras, 394, 443

\bibitem[{{Mullaney} {et~al.}(2009){Mullaney}, {Ward}, {Done}, {Ferland}, \&
  {Schurch}}]{2009MNRAS.394L..16M}
{Mullaney}, J.~R., {Ward}, M.~J., {Done}, C., {Ferland}, G.~J., \& {Schurch},
  N. 2009, \mnras, 394, L16

\bibitem[{{M{\"u}ller-S{\'a}nchez} {et~al.}(2011){M{\"u}ller-S{\'a}nchez},
  {Prieto}, {Hicks}, {et~al.}}]{2011ApJ...739...69M}
{M{\"u}ller-S{\'a}nchez}, F., {Prieto}, M.~A., {Hicks}, E.~K.~S., {et~al.}
  2011, \apj, 739, 69

\bibitem[{{Netzer}(1990)}]{1990agn..conf...57N}
{Netzer}, H. 1990, in Active Galactic Nuclei, ed. {R.~D.~Blandford, H.~Netzer,
  L.~Woltjer, T.~J.-L.~Courvoisier, \& M.~Mayor}, 57--160

\bibitem[{{Netzer} \& {Trakhtenbrot}(2007)}]{2007ApJ...654..754N}
{Netzer}, H. \& {Trakhtenbrot}, B. 2007, \apj, 654, 754

\bibitem[{{O'Donnell}(1994)}]{1994ApJ...422..158O}
{O'Donnell}, J.~E. 1994, \apj, 422, 158

\bibitem[{{Osterbrock} \& {Ferland}(2006)}]{2006agna.book.....O}
{Osterbrock}, D.~E. \& {Ferland}, G.~J. 2006, {Astrophysics of gaseous nebulae
  and active galactic nuclei}, ed. {Osterbrock, D.~E.~\& Ferland, G.~J.}

\bibitem[{{Osterbrock} \& {Mathews}(1986)}]{1986ARA&A..24..171O}
{Osterbrock}, D.~E. \& {Mathews}, W.~G. 1986, \araa, 24, 171

\bibitem[{{Osterbrock} \& {Pogge}(1985)}]{1985ApJ...297..166O}
{Osterbrock}, D.~E. \& {Pogge}, R.~W. 1985, \apj, 297, 166

\bibitem[{{Pagel} \& {Edmunds}(1981)}]{1981ARA&A..19...77P}
{Pagel}, B.~E.~J. \& {Edmunds}, M.~G. 1981, \araa, 19, 77

\bibitem[{{Penston} \& {Fosbury}(1978)}]{1978MNRAS.183..479P}
{Penston}, M.~V. \& {Fosbury}, R.~A.~E. 1978, \mnras, 183, 479

\bibitem[{{Penston} {et~al.}(1984){Penston}, {Fosbury}, {Boksenberg}, {Ward},
  \& {Wilson}}]{1984MNRAS.208..347P}
{Penston}, M.~V., {Fosbury}, R.~A.~E., {Boksenberg}, A., {Ward}, M.~J., \&
  {Wilson}, A.~S. 1984, \mnras, 208, 347

\bibitem[{{Pettini} \& {Pagel}(2004)}]{2004MNRAS.348L..59P}
{Pettini}, M. \& {Pagel}, B.~E.~J. 2004, \mnras, 348, L59

\bibitem[{{Phillips} {et~al.}(1983){Phillips}, {Charles}, \&
  {Baldwin}}]{1983ApJ...266..485P}
{Phillips}, M.~M., {Charles}, P.~A., \& {Baldwin}, J.~A. 1983, \apj, 266, 485

\bibitem[{{Proga} {et~al.}(2000){Proga}, {Stone}, \&
  {Kallman}}]{2000ApJ...543..686P}
{Proga}, D., {Stone}, J.~M., \& {Kallman}, T.~R. 2000, \apj, 543, 686

\bibitem[{{Richards} {et~al.}(2003){Richards}, {Hall}, {Vanden Berk},
  {et~al.}}]{2003AJ....126.1131R}
{Richards}, G.~T., {Hall}, P.~B., {Vanden Berk}, D.~E., {et~al.} 2003, \aj,
  126, 1131

\bibitem[{{Rodr{\'{\i}}guez-Ardila} {et~al.}(2011){Rodr{\'{\i}}guez-Ardila},
  {Prieto}, {Portilla}, \& {Tejeiro}}]{2011ApJ...743..100R}
{Rodr{\'{\i}}guez-Ardila}, A., {Prieto}, M.~A., {Portilla}, J.~G., \&
  {Tejeiro}, J.~M. 2011, \apj, 743, 100

\bibitem[{{Salviander} {et~al.}(2007){Salviander}, {Shields}, {Gebhardt}, \&
  {Bonning}}]{2007ApJ...662..131S}
{Salviander}, S., {Shields}, G.~A., {Gebhardt}, K., \& {Bonning}, E.~W. 2007,
  \apj, 662, 131

\bibitem[{{Schlegel} {et~al.}(1998){Schlegel}, {Finkbeiner}, \&
  {Davis}}]{1998ApJ...500..525S}
{Schlegel}, D.~J., {Finkbeiner}, D.~P., \& {Davis}, M. 1998, \apj, 500, 525

\bibitem[{{Schmitt} {et~al.}(2003){Schmitt}, {Donley}, {Antonucci},
  {et~al.}}]{2003ApJ...597..768S}
{Schmitt}, H.~R., {Donley}, J.~L., {Antonucci}, R.~R.~J., {et~al.} 2003, \apj,
  597, 768

\bibitem[{{Shang} {et~al.}(2005){Shang}, {Brotherton}, {Green},
  {et~al.}}]{2005ApJ...619...41S}
{Shang}, Z., {Brotherton}, M.~S., {Green}, R.~F., {et~al.} 2005, \apj, 619, 41

\bibitem[{{Sheinis} {et~al.}(2002){Sheinis}, {Bolte}, {Epps},
  {et~al.}}]{2002PASP..114..851S}
{Sheinis}, A.~I., {Bolte}, M., {Epps}, H.~W., {et~al.} 2002, \pasp, 114, 851

\bibitem[{{Shields}(1978)}]{1978Natur.272..706S}
{Shields}, G.~A. 1978, \nat, 272, 706

\bibitem[{{Sutherland} \& {Dopita}(1993)}]{1993ApJS...88..253S}
{Sutherland}, R.~S. \& {Dopita}, M.~A. 1993, \apjs, 88, 253

\bibitem[{{Tremaine} {et~al.}(2002){Tremaine}, {Gebhardt}, {Bender},
  {et~al.}}]{2002ApJ...574..740T}
{Tremaine}, S., {Gebhardt}, K., {Bender}, R., {et~al.} 2002, \apj, 574, 740

\bibitem[{{Tremonti} {et~al.}(2004){Tremonti}, {Heckman}, {Kauffmann},
  {et~al.}}]{2004ApJ...613..898T}
{Tremonti}, C.~A., {Heckman}, T.~M., {Kauffmann}, G., {et~al.} 2004, \apj, 613,
  898

\bibitem[{{Vanden Berk} {et~al.}(2001){Vanden Berk}, {Richards}, {Bauer},
  {et~al.}}]{2001AJ....122..549V}
{Vanden Berk}, D.~E., {Richards}, G.~T., {Bauer}, A., {et~al.} 2001, \aj, 122,
  549

\bibitem[{{Verner}(2000)}]{2000PhDT........10V}
{Verner}, E. 2000, PhD thesis, UNIVERSITY OF TORONTO (CANADA)

\bibitem[{{Vestergaard} \& {Wilkes}(2001)}]{2001ApJS..134....1V}
{Vestergaard}, M. \& {Wilkes}, B.~J. 2001, \apjs, 134, 1

\bibitem[{{Volonteri} \& {Natarajan}(2009)}]{2009MNRAS.400.1911V}
{Volonteri}, M. \& {Natarajan}, P. 2009, \mnras, 400, 1911

\bibitem[{{Wang} \& {Wei}(2009)}]{2009ApJ...696..741W}
{Wang}, J. \& {Wei}, J.~Y. 2009, \apj, 696, 741

\bibitem[{{Ward} \& {Morris}(1984)}]{1984MNRAS.207..867W}
{Ward}, M. \& {Morris}, S. 1984, \mnras, 207, 867

\bibitem[{{Whittle}(1985)}]{1985MNRAS.213...33W}
{Whittle}, M. 1985, \mnras, 213, 33

\bibitem[{{Whittle}(1992)}]{1992ApJS...79...49W}
---. 1992, \apjs, 79, 49

\bibitem[{{Xiao} {et~al.}(2011){Xiao}, {Barth}, {Greene},
  {et~al.}}]{2011ApJ...739...28X}
{Xiao}, T., {Barth}, A.~J., {Greene}, J.~E., {et~al.} 2011, \apj, 739, 28

\bibitem[{{York} {et~al.}(2000){York}, {Adelman}, {Anderson},
  {et~al.}}]{2000AJ....120.1579Y}
{York}, D.~G., {Adelman}, J., {Anderson}, Jr., J.~E., {et~al.} 2000, \aj, 120,
  1579

\bibitem[{{Zhu} {et~al.}(2009){Zhu}, {Zhang}, \& {Tang}}]{2009ApJ...700.1173Z}
{Zhu}, L., {Zhang}, S.~N., \& {Tang}, S. 2009, \apj, 700, 1173

\end{thebibliography}
\end{document}